\documentclass[11pt]{article}
\usepackage{UF_FRED_paper_style}
\usepackage{indentfirst}
\usepackage{enumitem}
\usepackage{multirow}

\usepackage{rotating}
\usepackage{geometry}
\usepackage{pdflscape}
  
\usepackage{graphicx} 
\usepackage[labelfont=bf]{caption,subcaption}
\usepackage{subcaption}

\title{Minimum Wage Pass-through to Wholesale and Retail Prices: Evidence from Cannabis Scanner Data}

\author{Carl Hase\textsuperscript{1}
    } 
    
\date{This version: October 2023}

\begin{document}
{\setstretch{.8}
\maketitle
\thispagestyle{empty}
\begin{abstract}

A growing empirical literature finds that firms pass the cost of minimum wage hikes onto consumers via higher retail prices. Yet, little is known about minimum wage effects on wholesale prices and whether retailers face a wholesale cost shock in addition to the labor cost shock. I exploit the vertically disintegrated market structure of Washington state's legal recreational cannabis industry to investigate minimum wage pass-through to wholesale and retail prices. In a difference-in-differences with continuous treatment framework, I utilize scanner data on \$6 billion of transactions across the supply chain and leverage geographic variation in firms' minimum wage exposure across six minimum wage hikes between 2018 and 2021. When ignoring wholesale cost effects, I find retail pass-through elasticities consistent with existing literature---yet retail pass-through elasticities more than double once wholesale cost effects are accounted for. Retail markups do not adjust to the wholesale cost shock, indicating a full pass-through of the wholesale cost shock to retail prices. The results highlight the importance of analyzing the entire supply chain when evaluating the product market effects of minimum wage hikes.

\medskip

\noindent
{\textbf{Keywords: }%
Minimum wages, inflation, wholesale prices, retail prices, price dynamics, price pass-through.} 
\medskip\\ 
\noindent
{\textbf{JEL Classification:} E31, J23, J38, L11, L81}

\end{abstract}
}
\footnotetext[1]{Goethe University Frankfurt and JGU Mainz. Email: carlhase@stud.uni-frankfurt.de

I am grateful to Thorsten Schank, Thomas Otter, Giulia Giupponi, Guido Friebel, Sebastian Siegloch, Katja Kaufmann, Georg Duernecker, Sharat Ganapati, and seminar participants at EALE and the Goethe University Frankfurt for helpful comments and discussions. I thank Tim Haggerty and Chuck Groom for invaluable insights on the cannabis industry and the traceability data. I also thank Scott Bailey, Spencer Cohen, and the Washington State Employment Security Department for access to data.}
\newpage
\clearpage
\pagenumbering{arabic}


\section{Introduction}

Minimum wage laws are a popular tool for combating poverty and reducing economic inequality.\footnote{Approximately 90 percent of countries worldwide have instituted some form of minimum wage \citep{ilo2021}.} Yet, despite their pervasiveness, the question of 'who pays' for the minimum wage---i.e. firms, workers, or consumers---remains hotly debated. The answer to this question depends in part on how firms react to the labor cost shock induced by the minimum wage.\footnote{Recent evidence suggests that in some settings, workers' reactions to the minimum wage may also be important (see e.g. \cite{ku2022}).} If firms reduce employment or non-wage compensation (e.g. vacation days or health benefits) for low-wage workers, then low-wage workers bear the brunt of the policy. If firms absorb the cost shock by reducing profits, then firms bear the cost of adjustment. Finally, firms may pass the labor cost shock on to consumers in the form of higher retail prices, in which case consumers pay for the minimum wage increase. Of these three margins of adjustment, the first has received the lion’s share of attention, and evidence on employment effects is conflicted.\footnote{See Neumark (2019) for a recent overview.} The second channel has received less attention, but existing findings point to small profit effects \citep{draca2011,lindner2019}. Instead, recent evidence indicates that the third channel---price adjustment---plays a key role. With the aid of high-frequency price scanner data, a small but growing empirical literature finds that firms pass the cost shock through to retail prices, which suggests that nominal wage increases from minimum wage hikes are partly offset by increases in the prices of goods and services \citep{renkin2020, leung2021}. While price scanner data exhibits unparalleled richness, however, it is largely confined to grocery, merchandise, and drug stores.\footnote{One way to overcome this limitation is to use internet-based pricing. \cite{reich2018}, for example, exploit internet-based restaurant menus to study the effects of a 25\% minimum wage increase in San Jose, California in 2013. They find an average price increase of 1.45\%, indicating that most of the cost increase was passed on to consumers.} As a result, less is known about minimum wage pass-through to prices in other sectors. Yet, retail scanner data carries an additional shortcoming in that it only conveys information on prices at the final point of the supply chain. In principle, minimum wage hikes may affect not only retail outlets, but firms higher up the supply chain as well. If upstream labor costs are passed on to retailers via wholesale prices, then retailers will face not one, but two cost shocks from a minimum wage hike. The first is the higher labor cost of the retailer's own minimum wage employees---a direct effect. The second is higher wholesale prices---an indirect effect. To the extent that retailers pass both cost shocks on to consumers, retail price adjustment reflects both \emph{direct} and \emph{indirect pass-through}. The latter may even eclipse the former since in many retail settings the cost of goods sold (COGS) accounts for over 80\% of retailers' variable costs, and retail prices have been shown to be sensitive to even small changes in COGS \citep{eichenbaum2011,nakamura2010,renkin2020}. Crucially, retail scanner data cannot distinguish between these two forms of pass-through because the data only captures point of sale prices. Moreover, as pointed out by \cite{renkin2020}, reduced form regressions using retail scanner data and local variation in minimum wages only reveal the full (i.e. direct and indirect) pass-through to retail prices in the special case that retailers purchase predominantly from local wholesalers. If, instead, wholesale goods are highly tradeable (as with e.g. drugstores and general merchandise stores), then indirect pass-through to retail prices is absorbed by time fixed effects \citep{renkin2020}.\footnote{If wholesale goods are highly tradable then a minimum wage hike will increase wholesale prices equally for stores everywhere \citep{renkin2020}.} In that case, estimates from retail scanner data only pick up direct pass-through effects, and hence, fail to capture the full effect of minimum wage hikes on retail prices. To assess the true impact of minimum wage hikes on real wages, it is therefore crucial to examine both direct and indirect pass-through to retail prices.

In this paper, I investigate the impact of minimum wage increases on wholesale and retail prices in the legal recreational cannabis industry. Cannabis is a major consumer market in the U.S. In 2021, sales exceeded \$25 billion and cannabis was consumed by 42\% of U.S. adults aged 19-30 and 25\% of adults aged 35-50 \citep{mtf2022,leafly2022}. I focus on Washington state's cannabis market because it is an ideal laboratory for studying minimum wage pass-through. Cannabis is one of the largest agricultural industries in the state and a major source of employment. Cannabis production is labor-intensive and wages are low at all points of the supply chain, meaning minimum wage hikes induce a sizeable cost shock for producers and retailers alike. In other markets, the distinction between upstream and downstream firms is often blurred by vertical integration, making it difficult to distinguish between pass-through at different points of the supply chain. In contrast, the cannabis supply chain comprises two distinct types of businesses: producers---who also serve as wholesalers---and retailers.\footnote{In this paper, I use the term "wholesaler", "producer", and "producer-processor" interchangeably. This reflects the vertically disintegrated structure of the cannabis market and is described in more detail in section \ref{section:2}.} Crucially, vertical integration is strictly prohibited, creating clearly defined vertical relationships between producers and retailers. Furthermore, the cannabis industry operates under statewide autarky, meaning producers and retailers are subject to the very same minimum wage hikes. This narrows the set of possible confounders by eliminating the influence of labor and product market shocks in other regions, with the result being an unusually clean set of labor cost shocks along the entire supply chain. Finally, rich scanner data provides a close-up of price dynamics for the universe of products at both the wholesale and retail levels. This enables straightforward estimation of direct and indirect pass-through using a reduced-form approach.\footnote{The data also bypasses reliability issues associated with internal firm prices. For example, \cite{hong2017} argue that intrafirm prices may be vulnerable to accounting fictions for tax avoidance or record-keeping purposes.}

To estimate minimum wage pass-through elasticities, I construct establishment-level price indexes using monthly scanner-level data on \$6 billion of wholesale and retail transactions. I apply a difference-in-differences approach with continuous treatment that exploits geographic variation in minimum wage exposure for 1,192 wholesale and retail cannabis establishments over a set of predetermined minimum wage hikes between 2018 and 2021. Using separate producer and retailer panels, I first estimate direct pass-through elasticities for wholesale and retail prices under the assumption that the minimum wage only induces a labor cost shock. I find that a 10\% increase in the minimum wage translates into a 0.77\% increase in retail prices, consistent with existing literature (see e.g. \cite{leung2021}). Importantly, I also find that a 10\% increase in the minimum wage corresponds to a 1.66\% increase in wholesale prices, which confirms that retailers face a wholesale cost shock in addition to a labor cost shock. 

I then investigate the relative importance of the labor and wholesale cost shocks to retail price pass-through. I am aided by unusually rich data on prices and quantities for the universe of retailers’ wholesale purchases. This allows me to construct a shift-share instrument quantifying each retailer's exposure to the wholesale cost shock, which I use to separately identify direct and indirect pass-through to retail prices. Crucially, once the wholesale cost shock is accounted for, pass-through to retail prices more than doubles from 0.77\% to 2.04\% (from a 10\% hike). This increase reflects a dominance of indirect over direct pass-through, with elasticities that are proportional to retailers' wholesale and labor cost shares. The finding that, at least in the cannabis industry, the majority of retail pass-through stems from changes in retailers’ wholesale costs (i.e. indirect pass-through) rather than labor costs (i.e. direct pass-through) underscores the importance of examining the entire supply chain when investigating the effects of minimum wage hikes on retail prices.

Next, I study the role of markup adjustment in pass-through to retail prices. In particular, I show that retailers do not adjust markups to the increase in wholesale prices, indicating a full pass-through of the wholesale cost shock to retail prices. 

One concern in the cannabis industry is that rules governing production capacity for producers have created an uneven playing field in which small producers operate on slim margins while large establishments enjoy a higher degree of market power \citep{lcbtier1}. This suggests that large producers may be able to absorb cost shocks along other margins (e.g. by adjusting profits) and thereby exhibit less price pass-through compared to smaller producers. I test this directly and find that pass-through elasticities monotonically decrease with the scale of production---and are zero for the largest producers---consistent with an increased ability of larger firms to adjust to the cost shock along other margins. 

Finally, I examine other potential margins of adjustment to minimum wage hikes. I find no evidence of employment effects for retailers or producers, and I document no effect on cannabis consumption. The latter finding precludes demand-induced feedback as a mechanism through which minimum wage hikes affect retail cannabis prices.\footnote{There is some debate about the potential for demand-induced feedback from minimum wage hikes to retail prices. Using similar data and time periods, \cite{leung2021} finds evidence in favor of such an effect while \cite{renkin2020} do not.}

Taken together, my results highlight the importance of examining the entire supply chain---beyond the final point of sale---when investigating the price level effects of minimum wage hikes. Since minimum wage pass-through to retail prices attenuates the increase in real wages desired by policymakers, it is important to consider both direct and indirect pass-through when evaluating the efficacy of minimum wages as a policy tool.

I make three main contributions in this paper. First, I provide evidence that minimum wages affect retail \emph{and} wholesale prices. While previous studies have estimated minimum wage pass-through to retail prices, to the best of my knowledge, pass-through to wholesale prices has not been studied before. Importantly, my results imply that retailers face a direct and an indirect cost shock from minimum wage hikes. Since wholesale costs typically dominate labor costs in retailers' variable cost structure, measuring the indirect cost shock is crucial to understanding the total effect of minimum wages on retailers' costs.

Second, I explicitly quantify direct and indirect pass-through effects. This allows me to compare their magnitudes and relate them to retailers' wholesale and labor cost shares. Since data on wholesale costs are rarely available the literature does not distinguish between these two forms of pass-through, meaning it is generally not possible to determine whether previous estimates capture combined (i.e. direct and indirect) effects or direct pass-through only. To see this, note that it is common to estimate pass-through using a difference-in-differences model that relates store-level monthly inflation rates to the percent change in state-level minimum wages.\footnote{In such a setup, inflation in states with no minimum wage hike serves as the counterfactual for inflation in states with a minimum wage hike.} \cite{renkin2020} show that such a model delivers different pass-through measures depending on implicit assumptions regarding the nature of retailers' wholesale purchases. If retailers predominantly purchase from local producers (e.g. as is the case with grocery stores), then reduced-form regressions from statewide hikes will capture both direct and indirect pass-through to retail prices. However, in industries with highly tradeable goods (e.g. drugstores and general merchandise stores), the wholesale cost shock is common across all states and indirect pass-through is absorbed by time fixed effects. In that case, reduced form estimates only reflect direct pass-through and fail to capture the full impact of minimum wages on retail prices, and hence, real wages. I overcome this issue by exploiting variation in the minimum wage bite at the industry-by-county level. Since cannabis retailers and producers belong to different three-digit industries, estimates of direct pass-through to retail prices are independent of wholesale cost effects. Moreover, full information on retailers' wholesale transactions allows me to quantify retailers' exposure to the wholesale cost shock, which in turn enables me to estimate indirect pass-through independent of direct pass-through.

Third, the literature on minimum wage pass-through focuses on restaurants, grocery, drug, and merchandise stores. By investigating pass-through to cannabis prices, I provide novel insight on firms' margins of adjustment in a large agricultural market of growing importance.

This paper primarily relates to three strands of literature. The first is the small but growing literature on the product market effects of minimum wages, which until recently has centered on the restaurant industry (see e.g. \cite{aaronson2001, reich2018, fougere2010}). Most closely related are two papers by \cite{renkin2020} and \cite{leung2021}, who use high frequency scanner data to study the impact of a large number of state-level minimum wage hikes on consumer prices in the U.S. Both studies employ difference-in-differences with continuous treatment and find full and more than full pass-through to grocery prices, respectively, but no effect on prices at merchandise stores. I deviate from these studies by adopting an identification strategy that exploits geographic variation in the minimum wage bite at the industry level. Since cannabis retailers and producers belong to different industries, this allows me to separately identify direct pass-through to retail and wholesale prices. My paper is a natural extension of \cite{renkin2020}, as they consider the possibility that minimum wage hikes induce a wholesale cost shock, but they cannot test for it because their data does not include information on wholesale cost. Instead, they calculate an upper bound for indirect pass-through using input-output tables under the assumption of full pass-through to wholesale prices. In contrast, I directly observe wholesale cost because my data contains prices and quantities for the universe of retailers' wholesale transactions. I construct a measure of each retailer's exposure to the wholesale cost shock and estimate indirect pass-through to retail prices using a reduced-form shift-share approach.

By leveraging full information on retailers' wholesale costs, I also deviate from \cite{renkin2020} in how I quantify the degree of cost pass-through.\footnote{Since only a fraction of workers are affected by the minimum wage and since labor is only one contributor to marginal cost, a \$1.00 increase in the minimum wage does not translate to a \$1.00 increase in marginal cost. Therefore, minimum wage pass-through to prices is in and of itself not informative about the degree of cost pass-through.} \cite{renkin2020} divide the minimum wage pass-through elasticity by an estimate of the minimum wage elasticity of marginal cost, where the latter requires estimating the minimum wage elasticity of average wages. In contrast, since I observe the wholesale prices paid by each retailer---and hence markup over marginal input cost---I can directly test whether markups adjust to the wholesale cost shock. This provides a straightforward and precise method of quantifying the degree of cost pass-through. I find that markups do not adjust to the wholesale cost shock, indicating full cost pass-through to retail prices.


Second, the paper contributes to the literature on the transmission of cost shocks to firm pricing, much of which concerns exchange rate pass-through in specific industries (see \cite{burstein2014} for an overview). These papers typically combine separate wholesale and retail data sets and use structural models to infer pass-through of wholesale cost shocks to retail prices (see e.g. \citep{nakamura2010,bonnet2013}.\footnote{An exception is \cite{eichenbaum2011} who use data on prices and costs from a single U.S. retailer and find that retail price changes largely reflect changes in wholesale cost. \cite{hong2017} uses similar data to investigate the role of market structure on retail pass-through.} In contrast, my data uniquely identify both parties to each wholesale transaction and allow me to trace each product as it moves across the supply chain. As a result, I can estimate indirect pass-through directly from the data using a reduced form approach. More generally, I add to the literature on the transmission of upstream cost shocks by extending it to the minimum wage context.

Third, the paper contributes to the small but growing literature that uses the cannabis industry to investigate topics in industrial organization. Most closely related are two papers that study the role of the market structure on cannabis firm pricing. \cite{hollenbeck2021} consider the impact of cannabis license restrictions on retail market power while \cite{hansen2021} examine how a change in Washington's cannabis tax affected vertical integration among cannabis producers. I build on this literature by investigating the effects of minimum wages on cannabis pricing. In addition, I use scanner data from a newer administrative data software system that was introduced in early 2018. The newer data identifies products at the level of the stock keeping unit (SKU), which allows me to construct price indexes at a more granular level than was previously possible.

One issue with this type of analysis is the degree to which results from one industry can be used to infer pricing dynamics in other industries. I show that retail cannabis is remarkably similar to other industries studied in the literature. The variable cost structure for cannabis retailers is typical of grocery stores and other conventional retail settings, making the relative magnitudes of direct and indirect retail pass-through elasticities broadly applicable. Moreover, the price elasticity of demand---a key determinant of cost pass-through---has been shown to be similar for cannabis as for other industries studied in the literature \citep{hollenbeck2021}. Nevertheless, it is important to note that cannabis production is potentially more labor intensive than other agricultural industries due to the prevalence of small-scale indoor cultivation. Accordingly, the labor share of variable cost for cannabis producers is expected to be higher, and hence, the upstream labor cost shock induced by the minimum wage may be larger compared to other agricultural industries. Indeed, I chose to analyze the cannabis industry partly due to the labor intensive nature of cannabis cultivation, since identifying an upstream cost shock is a prerequisite for tracing pass-through of wholesale costs to retail prices.

This paper proceeds as follows. Section 2 describes the institutional context for the study. Section 3 details the data and the main empirical strategy. Section 4 presents direct pass-through estimates to wholesale and retail prices and discusses robustness checks. Section 5 investigates indirect pass-through to retail prices and compares them to the direct pass-through estimates from section 4. Section 6 further dissects pass-through by examining markups over marginal input cost and price effect heterogeneity. Section 7 investigates other possible margins of adjustment to minimum wage hikes including employment, productivity, and demand effects. Section 8 concludes.

\section{Institutional context}\label{section:2}

\subsection{The cannabis industry in Washington state}

In November 2012, voters in Washington state approved the creation of a legal recreational marijuana market for adults 21 years and older.\footnote{Cannabis production and consumption remains prohibited at the federal level. However, in August 2013, the United States Department of Justice announced that it would not interfere with state-level legalization as long as distribution and sales were strictly regulated by states. This effectively green-lit legalization for U.S. states.} Cannabis has since become a major agricultural industry in the state. In 2020, retail sales topped \$1.4 billion and the industry contributed \$1.85 billion to gross state product, making it the fourth most valuable agricultural crop in the state behind apples, wheat, and potatoes but ahead of timber, cherries, and hay \citep{wsu2020}. Cannabis is an important source of employment as the sector supports approximately 18,700 full-time equivalent (FTE) jobs in Washington \citep{wsu2020}. This mirrors the growing importance of cannabis employment in the U.S. more generally, where, according to one industry report, cannabis employs more than 428,000 workers \citep{leafly2022}.\footnote{To add perspective, there are more cannabis workers than hair stylists, barbers, and cosmetologists combined \citep{leafly2022}.} 

\subsubsection*{Cannabis labor}
Several features of cannabis labor make the industry particularly well-suited for investigating the effects of minimum wage hikes. First, cannabis is primarily grown in small indoor facilities in a setting that is averse to mechanization and more labor intensive than outdoor cultivation \citep{caulkins2010}. Most harvesting, drying, trimming, and packaging is done by hand, as this allows growers to produce higher quality buds that sell at a higher price point \citep{miller2022}. Second, wages in cannabis are very low---less than 1/3 to 1/2 of the statewide average wage---reflecting the low-skill nature of cannabis labor. Cannabis producers typically employ 1-2 `master growers', who manage cultivation systems and oversee harvesting, along with a much larger number of low-skill workers who harvest, trim, and package cannabis. At the retail level, a service counter forms a physical barrier between customers and the products, and customers can only make a purchase with the help of a sales representative known as a `budtender'. Budtending requires no formal training and the job resembles low-skilled retail employment in other industries. As a result, establishments at all points of the cannabis supply chain have a high degree of minimum wage exposure. Appendix \ref{appendix:g} describes labor and wages in cannabis in further detail. 

\subsubsection*{The cannabis market structure}
\begin{figure}[!htbp]
\caption{The supply chain in the Washington state cannabis market}
\centering
	\includegraphics[width=0.6\textwidth]{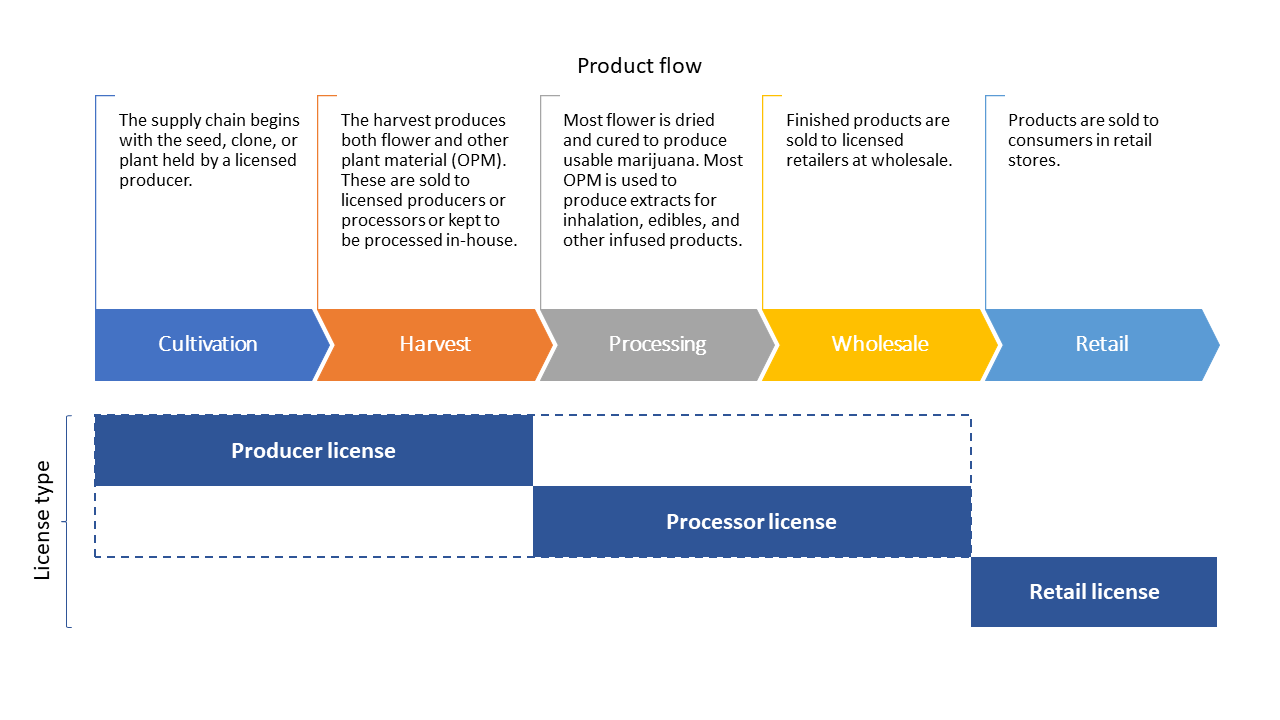}
\label{fig:0a}
\par \bigskip
\rule{\textwidth}{0.8pt}
\begin{minipage}[h]{\textwidth}
\medskip
\small \emph{Notes:} This figure depicts the flow of cannabis products, from left to right, as they move through the supply chain. Only licensed producers are permitted to cultivate and harvest cannabis plants; producers can only sell to licensed processors, who in turn are permitted to process products; only processors can sell finished products at wholesale to retailers; licensed retailers can sell finished products to end consumers. An establishment can jointly hold producer and processor licenses, so the overwhelming majority of upstream establishments hold both licenses (i.e. producer-processors). Retailers may not hold a producer or a processor license and vice versa. As a result, production and retail activities are legally separated.
\end{minipage}
\end{figure}

A defining feature of Washington state's cannabis industry is its unique market structure, depicted in Figure \ref{fig:0a}. The industry is regulated by the Washington State Liquor and Cannabis Board (LCB) which offers three separate licenses for cannabis businesses, each representing a different stage of the supply chain. The first license is for producers and it allows an establishment to cultivate, harvest, and package cannabis to be sold at wholesale to other licensed producers and processors. The second license is for processors and it permits an establishment to purchase cannabis from producers and process it into derivative subproducts (e.g. concentrate, edibles, etc). While there is some overlap between the producer and the processor licenses, the key distinction is that processors cannot cultivate plants and producers cannot sell to retailers. The third license is for retailers; they are permitted to sell usable cannabis products in retail stores. A key stipulation is that producer and processor licenses can be held simultaneously but retailers cannot obtain either producer or processor licenses.\footnote{Such `tied-house' rules are a remnant of the early days of U.S. alcohol regulation. They were imposed by states to limit the market power of brewers and distillers and prevent monopolies from preying on consumers' ``worst habits" \citep{wallach2014}. Washington state lawmakers adopted similar rules out of an abundance of caution and to increase the likelihood of legalization passing the state legislature \citep{wallach2014}.} As a result, the vast majority of upstream establishments own both producer and processor licenses and are commonly referred to as 'producer-processors'. Importantly, producer-processors may only sell cannabis to licensed retailers---they cannot sell directly to consumers. Retailers, moreover, can only sell to consumers. This creates a complete vertical separation between producer-processors on the one hand, and retailers on the other. 

Another feature of the cannabis market is that it operates under statewide autarky. That is, retailers can only buy from producer-processors located in Washington state, and producer-processors can only sell to retailers in the state. This seals off the core of the supply chain from other U.S. states with legal recreational markets.\footnote{The supply chain is not 100 percent sealed off, since consumers from other states can travel to Washington to purchase cannabis at retail stores, and producer-processors can purchase certain inputs such as grow lights, soil, and fertilizers from businesses in other states.} As a result, producer-processors and retailers are subject to the same minimum wage hikes, and hence, the same set of labor cost shocks.

Before moving on, it is worth noting several points. First, since the number of establishments with only a processor license (as opposed to a joint producer-processor license) is very small, I drop these from my analysis.\footnote{I keep establishments with only a producer license since these belong to the same industrial classification as producer-processors (see section \ref{section:3}).} Second, I use the term 'producer', 'producer-processor' and 'wholesaler' interchangeably throughout the paper to refer to upstream establishments. This reflects the dual role played by these establishments in the cannabis market since, besides being producers, they also act as wholesalers when viewed from the perspective of retailers. Third, since producers occupy the upstream portion of the supply chain, I assume that the minimum wage only induces a labor cost shock for producers---that is, the minimum wage does not affect material input prices for these firms. In principle, this assumption may not hold entirely and producers may be subject to minimum wage pass-through from their input suppliers. However, producer inputs like hydroponic systems, grow lights, and raw materials (e.g. soil or fertilizer) can be purchased from suppliers outside of Washington state, meaning minimum wage pass-through to producers' input prices is likely small. Therefore, for wholesale prices I only estimate direct pass-through, whereas for retail prices I estimate both direct and indirect pass-through.

\subsection{The minimum wage in Washington state}

\begin{figure}[!htbp]
\caption{Minimum wage hikes in Washington state, August 2018-July 2021}
\centering
	\includegraphics[width=0.4\textwidth]{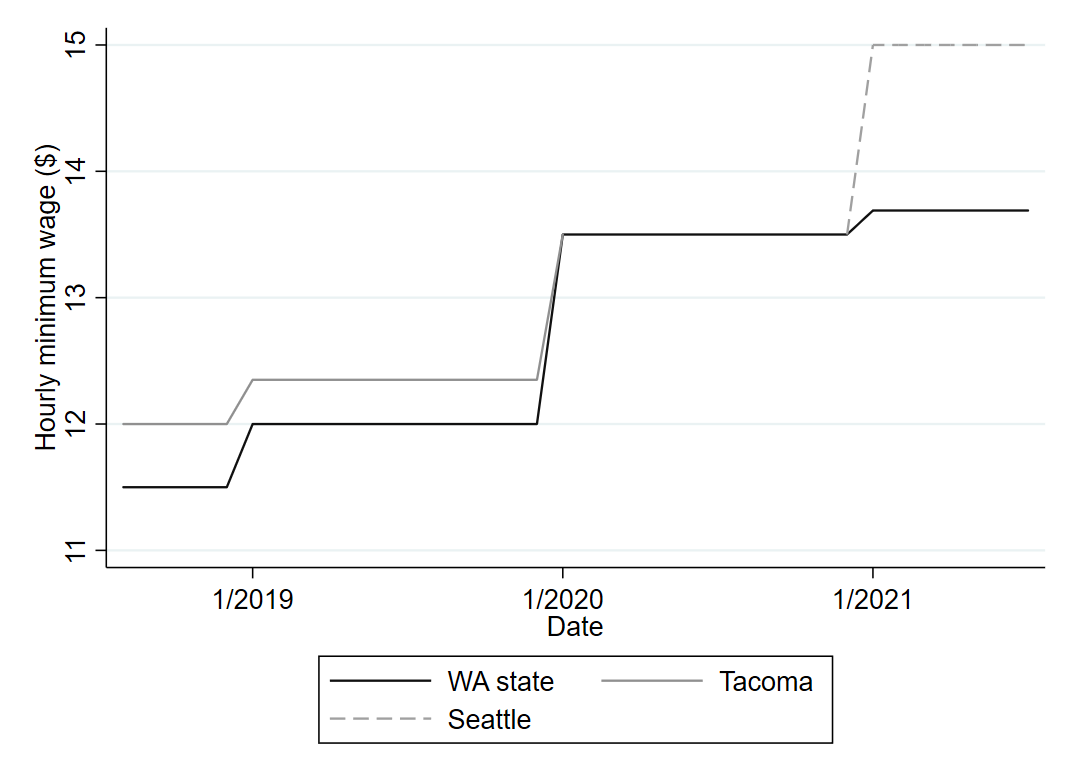}
\label{fig:0b}
\par \bigskip
\rule{\textwidth}{0.4pt}
\begin{minipage}[h]{\textwidth}
\medskip
\small \emph{Notes:} The figure depicts the minimum wage hikes for the sample period in my analysis (August 2018 through July 2021). The state minimum wage applies to all cities except Seattle and Tacoma. Tacoma's minimum wage converged with the state minimum wage on January 1, 2020. Seattle's minimum wage is depicted under the assumption that employers paid at least \$2.19/hour in benefits (the alternative schedule is depicted in appendix Figure \ref{fig:d.1}).
\end{minipage}
\end{figure}

Figure \ref{fig:0b} summarizes the minimum wage hikes used in my main analysis. In November 2016, Washington voters approved a ballot measure to scale up the state minimum wage from \$9.47 to \$13.50 by the year 2020. The measure spelled out predetermined, stepwise increases for January 1st of each year, with an initial increase to \$11.00 in 2017, then \$11.50 in 2018, \$12.00 in 2019, followed by the final increase to \$13.50 in 2020. Then, starting January 1st, 2021, the minimum wage was to adjust with the federal Consumer Price Index for Urban Wage Earners and Clerical Workers (CPI-W) on an annual basis. Besides the state minimum wage, there are two cities in Washington state with a binding citywide minimum wage. The city of Tacoma's minimum wage took effect in early 2016 with a predetermined schedule of annual increases designed such that the city and state minimum wages converged in 2020, with the latter binding for all subsequent years. Seattle's minimum wage went into effect in April 2015 and contained two sets of hikes depending on whether an employer paid towards an individual employee's medical benefits.\footnote{Firms with over 501 employees are subject to a higher minimum wage than small employers. Since no cannabis business in Seattle has more than 500 employees, the large employer minimum wage does not apply.} For employees earning \$2.19 per hour in benefits (on top of their hourly wage), the minimum wage was identical to the state minimum wage except for a larger (predetermined) jump to \$15 in 2021. In my main analysis, I assume that this is the schedule of hikes applicable to cannabis establishments in Seattle. However, in a series of robustness checks, I also consider the alternative schedule for employees earning less than \$2.19 in benefits. In that schedule, the minimum wage increased more steeply and reached \$15.75 in 2020, while in 2021 it adjusted according to a local CPI (this feature was written into the law in 2015). Due to the potential for reverse causality in that scenario, I drop Seattle establishments from the sample for the 2021 hike and find that results are unaffected (see appendix \ref{appendix:d} for details).\footnote{I also consider potential wage spillovers from Seattle to surrounding areas, in which case estimates for establishments in neighboring cities would also suffer from reverse causality in 2021. See appendix \ref{appendix:d} for details.} For both Seattle and Tacoma, the citywide hikes occurred on the same day of the year as the statewide hikes (January 1st).

\section{Data and empirical strategy}\label{section:3}

\subsection{Price data}

To monitor developments in the cannabis market, legalization came with stringent data reporting and sharing requirements for all licensed cannabis businesses. Producers and retailers are required to track every step of production from `seed to sale' and they must regularly upload data feeds about plants, harvests, processing, transfers between businesses, and retail sales to the LCB. The data, which is usually reported weekly, contains detailed information on the price and quantity of each product sold by a producer-processor to a retailer, and the subsequent price and quantity of that very same product sold at the retail level.\footnote{Compliance with seed-to-sale traceability is strictly enforced by the LCB. When a business is issued a violation, it can receive a fine, a temporary license suspension, or both. In cases of repeated violations, a license can be revoked by the LCB board. Given such strict enforcement, violations are uncommon. In 2021 for example, the LCB issued 66 violations among approximately 1,192 licensees. See: https://lcb.wa.gov/enforcement/violations-and-due-process} The LCB switched providers for its traceability system in October 2017 and again in December 2021, creating two structural breaks in the price data. My sample period lies between these breaks and spans August 2018 through July 2021, a period that covers three statewide and three citywide minimum wage hikes (see Figure \ref{fig:0b}). I obtained the data from Top Shelf Data, a data analytic firm that ingests the raw tracking data from the LCB and matches it with additional product information. The estimation sample covers sales from 1,192 distinct retail and producer establishments and contains an industry-wide average of 31,800 unique retail products and 18,268 unique wholesale products per month (see Table \ref{tab:1}). To give an example, a 1.0 gram package and a 2.0 gram package of Sunset Sherbert usable marijuana (dried flower) produced by Northwest Harvesting Co are treated as different products in the data.\footnote{Similar to how wines can be distinguished by the grape (e.g. Riesling, Chardonnay, etc), cannabis comes in many strains, which is 'Sunset Sherbert' in the given example.} The LCB classifies products into 12 categories. As Table \ref{tab:1} illustrates, usable marijuana and concentrate for inhalation account for more than 80\% of all retail sales. Another 14\% of retail sales comes from solid edibles (chocolate bars, cookies, etc), liquid edibles (soda and other infused drinks), and infused mix (e.g. pre-roll joints infused with concentrates). The remaining categories make up less than 2\% of total revenue; these are topical products (e.g. creams and ointments), packaged marijuana mix (e.g. pre-roll joints), capsules, tinctures, transdermal patches, sample jar, and suppository. Retailers are located in 37 counties while producers are located in 35 counties in Washington state (the state has 39 counties in total). Due to restrictions on the number of licenses a firm can hold, the vast majority are single-establishment firms. Over the entire sample period, the data contain \$4.47 billion and \$1.46 billion in retail and wholesale sales, respectively. Note that all retail cannabis sales are subject to a 37\% excise tax, and retail prices and revenues reported in this paper are tax-inclusive. In contrast, there is no tax on wholesale transactions.

\begin{table}[!htbp] 
\centering
\caption{Descriptive statistics for cannabis establishments}
\label{tab:1}
{
\def\sym#1{\ifmmode^{#1}\else\(^{#1}\)\fi}
\begin{tabular*}{\hsize}{@{\hskip\tabcolsep\extracolsep\fill}l*{2}{c}}
\toprule
\addlinespace
\addlinespace
\addlinespace

\multicolumn{3}{c}{(a) Sample totals} \\
\addlinespace
\midrule
&\multicolumn{1}{c}{Retail}&\multicolumn{1}{c}{Wholesale}\\
\midrule
Establishments     & 500    &  692  \\

\addlinespace
Units sold       &    232,133,427 & 228,423,415$^{+}$ \\

\addlinespace
Distinct products       &   172,688 & 147,273 \\

\addlinespace
Total revenue & \$4.47 billion & \$1.46 billion \\

\midrule
\addlinespace
\addlinespace
\addlinespace
\multicolumn{3}{c}{(b) Establishment monthly averages} \\
\addlinespace
\midrule
&\multicolumn{1}{c}{Retail}&\multicolumn{1}{c}{Wholesale}\\
\midrule
Distinct products      &   471  &   55* \\
\addlinespace
Revenue      &  \$304,032   &    \$106,634 \\
\addlinespace
Units sold      & 15,844    &  16,735*  \\

\midrule
\addlinespace
\addlinespace
\addlinespace
\multicolumn{3}{c}{(c) Market share by product category} \\
\addlinespace
\midrule
            &\multicolumn{1}{c}{Retail}&\multicolumn{1}{c}{Wholesale}\\
\midrule
\addlinespace
Usable marijuana      &     0.53 &     0.61  \\

\addlinespace

Concentrate for inhalation      &   0.31  & 0.28 \\
\addlinespace
Solid edible    &   0.07   &     0.03 \\
\addlinespace
Liquid edible       &   0.03  &  0.02  \\
\addlinespace
Infused mix     & 0.04    & 0.04  \\

\addlinespace
Other       &   0.02 & 0.02 \\

\bottomrule
\end{tabular*}
\begin{minipage}[h]{\textwidth}
\medskip
\small \emph{Notes:} This table displays summary statistics for the estimation sample. The sample period is August 2018 through July 2021. Panel (a) reports totals across all establishments and months in the sample. Panel (b) reports monthly averages at the establishment level. Sales from processor-only establishments are excluded. Sales between producer-processor establishments are included. Retail revenue is based on tax-inclusive prices. Panel (c) shows market shares for the product categories defined by the LCB. "Other" includes any category with less than 1 percent market share. These are: topical, packaged marijuana mix, capsules, tinctures, transdermal patches, sample jar, and suppository. Sales from processor-only establishments are excluded. Sales between producer-processor establishments are included. Data source: Top Shelf Data.Data source: Top Shelf Data.\\
$^{+}$ For producers, the LCB reports the unit weight for some product types (e.g. flower lots) in 1g units regardless of how the product is actually bundled. For such items, the number of units is the weight of the product in grams. As a result, the number of distinct products visible in the wholesale data is artificially low (since different unit weights are treated as a single product), and the number of units sold is artificially high.
\end{minipage}

}

\end{table}

\begin{figure}[!htbp]
\caption{Establishment-level inflation rates for cannabis, August 2018-July 2021}
\centering
	\begin{subfigure}{0.5\textwidth}
	\centering
			\includegraphics[width=\linewidth]{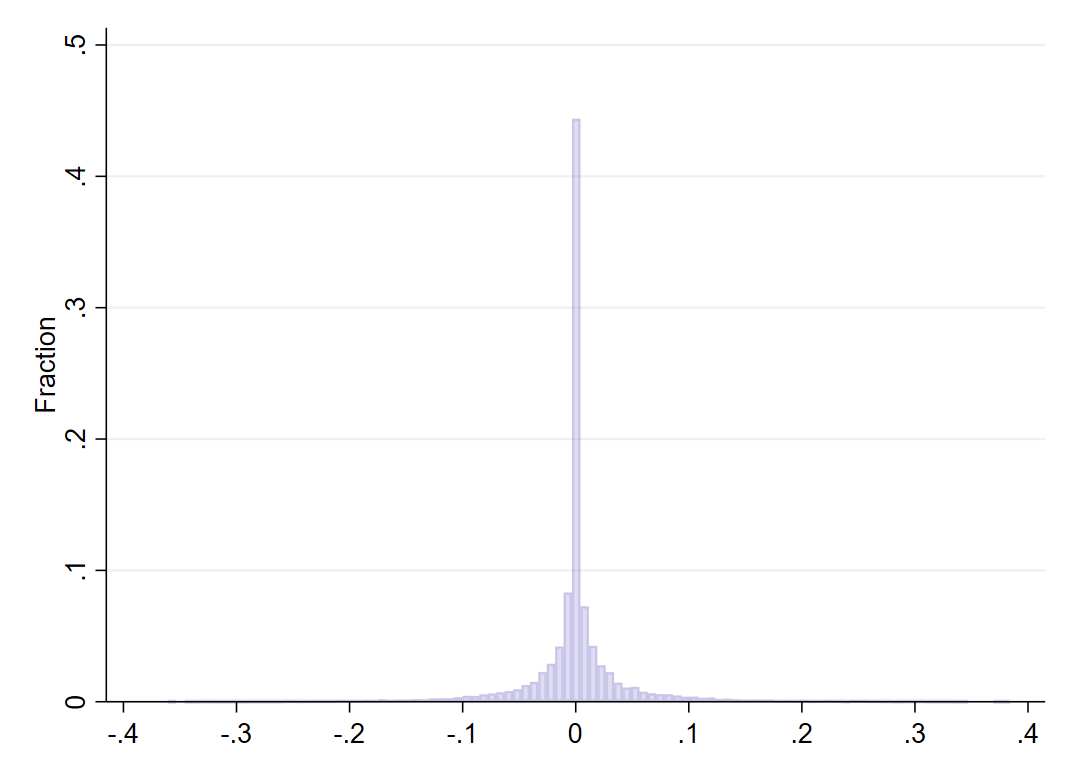}
		\caption{Monthly wholesale inflation rates}
		\label{fig:1a}		
	\end{subfigure}\hfil
	\begin{subfigure}{0.5\textwidth}
		\centering
        \includegraphics[width=\linewidth]{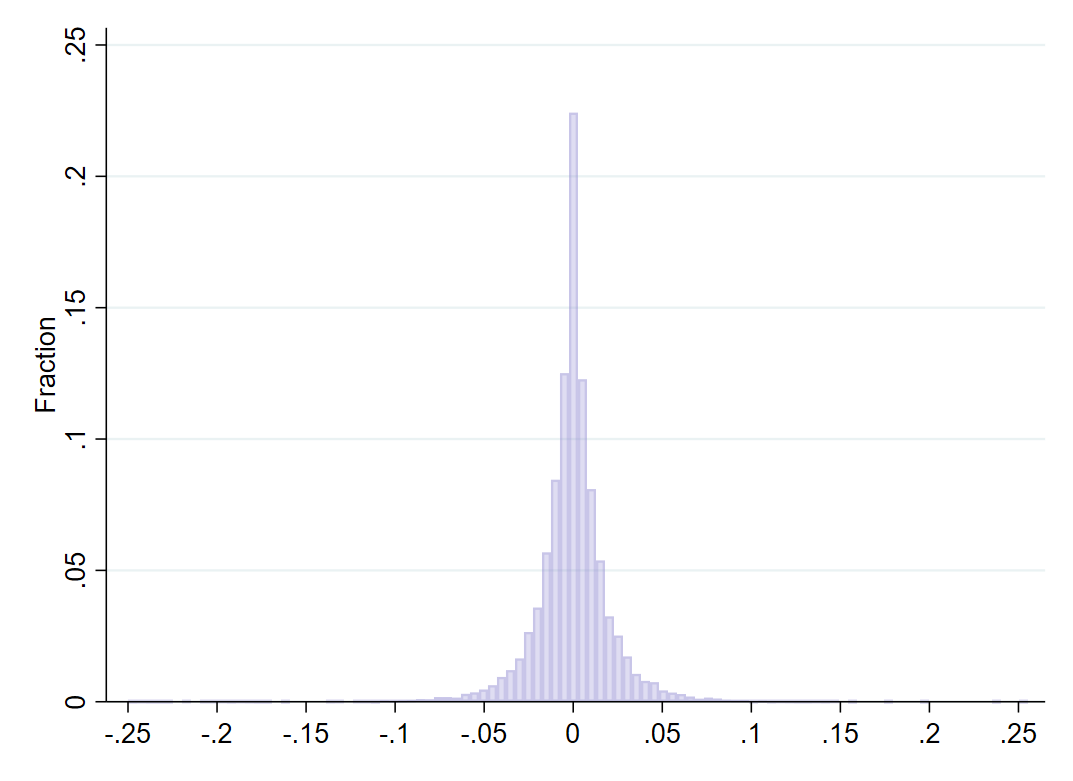}
		\caption{Monthly retail inflation rates}\label{fig:1b}
	\end{subfigure}
\label{fig:1}
\par \bigskip
\rule{\textwidth}{0.4pt}
\begin{minipage}[h]{\textwidth}
\medskip
\small Notes: The figures show the distribution of monthly establishment-level inflation rates for cannabis producerrs (Figure a) and retailers (Figure b) in the estimation sample. Data: Top Shelf Data, August 2018-July 2021.
\end{minipage}
\end{figure}


To estimate pass-through elasticities I follow previous studies (e.g. \cite{renkin2020,leung2021}) and define the dependent variable as the natural logarithm of the monthly establishment-level price index:
\begin{equation}
    \pi_{j,t} = \ln I_{j,t}, \, \text{with} \, I_{j,t} = \prod_c I_{c,j,t}^{\omega_{c,j,y(t)}}
\end{equation}
$\pi_{j,t}$ is the inflation rate for establishment $j$ in month $t$; $I_{j,t}$ is an establishment-level Lowe price index that aggregates price changes across product subcategories $c$; the weight $\omega_{c,j,y(t)}$ is the revenue share of subcategory $c$ in establishment $j$ during the calendar year of month $t$.\footnote{As pointed out by\cite{renkin2020}, price indexes are often constructed using lagged quantity weights. Since product turnover is high in cannabis retail, lagged weights would limit the number of products used in constructing the price indexes. Thus, contemporaneous weights are used.} To limit the potential impact of outliers, I trim inflation rates above the 99.5th and below the 0.5th percentile of the monthly distribution in my main specification (results are robust to keeping outliers). I describe the establishment-level price index in more detail in Appendix \ref{appendix:a}.

\subsection{Wage data}

\begin{figure}[!htbp]
\caption{Average minimum wage bite, 2018-2021}

\centering
	\begin{subfigure}{0.5\textwidth}
	\centering
			\includegraphics[width=\linewidth]{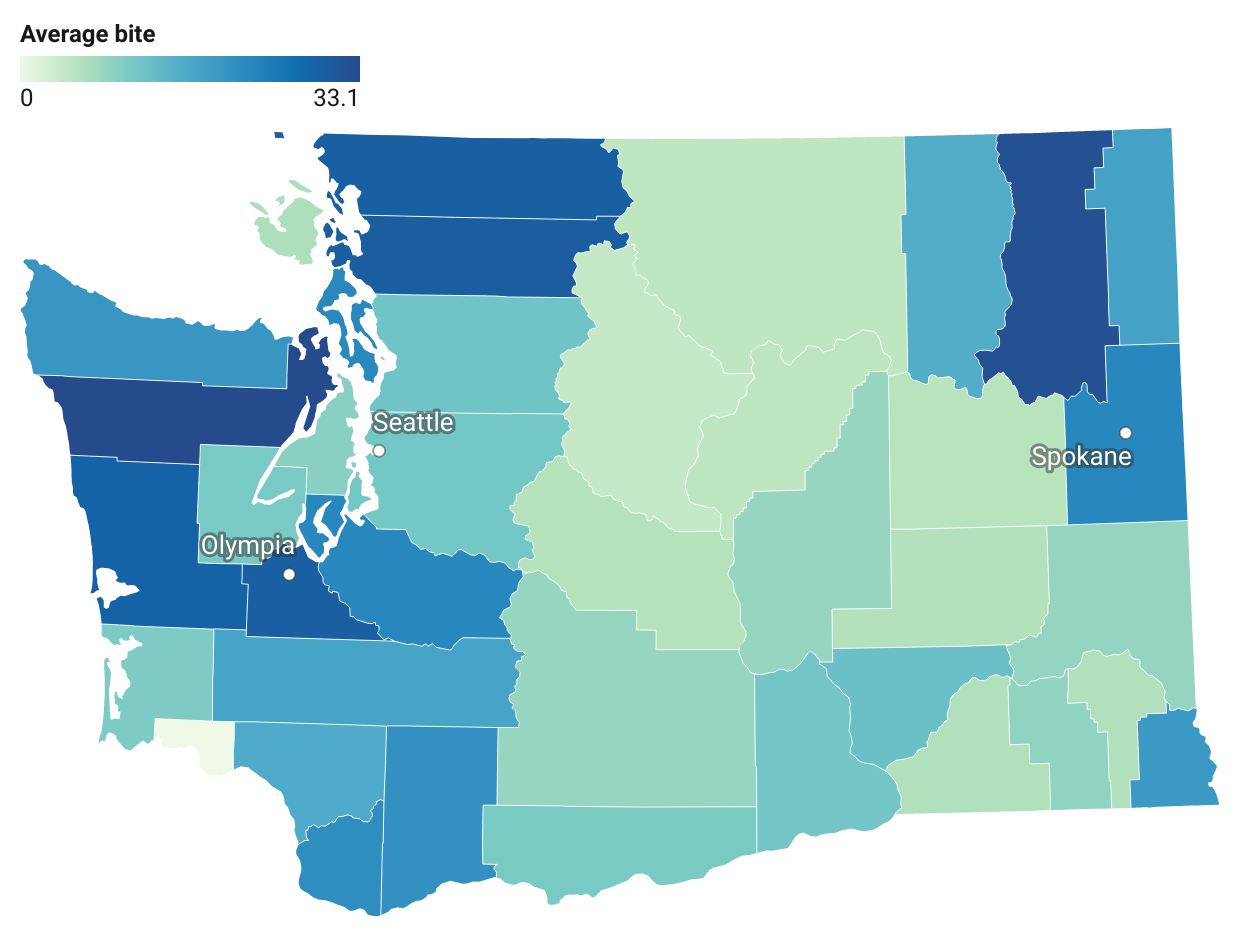}
		\caption{Crop production}
		\label{fig:2a}		
	\end{subfigure}\hfil
	\begin{subfigure}{0.5\textwidth}
		\centering
        \includegraphics[width=\linewidth]{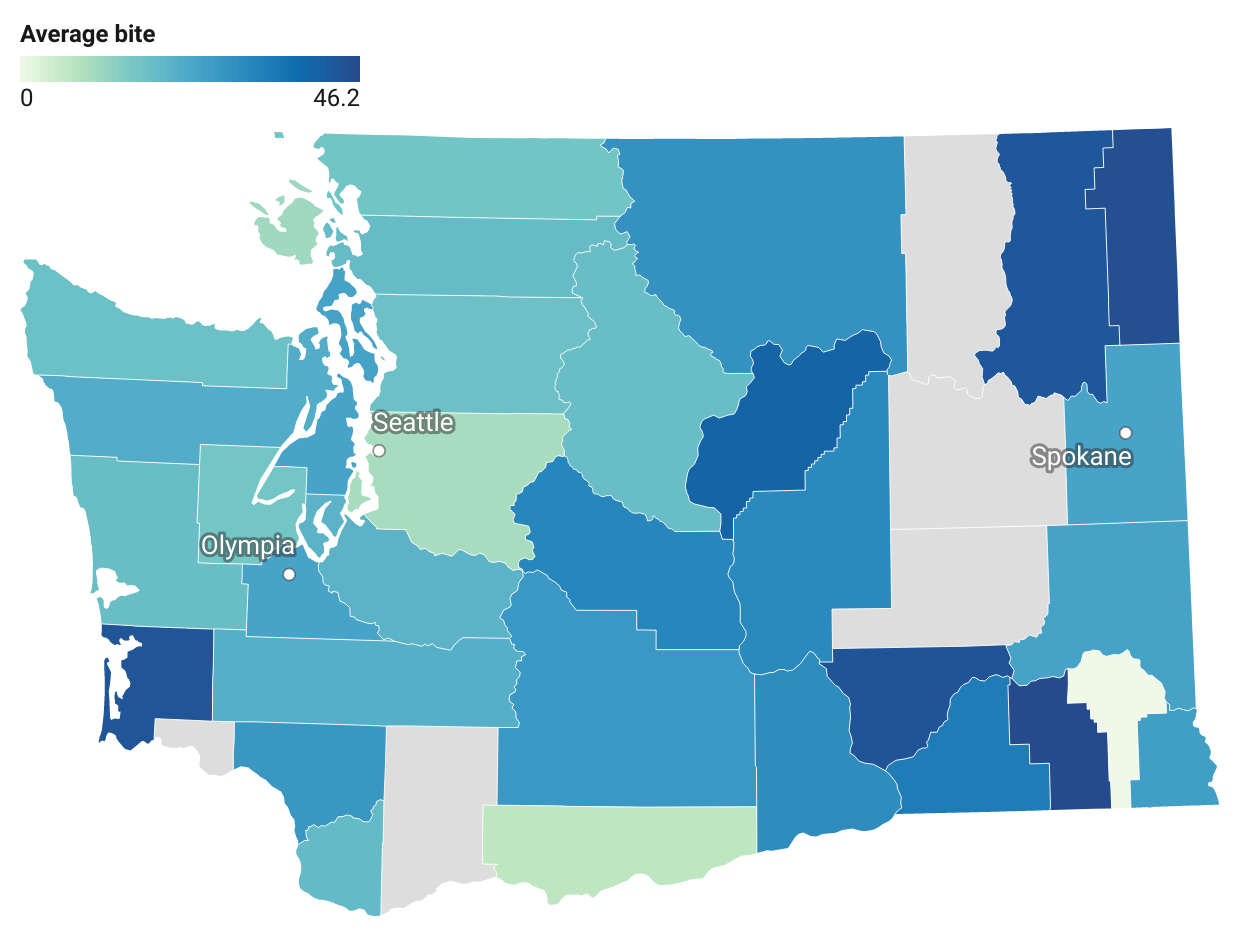}
		\caption{Misc. retail}\label{fig:2b}
	\end{subfigure}
\label{fig:2}
\par \bigskip
\rule{\textwidth}{0.4pt}
\begin{minipage}[h]{\textwidth}
\medskip
\small Notes: The figure shows average minimum wage bite for counties in Washington state over three statewide minimum wage hikes spanning 2019-2021. Bite is computed as the share of FTE earning below the new minimum wage two quarters prior to the hike. The panel on the left shows bite for crop production (NAICS 111), the industry that includes cannabis producers. The panel on the right shows bite for miscellaneous store retailers (NAICS 453), the industry that includes cannabis retailers. Counties in grey indicate the data do not meet ESD confidentiality standards---these counties are not included in my analysis. Data source: Washington ESD.
\end{minipage}
\end{figure}

My identification strategy rests on the idea that minimum wage hikes affect establishments with a high share of minimum wage workers more than those with a low share. Since wages are not observable at the establishment level, I follow previous studies and use geographic variation in the minimum wage bite as a proxy (see e.g. \cite{card1992,schank2022,leung2021,renkin2020,lindner2022}).\footnote{Firm-level wages are almost never observed in the minimum wage literature. An exception is \cite{ash2022}.} I define bite as the share of FTE workers in an industry-county earning below the new minimum wage two quarters prior to the hike. The industries are based on the North American Industrial Classification System (NAICS) which explicitly spells out classification for cannabis establishments of various types. NAICS 453 ("Miscellaneous store retailers") captures all cannabis retailers since NAICS 453998 includes "All Other Miscellaneous Store Retailers (except Tobacco Stores), including Marijuana Stores, Medicinal and Recreational" \citep{naics2007}. NAICS 111 ("Crop production") captures cannabis producers, since NAICS 111998 includes "All Other Miscellaneous Crop Farming, including Marijuana Grown in an Open Field" and NAICS 111419 includes "Other Food Crops Grown Under Cover, including Marijuana Grown Under Cover" \citep{naics2007}.\footnote{Recall that in addition to growing cannabis, most producers are also processors (i.e. producer-processors). Processing falls under NAICS 424 which includes as a subcomponent "Other Farm Product Raw Material Merchant Wholesalers, including Marijuana Merchant wholesalers" (NAICS 424590). However, NAICS classifies an establishment based on its primary activity, meaning that a cannabis producer only belongs to NAICS 424 if its revenue from processing activities exceeds that of its own crop production \citep{naics2007}. Table \ref{tab:1} shows that unprocessed "Usable Marijuana" accounts for the majority of revenue for producers, indicating that they belong to NAICS 111. Further evidence comes from \cite{miller2022}, who show that when cannabis was first legalized, the establishment count for NAICS 1114 in Washington increased by a similar count as the number of producer cannabis licenses. Moreover, the state saw a proportional increase in the number of workers and the total wages paid in NAICS 1114 \citep{miller2022}.} Figure \ref{fig:2} depicts the average industry-by-county bite in the sample period for the NAICS industries containing cannabis establishments. 


By defining bite at the level of the three-digit industry, I assume that variation in wages at cannabis establishments resembles variation in the corresponding NAICS industries.\footnote{In Appendix \ref{appendix:d}, I construct an alternative bite variable at the five-digit NAICS level and show that my results do not depend on the chosen level of industrial classification.} I document several facts to support this assumption. First, in Appendix \ref{appendix:g} I show that average wages for cannabis retailers and producers are very similar to those in the corresponding NAICS industries.\footnote{I also find that average wages are homogenous among the industries contained in the relevant 3-digit NAICS industries.} Moreover, for both the cannabis industry and the NAICS industries, average wages are remarkably close to the wage floor imposed by the minimum wage.\footnote{For producers, the gross average wage is 5\%-10\% above the minimum wage for the years 2018 to 2020, while for retailers it ranges from 15\%-19\% above the minimum wage. See appendix \ref{appendix:g}.} Thus, to the extent that the wage distributions differ between cannabis establishments and their NAICS industries, these differences should come from the upper part of the wage distributions rather than the lower part (since outliers are bounded from below by the minimum wage but unbounded from above). Furthermore, my regressions control for local labor market conditions (county-level average wage and unemployment) as well as establishment fixed effects, which implies that any remaining measurement error is likely to be random and will lead to conservative treatment effect estimates. Finally, the dynamic difference-in-differences framework allows me to closely examine treatment effect timing, meaning that for estimates to be biased, non-random measurement error would have to induce bias in the exact period that the minimum wage hike occurs, a scenario which I consider unlikely.

I obtained the bite data from the Washington Employment Security Department (ESD) which collects data on employment and wages in industries covered by unemployment insurance (about 95\% of U.S. jobs).\footnote{The ESD data feeds into the better-known Quarterly Census of Employment and Wages (QCEW), a federal/state cooperative program that measures employment and wages in industries covered by unemployment insurance at the detailed-industry-by-county level.} A similar dataset has been used in the recent minimum wage literature (see e.g. \cite{dube2016,renkin2020,leung2021}). It is important to note that, while the treatment intensity varies across time and space in my sample, the timing of the treatment does not vary (i.e. no staggered treatment). Thus, the six minimum wage hikes (three citywide and three statewide hikes) in the sample period amount to three minimum wage events, each spaced 12 months apart. 

\section{Direct pass-through to wholesale and retail prices}\label{section:4}

In this section, I estimate direct pass-through to wholesale and retail prices under the assumption that the minimum wage hike only induces a labor cost shock.

\subsection{Main identification strategy}

To estimate direct pass-through to wholesale and retail prices, I employ a difference-in-differences specification with a continuous treatment. This strategy, which is commonly applied in the literature investigating the effects of minimum wage hikes (e.g. \cite{card1992}), identifies the effects of aggregate shocks through cross-sectional variation in the fraction affected (the “treatment intensity”).\footnote{Other examples of this approach beyond the minimum wage literature are \cite{lucca2019}, who study the link between student loan credit expansion and college tuition, and \cite{cunningham2020}, who investigate the effect of abortion clinic closures on abortion rates.} The treatment intensity in my setting is the minimum wage bite variable constructed at the industry-by-county level. In contrast to a binary treatment, the continuous treatment setup identifies an average causal response parameter capturing the overall causal response of a small change in the minimum wage bite. \cite{callaway2021} show that a key identifying assumption in this setting is no "selection bias" among units with similar treatment intensities.\footnote{This is in addition to the standard parallel trends assumption.} This implies that conditional on a set of controls and fixed effects, inflation in counties with slightly lower minimum wage bite provides a valid counterfactual for inflation in counties with slightly higher bite.\footnote{It is important to emphasize the local nature of the selection bias assumption: counties with vastly different minimum wage bite need not have the same potential outcomes.} In appendix \ref{appendix:l}, I describe the average causal response parameter in more detail and provide supportive evidence of the zero selection bias assumption.

Since firms may be forward-looking in their price setting, it is important to consider anticipatory effects that may cause price increases in the months leading up to the hike. Alternatively, firms may smooth price changes across several periods before and after a hike. The high frequency of the price data allows me to capture such dynamics, and I specify a nonparametric distributed lag model with leads and lags before and after each hike. Since the establishment-level price index and the minimum wage bite are first-differenced by construction, I specify the model in first differences. I estimate the following equation separately for retailers and producers:
\begin{equation}
   \pi_{j,t} = \sum_{l =-5}^{6} \beta_l \Delta MW_{j,t-l} \times Bite_{k(j),t-l} + X_{k(j),q(t)} + \gamma_t + \epsilon_{j,t}.\label{eq:2}
\end{equation}
Equation \ref{eq:2} relates the monthly establishment-level inflation rate, $\pi_{j,t}$, to the treatment intensity in county $k$, which is defined as the interaction between the percent change in the minimum wage applicable to establishment $j$, $\Delta MW_{j,t-l}$, and the minimum wage bite in the industry-county $k$ that establishment $j$ belongs to, $Bite_{k(j),t-l}$.\footnote{For establishments subject to a citywide minimum wage, $\Delta MW_{j,t-l}$ corresponds to the citywide hike.} Note that $\Delta MW_{j,t-l}$ does not contribute to the identifying variation and simply scales the bite variable (the main identifying variation) such that the estimated coefficients are interpretable as pass-through elasticities at a given $Bite_{k(j),t-l}$.\footnote{I show in appendix \ref{appendix:d} that results are similar when $Bite_{k(j),t-l}$ is not scaled by $\Delta MW_{j,t-l}$.} The vector of control variables, $X_{k(j),q(t)}$, contains the average wage and unemployment rate for county $k$ in the quarter $q$ of month $t$. I include these to absorb variation in cannabis prices related to local macroeconomic factors that may covary with the minimum wage bite. Time fixed effects $\gamma_t$ account for monthly industry-wide changes in cannabis prices. Since the identifying variation is at the county level, standard errors are clustered by county to allow for autocorrelation in unobservables within counties, as in \cite{bertrand2004}.

For a given minimum wage hike, the parameter $\beta_l$ measures the percent change in establishment $j$'s prices resulting from a percentage point increase in minimum wage exposure $l$ months after the minimum wage hike (or $l$ months before when $l$ is negative). Though inflation is the dependent variable, I follow previous studies and present the estimates as the effect of the minimum wage on the price level (see e.g. \cite{renkin2020,leung2021}). I thus normalize the effect to zero in a baseline period $m$ months before each hike and report the cumulative treatment effect as the sum of $\beta_l$ at various lags: $E_L = \sum_{l=-m}^{L}\beta_l$. The pre-treatment coefficients are reported in a similar manner with $P_L = -\sum_{l=m}^{-L-1}\beta_{-l}$.\footnote{\cite{siegloth2020} show that cumulative distributed lag coefficients are numerically equivalent to the parameter estimates from an event study design with binned endpoints. Since distributed lag coefficients measure treatment effect changes, one fewer lead has to be estimated compared to an event study specification. Thus, a 12 month event window requires estimating 11 distributed lag coefficients.}

My approach resembles strategies that attempt to identify effects of aggregate shocks through cross-sectional variation in the fraction affected (see e.g. \cite{bartik1991,lucca2019,goldsmith2020}). This strategy is useful in the context of Washington's cannabis market as it enables me to estimate pass-through despite the relatively small number of minimum wage hikes. Another advantage is that the estimated price level effects $E_L$ can be reformulated as pass-through elasticities at the average bite, allowing for direct comparison to elasticities found in the literature on minimum wage pass-through (e.g. \cite{leung2021,renkin2020}).

An important consideration is the number of leads and lags to include in equation \ref{eq:2}. One limitation is that minimum wage hikes occur in exact 12 month intervals, meaning event variables get highly collinear when $l$ is large. Another issue is that the establishment panel is not balanced, meaning that changes in the underlying sample may affect estimates when $l$ is large \citep{renkin2020}. Therefore, in my baseline estimation I opt for a non-overlapping 12-month event window. While this may seem restrictive, I show in appendix \ref{appendix:d} that treatment effects remain stable over a longer event window such that the 12-month event window adequately captures the short-run impact of the minimum wage on prices.\footnote{This mirrors the results from \cite{renkin2020} and \cite{leung2021}, who find that firms adjust prices at most three months prior to an event and that effects plateau within 1-2 months after the hike \citep{renkin2020, leung2021}.}

A central concern with this research design is possible reverse causality. Since the treatment intensity is the product of two variables, $\Delta MW_{j,t-l} \times Bite_{k(j),t-l}$, the potential for reverse causality must be addressed for each of these variables in turn. $\Delta MW_{j,t-l}$ would suffer from reverse causality if policymakers were to increase the minimum wage in response to local inflation (e.g. in an effort to keep real wages constant). This is clearly not the case with the statewide hikes in my sample, since they are either predetermined or linked to the CPI-W, a national---not local---price index.\footnote{The city of Seattle has a citywide minimum wage that could be endogenous for some businesses for event 3 (January 1st, 2021). I address this possibility in Appendix \ref{appendix:d.2} and show that the main results are robust to dropping Seattle establishments for event 3.} $Bite_{k(j),t-l}$ would suffer from reverse causality if county-level inflation drove wages. To account for this possibility, in a robustness check I include county fixed effects to absorb county-level differences in trend inflation.\footnote{Note that the baseline specification controls for establishment fixed effects because both the treatment and outcome variables are first-differenced by construction.} Moreover, the distributed lag specification allows me to closely examine effect timing so that to the extent that differences in inflation trends remain, these can be easily distinguished from treatment effects.

It is important to highlight that when estimated for retailers, equation \ref{eq:2} uniquely identifies direct pass-through to retail prices and avoids picking up indirect pass-through effects. To see this, note that two conditions must be jointly met for the direct pass-through estimates to be contaminated by indirect pass-through. First, retailers would need to purchase predominantly from producers located in the retailer’s own county. Second, the bite variable for retailers would need to correlate with bite for producers within each county.\footnote{Importantly, both conditions must hold for the direct pass-through estimates to be contaminated by indirect pass-through effects. If the first condition is met but the second condition doesn't hold, then the minimum wage effect on wholesale prices is part of the error term, but it is orthogonal to retail bite, and hence does not bias direct pass-through estimates. If the second condition holds but not the first, then producer bite and retail bite are not independent, but the minimum wage effect on wholesale prices in a given county has no impact on retail prices in that county since retailers don't purchase from local producers.} In appendix \ref{appendix:g}, I show that the first condition does not hold since over 85\% of retailers’ wholesale purchases are from producers located in other counties. Moreover, the within-county correlation between producer and retail bite is low: 0.18 (unconditional) and -0.03 (conditional on covariates). In other words, there is no systematic relationship between producer and retail bite, which implies that the second condition also does not hold. 

One limitation is that my research design cannot distinguish between the effects of minimum wage legislation and implementation. If firms are forward-looking in their price setting, prices may adjust when a minimum wage hike is announced rather than when the hike actually takes effect.\footnote{\cite{renkin2020}, for example, find that price effects occur primarily in the three months following the passage of minimum wage legislation rather than after the hike itself.} The first two hikes in my sample period were announced in 2016, two and three years prior to implementation, respectively. Because my sample runs from August 2018 through July 2021, any price effects from that announcement fall outside of the sample window and cannot be estimated. For the third event, the magnitude of the hike was announced three months prior to implementation, meaning price effects at announcement can be directly observed using my event study framework. As detailed in appendix \ref{appendix:h.3}, I find no evidence of price effects at announcement but large effects at implementation for both wholesale and retail prices. This indicates that cannabis establishments wait until the cost shock hits before adjusting prices even if they have full prior knowledge about the magnitude of the shock.

\subsection{Direct pass-through to wholesale prices}\label{section:5.1}

\begin{figure}[!htbp]
\caption{Direct pass-through of minimum wage hikes to wholesale prices}
\centering
    \includegraphics[width=.5\textwidth]{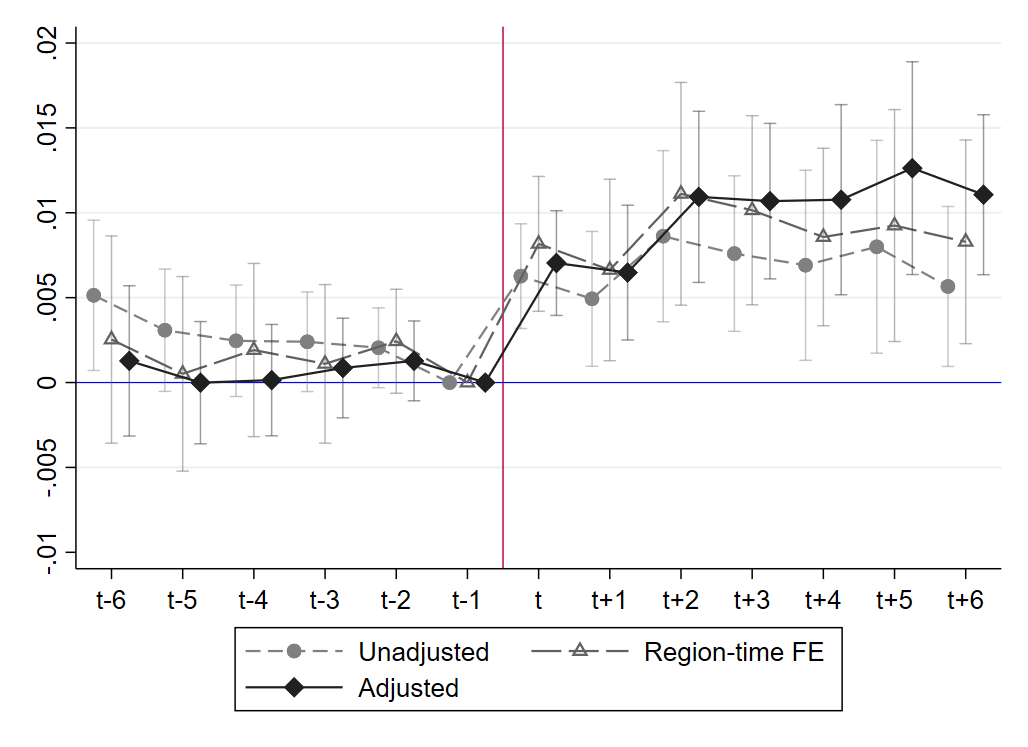}
		\label{fig:5.1b}
\label{fig:5.1}
\par \bigskip
\rule{\textwidth}{0.5pt}
\begin{minipage}[h]{\textwidth}
\medskip
\small \emph{Notes:} The figure shows estimates from equation \ref{eq:2} under three different specifications: unadjusted, trend-adjusted, and region-time FE. The dependent variable is the establishment-level inflation rate for cannabis producers. The figure depicts cumulative price level effects ($E_L$) relative to the baseline period in $t-1$. Cumulative effects $E_L$ are obtained by summing the distributed lag coefficients to lag $L$ as detailed in the main text. The figure shows 90\% confidence intervals of the sums based on SE clustered at the county level.  Data source: Top Shelf Data and Washington ESD, July 2018 to August 2021.
\end{minipage}
\end{figure}

I begin by estimating the direct effect of minimum wage hikes on wholesale prices and depict the results in Figure \ref{fig:5.1b}. In my preferred specification I include time FE but no county controls (results are robust to including controls). One question regarding the wholesale estimates is whether to control for a treatment-specific pre-trend since the baseline specification reveals a slight negative trend in the pre-treatment period. Though the trend is interrupted by a large and highly statistically significant treatment effect in the period that the minimum wage hike occurs, the contemporaneous treatment effect is slightly undone in subsequent periods as the pre-trend continues into the post-treatment period. Thus, while the trend does not mask the effect in period $t$, failure to account for the trend changes the interpretation of the results over a longer time horizon.\footnote{In appendix \ref{appendix:c} I show that the pre-trend is entirely driven by event 2, a period corresponding to a wholesale supply glut and falling wholesale prices across the industry. It is therefore plausible that for event 2 unobserved confounders covary with treatment intensity and wholesale cannabis deflation. The trend persists despite the inclusion of county FE because county means are based on all three events and the trend is only present for a single event.}

I apply two common strategies to control for the pre-trend, both of which yield similar results. First, I include region-time FE (i.e. interactions between time and region dummy variables) to account for regional economic trends that may covary with bite and inflation.\footnote{The regions are based on the three major socioeconomic regions in Washington state, where each region includes a subset of counties. See Appendix \ref{appendix:i} for details.} To the extent that unobserved time-variant heterogeneity is common within regions, region-time FE will control for the treatment-specific trend \citep{neumark2014}.\footnote{Unlike the baseline specification, I include county controls to further capture time-varying heterogeneity. Results are robust to including both county FE and controls or neither.} This assumes that the proper counterfactual for inflation in counties with higher treatment intensity is inflation in counties with slightly lower treatment intensity \emph{located in the same region}; that is, the identifying information comes from within-region variation in the treatment intensity. 

Second, I apply the two-step procedure from \cite{gb2021} and re-estimate equation \ref{eq:2} using a trend-adjusted dependent variable. Specifically, I calculate the average of the distributed lag estimates (from equation \ref{eq:2}) in the pre-baseline period and then extrapolate this pre-trend through the 12-month event window to obtain the treatment-specific linear trend $\hat \pi_{j,t}$. I then subtract the linear trend from the original dependent variable to get the trend-adjusted variable $\tilde \pi_{j,t(e)} = \pi_{j,t} - \hat \pi_{j,t}$. As argued by \cite{roth2022a}, this assumes that the observable linear pre-trend is a valid counterfactual for the unobservable post-trend. I view this as a valid assumption since the mean observable post-treatment trend is -0.00074 (95\% CI: -0.00218, 0.00070) which is nearly identical to---and not statistically significantly different from---the pre-treatment trend of -0.00077, (95\% CI: -0.00213, 0.00059).\footnote{In appendix \ref{appendix:c}, I show that for all three specifications (unadjusted, trend-adjusted, region-time FE) the distributed lag coefficients are not statistically significantly different from zero for $t-5$ through $t-2$, and the period $t$ treatment effects are large and not statistically significantly different from each other.} 

Figure \ref{fig:5.1b} illustrates that the period $t$ treatment effects are large and statistically significant at the 1-5\% level for all three specifications. At the average bite (17.20\%), a 10\% increase in the minimum wage corresponds to a 1.07\% increase in wholesale prices with the unadjusted dependent variable; 1.21\% for the trend-adjusted specification; and 1.40\% with region-time FE. In the latter two specifications, the pre-treatment period shows no significant trend and the large contemporaneous inflationary effect is no longer undone by the continuation of the pre-trend into the post-treatment period.\footnote{At higher lags, the price level effects from the specification with region-time FE are slightly lower than the trend-adjusted regression, but the difference is not statistically significant.} Thus, it matters little how one controls for the trend, as the linear trend adjustment and region-time FE specifications both lead to a permanently higher wholesale price level effect.

\subsection{Direct pass-through to retail prices}\label{section:5.2}

\begin{figure}[!htbp]
\caption{Direct pass-through of minimum wage hikes to retail prices}
\centering
	\includegraphics[width=.5\textwidth]{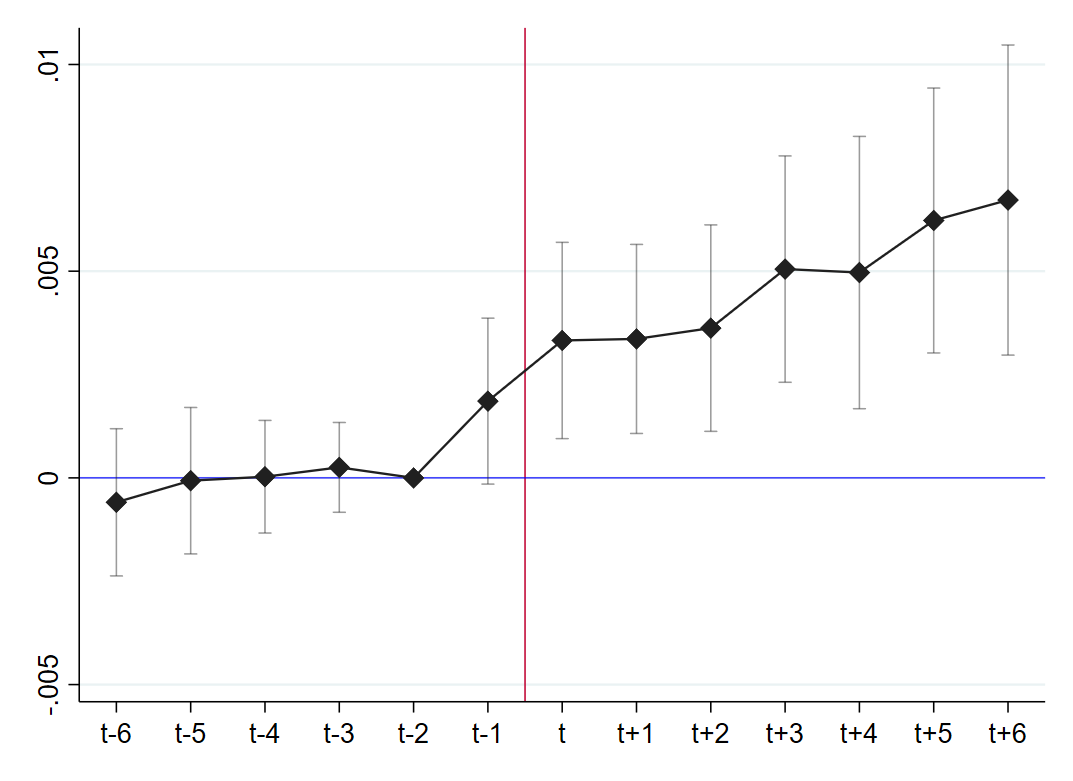}
\label{fig:5.2}
\par \bigskip
\rule{\textwidth}{0.5pt}
\begin{minipage}[h]{\textwidth}
\medskip
\small \emph{Notes:} Estimates are from equation \ref{eq:2} with time fixed effects and county-level controls. The dependent variable is the establishment-level inflation rate for cannabis retailers. The figure depicts cumulative price level effects ($E_L$) relative to the baseline period in $t-2$. Cumulative effects $E_L$ are obtained by summing the distributed lag coefficients to lag $L$ as detailed in the main text. The figure shows 90\% confidence intervals of the sums based on SE clustered at the county level. Data source: Top Shelf Data and Washington ESD, July 2018 to August 2021.
\end{minipage}
\end{figure}

Having obtained estimates for producers, I next estimate equation \ref{eq:2} for retail establishments. In my preferred specification for retailers, I include time FE and county controls (effect sizes are similar with county FE but estimates tend to be less precise at higher lags). Figure \ref{fig:5.2} illustrates that the effects for retailers differ from those of producers in several respects. First, effects for retailers show no pre-trend. Second, the treatment effect appears in $t-2$, i.e. one period prior to that for producers, suggesting that retailers may be more forward-looking in their pricing than producers.\footnote{This is consistent with the findings of \cite{hollenbeck2021}, who find that Washington's cannabis retailers in have substantial market power and behave like local monopolists. Though producers' market power has not been formally investigated in the literature, a common complaint among producers is their lack of market power compared to retailers \citep{lcbtier1,schaneman2021,cbt2021}.} Given the earlier treatment effect, I normalize the baseline period in $t-2$ when calculating cumulative effects on retail prices. For retailers at the average bite (19.37\%), a 10\% increase in the minimum wage corresponds to a 0.64\% jump in prices in period $t$. 

\subsection{Robustness checks} \label{section:5.3}
\subsubsection*{Alternative specifications}
The results from the previous subsection stand up to a multitude of robustness checks. In Table \ref{tab:5.1}, I present several variants of my empirical strategy for producers. I use the linear trend-adjustment as my preferred specification as this enables direct comparison to the indirect pass-through estimates later on (see section \ref{section:5}). Moreover, I normalize the baseline period in $t-2$ so that cumulative wholesale and retail results line up temporally. Note that changing the baseline period has no bearing on the estimated coefficients from equation \ref{eq:2} and simply amounts to a (downward) level shift in cumulative wholesale price level effects. For the baseline specification (column 1), I estimate equation \ref{eq:2} with time fixed effects but no controls. Column 2 shows that the estimated effects are virtually identical when including county-level controls. Column 3 shows that effects decrease slightly when county FE are included to capture price trends (since equation \ref{eq:2} is in first differences). Effect sizes are similar when the dependent variable is winsorized (column 4) or includes outliers (column 5) but standard errors tend to be larger in both cases compared to the baseline specification.\footnote{Recall that I trim the top and bottom 0.5\% of inflation per month in the baseline specification.} Column 6 shows that results are unchanged when the bite variable is trimmed by 0.5\%, indicating that the main results are not driven by treatment intensity outliers. Columns 7-9 show results when the dependent variable is not adjusted for a linear pre-trend. Column 7 shows price level effects when region-time FE are used to control for the pre-trend.\footnote{In column 7 I include time-variant county controls as these should capture additional confounding differences in price trends during event 2. Results are robust to including county FE, omitting controls, and a variety of other specifications. See Appendix \ref{appendix:d}.} Column 8 shows that with no linear trend adjustment, effects for the baseline specification are statistically significant, though smaller, through $t+2$, but effects are offset at $t+4$ by the continuation of the pre-trend. Column 9 shows an upward shift in effect sizes when the baseline period is set to one month before effects appear rather than two months before.

As Table \ref{tab:5.2} illustrates, retail price level effects are similarly stable across specifications. Column 2 shows that effect sizes do not depend on the inclusion of county controls; columns 3 and 4 show similar effect sizes with county FE and region-time FE, respectively. As with the wholesale regressions, retail price effects are not affected by winsorizing (column 5) or including outliers (column 6), though standard errors tend to be larger.

\begin{table}[!htbp] 
\centering
\caption{Direct pass-through of minimum wage hikes to the wholesale price level}
\label{tab:5.1}
\renewcommand{\tabcolsep}{1pt}{
\def\sym#1{\ifmmode^{#1}\else\(^{#1}\)\fi}
\begin{tabular*}{\hsize}{@{\hskip\tabcolsep\extracolsep\fill}l*{9}{c}}
\toprule
 \addlinespace

            &\multicolumn{6}{c}{Trend-adjusted} &\multicolumn{3}{c}{Unadjusted} \\
            \addlinespace
            \cline{2-7} \cline{8-10}
            \addlinespace

            &\multicolumn{1}{c}{(1)}&\multicolumn{1}{c}{(2)}&\multicolumn{1}{c}{(3)}&\multicolumn{1}{c}{(4)}&\multicolumn{1}{c}{(5)}&\multicolumn{1}{c}{(6)}&\multicolumn{1}{c}{(7)}&\multicolumn{1}{c}{(8)} &\multicolumn{1}{c}{(9)}\\
            
            &\multicolumn{1}{c}{\centering Baseline} & \multicolumn{1}{c}{\centering Controls} & \multicolumn{1}{c}{\parbox{1cm}{\centering County trends}} & \multicolumn{1}{c}{\parbox{1cm}{\centering Wins-orized}} & \multicolumn{1}{c}{\parbox{1cm}{\centering Outliers}} & \multicolumn{1}{c}{\parbox{1cm}{\centering Trim-med bite}} & \multicolumn{1}{c}{\parbox{1cm}{\centering Reg.-time FE}} &\multicolumn{1}{c}{\parbox{1cm}{\centering Baseline}} &\multicolumn{1}{c}{\parbox{1cm}{\centering t-1 base}} \\
\midrule
\midrule
\addlinespace
$E_0$       &   0.006*** & 0.006*** & 0.004** & 0.006*** & 0.005** & 0.006*** & 0.006** & 0.004** & 0.006***  \\
            &   (0.002) & (0.002) & (0.002) & (0.002) & (0.002) & (0.002) & (0.002) & (0.002) & (0.002)  \\
\addlinespace
$E_2$       &   0.010*** & 0.010*** & 0.006** & 0.010*** & 0.010*** & 0.010*** & 0.009** & 0.007**  & 0.009*** \\
            &   (0.003) & (0.003) & (0.003) & (0.003) & (0.004) & (0.003) & (0.004)  & (0.003) & (0.003) \\
\addlinespace
$E_4$       &    0.009*** & 0.010*** & 0.005 & 0.010*** & 0.010*** & 0.010*** & 0.006  & 0.005 & 0.007** \\
            &   (0.004) & (0.004) & (0.004) & (0.004) & (0.004) & (0.004) & (0.004)  & (0.004) & (0.003) \\
\midrule
$\sum \text{Pre-event}$  
            &   1.0e-07 & -2.0e-04 & 0.003 & 1.0e-07 & 2.0e-07 & 2.0e-07 & 1.0e-04  & 0.003 & 0.005* \\
            &   (0.003) & (0.003) & (0.003) & (0.004) & (0.004) & (0.003) & (0.003)  & (0.003) & (0.003) \\
\midrule
\(N\)       &       14,777 & 14,777 & 14,777 & 14,932 & 14,932 & 14,735 & 14,777 & 14,777 & 14,777 \\
Time FE     &        YES          &   YES               &   YES                     &   YES               &         YES             &   YES  & YES & YES & YES\\
Controls    &        NO           &   YES               &   NO                      &   NO                &         NO              &   NO & YES & NO & NO \\
Store FE   &        YES           &   YES                &   NO                      &   YES               &         YES              &   NO  & YES & NO & NO \\
\bottomrule
\end{tabular*}
\begin{minipage}[h]{\textwidth}
\medskip
\small \emph{Notes:} The dependent variable is the establishment-level inflation rate for cannabis wholesalers, adjusted for a bite-specific trend as detailed in section \ref{section:5.1}. The listed coefficients are the sum of the distributed lag coefficients $E_L$, $L$ months after the minimum wage hikes, relative to the normalized baseline period in $t-2$. The distributed lag coefficients are estimated from equation \ref{eq:2}. The control variables in (2) are the monthly unemployment rate and monthly average wage, both at the county level. (3) includes price trends at the county level. (4) uses a winsorized dependent variable (99\% windsorization). (5) does not trim or winsorize the dependent variable. In (6) the bite variable is trimmed by 1\%. In (7) the dependent variable is not trend-adjusted but region-time FE are included; county controls are included to further account for time-variant confounders. (8) shows the baseline specification when the dependent variable is not adjusted for a linear pre-trend. In (9) the dependent variable is not trend-adjusted and the normalized baseline period is one month before to the hike rather than two months before. Standard errors are clustered at the county level and are shown in parentheses. \sym{*} \(p<0.10\), \sym{**} \(p<0.05\), \sym{***} \(p<0.01\). Data from Washington ESD and Top Shelf Data, July 2018-August 2021.
\end{minipage}
}

\end{table}

\begin{table}[!htbp] 
\centering
\caption{Direct pass-through of minimum wage hikes to the retail price level}
\label{tab:5.2}
\renewcommand{\tabcolsep}{1pt}{
\def\sym#1{\ifmmode^{#1}\else\(^{#1}\)\fi}
\begin{tabular*}{\hsize}{@{\hskip\tabcolsep\extracolsep\fill}l*{7}{c}}
\toprule
            &\multicolumn{1}{c}{(1)}&\multicolumn{1}{c}{(2)}&\multicolumn{1}{c}{(3)}&\multicolumn{1}{c}{(4)}&\multicolumn{1}{c}{(5)}&\multicolumn{1}{c}{(6)}&\multicolumn{1}{c}{(7)}\\
            
            &\multicolumn{1}{c}{\parbox{1cm}{\centering Baseline}}&\multicolumn{1}{c}{\parbox{1cm}{\centering No controls}}&\multicolumn{1}{c}{\parbox{1cm}{\centering County trends}}&\multicolumn{1}{c}{\parbox{1cm}{\centering Reg.-time FE}}&\multicolumn{1}{c}{\parbox{1cm}{\centering Winsor-ized}}&\multicolumn{1}{c}{\parbox{1cm}{\centering Outliers}} &\multicolumn{1}{c}{\parbox{1cm}{\centering Trim-med bite}} \\
\midrule
\midrule
\addlinespace
$E_0$       &      0.003** & 0.003** & 0.003* & 0.003* & 0.003** & 0.004** & 0.003**  \\
            &    (0.001) & (0.001) & (0.002) & (0.002) & (0.001) & (0.002) & (0.002)  \\
\addlinespace
$E_2$       &     0.004** & 0.003** & 0.003 & 0.004** & 0.004** & 0.005** & 0.004** \\
            &    (0.002) & (0.001) & (0.002) & (0.002) & (0.002) & (0.002) & (0.002)  \\
\addlinespace
$E_4$       &      0.005** & 0.004** & 0.005* & 0.006** & 0.004* & 0.005 & 0.005** \\
            &   (0.002) & (0.002) & (0.003) & (0.002) & (0.002) & (0.003) & (0.002) \\
\midrule
$\sum \text{Pre-event}$   
            &  -6.0e-04 & 2.0e-04 & -0.001 & 1.0e-04 & 5.0e-04 & 0.001 & -8.0e-04 \\ 
            &   (0.001) & (0.0009) & (0.001) & (0.001) & (0.001) & (0.001) & (0.001) \\ 
\midrule
\(N\)       &       14,044 & 14,044 & 14,044 & 14,044 & 14,189 & 14,189 & 13,963  \\
Time FE     &        YES          &   YES               &   YES                     &   YES                 &         YES               &   YES  & YES \\
Controls    &        YES          &   NO               &   YES                      &   YES                  &         YES                &   YES & YES \\
County FE   &        NO           &   NO                &   YES                      &   NO                 &         NO                &   NO  & NO \\
Region-time FE    &        NO           &   NO                &   NO                     &   YES                  &         NO                &   NO  & NO \\
Trimmed     &        YES          &   YES               &   YES                     &   YES                 &         NO                &   NO  & NO \\
Winsorized &        NO           &   NO                &   NO                      &   NO                  &   YES                     &   NO & NO \\

\bottomrule
\end{tabular*}
\begin{minipage}[h]{\textwidth}
\medskip
\small \emph{Notes:} The dependent variable is the establishment-level inflation rate for cannabis retailers. The listed coefficients are the sum of the distributed lag coefficients $E_L$, $L$ months after the minimum wage hikes, relative to the normalized baseline period in $t-2$. The distributed lag coefficients are estimated from equation \ref{eq:2}. The baseline specification in (1) includes as controls the monthly unemployment rate and monthly average wage, both at the county level. (2) excludes county controls. (3) controls for county-level price trends. (4) includes region-time FE but not county FE. (5) uses a winsorized outcome (99\% winsorization). (6) does not trim or winsorize the outcome. In (7) the bite variable is trimmed by 1\%. Standard errors are clustered at the county level and are shown in parentheses. \sym{*} \(p<0.10\), \sym{**} \(p<0.05\), \sym{***} \(p<0.01\). Data from Washington ESD and Top Shelf Data, July 2018-August 2021.
\end{minipage}

}
\end{table}

\subsubsection*{Further robustness checks}
Besides testing different specifications, it is also important to consider the institutional assumptions underlying identification in my research design. In this section, I discuss these assumptions and the implications of them being violated. I report results from these additional robustness checks in appendix \ref{appendix:d}. 

First, since Washington's primary crop harvesting season is in Q3---the same quarter that the bite variable is calculated---it is important to ensure that seasonal labor fluctuations do not cause endogeneity in the bite variable. Therefore, I check whether results change if the bite variable is based on Q4 wages, i.e. outside of the main harvesting season. As appendix Tables \ref{tab:d.1} and \ref{tab:d.2} illustrate, results are robust to using this alternative bite variable.

I also consider the possibility that firms may not fully comply with the new minimum wage. If that were the case, the bite variable would not accurately measure minimum wage exposure since higher bite would not translate into a larger cost increase for firms. To account for such non-compliance, I redefine the bite variable as the difference between bite two quarters before and one quarter after the hike,
\begin{equation}
    \Delta Bite_{k(j)} = Bite_{k(j),Q3,y}-Bite_{k(j),Q1,y+1}
\end{equation}
This effectively nets out non-compliance at the county level. Appendix Tables \ref{tab:d.1} and \ref{tab:d.2} show that results are robust to this alternative bite variable.

An obvious concern is that policymakers may set minimum wage policy according to local price trends. If that were the case, then the treatment intensity would be endogenous due to reverse causality. Luckily, this concern does not apply to the statewide hikes in the sample since they are either predetermined (events 1 and 2) or linked to a national price index (event 3). However, the city of Seattle has a citywide minimum wage that could, under certain circumstances, be endogenous for some businesses in event 3. In Appendix \ref{appendix:d.2}, I consider the scenarios under which Seattle's minimum wage could be endogenous and show that the main results are unchanged when Seattle establishments are dropped from the sample for event 3.

Next, to ensure that my results are not driven by market entry or exit, I restrict the sample to establishments that are present at least 10 months for a given 12-month event. Results are robust to using this more balanced sample.

Since the establishment-level price indexes are constructed using annual product and subcategory weights, the weights change at the same time as the minimum wage hike. To ensure that effect sizes are not an artifact of this weighting scheme, I use alternate weights based on a fiscal year starting in July and ending in June each year (i.e. six months offset from the weights in the baseline model).\footnote{For the weights to cause endogeneity, the change in product and subcategory revenue shares within an establishment would need to covary with bite.} Results are unaffected by this alternate weighting scheme.

Next, I set the treatment intensity equal to the minimum wage bite itself to show that results do not rely on interacting bite with the size of the minimum wage hike.

It is also important to test whether the results are impacted by the level of industry classification used to measure minimum wage bite. Therefore, I construct an alternative bite variable based on 5-digit NAICS codes to show that the main results do not depend on the level of industrial classification used.\footnote{Though based on a more detailed level of industry classification, the alternate bite variable contains measurement error of a different sort, meaning it is not apparent that it is a better measure of minimum wage exposure. See appendix \ref{appendix:d} for details.}

\section{Indirect pass-through to retail prices}\label{section:5}

Since producers occupy the upstream portion of the supply chain, the pass-through estimates from equation \ref{eq:2} provide a complete measure of wholesale price adjustment in response to minimum wage hikes.\footnote{In practice, producers may also be subject to minimum wage pass-through from their input suppliers. However, producer inputs like hydroponic systems, grow lights, and raw materials can be purchased from suppliers outside of Washington state. Therefore, minimum wage pass-through to producer input prices is likely small. To simplify the analysis, I do not consider indirect pass-through for producers.} For retail prices, however, equation 2 only estimates direct pass-through and therefore fails to capture indirect pass-through. Thus, an analysis based solely on equation 2 may not reflect the full impact of the minimum wage on retail prices. To capture both direct and indirect effects, I estimate the following equation for retailers only:
\begin{equation}\label{eq:5}
   \pi_{r,t} = \sum_{l =-5}^{6} \beta_l \Delta MW_{r,t-l} \times Bite_{k(r),t-l} + \sum_{l =-5}^{6} \psi_l IB_{r,P,t-l} + X_{k(r),q(t)} + \gamma_t + \epsilon_{r,t}.
\end{equation}
In contrast to equation \ref{eq:2}, equation \ref{eq:5} contains not one, but two treatment variables. The first, $\Delta MW_{r,t-l} \times Bite_{k(r),t-l}$, is identical to that from equation \ref{eq:2} except that the index $r$ replaces $j$ to emphasize that the bite corresponds to retailer $r$. The second treatment variable is indirect bite, $IB_{r,P,t-l}$, a shift-share instrument that measures the weighted average minimum wage exposure of the producers that retailer $r$ purchases from.\footnote{Shift-share (or "Bartik") instruments are common in the empirical literature. For recent examples, see e.g. \cite{Jaravel2019,hummels2014,xu2022}. I discuss and provide supportive evidence for instrument validity in appendix \ref{appendix:l}.} $IB_{r,P,t-l}$ is calculated as follows:
\begin{equation}\label{eq:6}
    IB_{r,P,t-l} = \sum_{p = s}^{S} \alpha_{r,p} \Delta MW_{p,t-l} \times \sum_{p=s}^{S} \alpha_{r,p} Bite_{k(p),t-l}
\end{equation}
Here, $\Delta MW_{p,t-l}$ is the size of the minimum wage hike for producer $p$; $\alpha_{r,p}$ is the average share of retailer $r$'s wholesale expenditures going to producer $p$ from $t-4$ through $t-2$, i.e. in the months leading up to the hike; and $Bite_{k(p),t-l}$ is the minimum wage bite for the industry-county $k$ that producer $p$ is located in.\footnote{Note that a retailer and a producer located in the same county will have different bites since they belong to different three-digit industries.} Thus, the first term in equation \ref{eq:6} measures the average minimum wage hike for the set of producers that retailer $r$ purchases from, while the second term measures the average bite for that same set of producers.\footnote{One could instead directly interact producer expenditure share, hike size, and producer bite as follows: $IB_{r,P,t-l} = \sum_{p = s}^{S} \alpha_{r,p} \Delta MW_{p,t-l} Bite_{k(p),t-l}$. Results are virtually identical under this definition of indirect bite. However, the advantage of averaging before interacting (as in equation \ref{eq:6}) is that the coefficient $\psi_l$ can be interpreted as a pass-through elasticity.} It is worth emphasizing that $\alpha_{r,p}$ contains no time index and is therefore fixed for each retailer-producer-event. In practice, retailers may react to the increase in wholesale prices by recalibrating their wholesale bundles (e.g. by substituting out of high pass-through products), in which case $\alpha_{r,p}$ would change from month to month. However, allowing $\alpha_{r,p}$ to vary within an event could result in reverse causality since a retailer's wholesale substitution patterns may reflect its own inflation. Defining $\alpha_{r,p}$ as the average expenditure share from $t-4$ through $t-2$ avoids this endogeneity, particularly since the results from the previous section indicate that wholesale price effects do not emerge until $t-1$. In other words, the expenditure shares $\alpha_{r,p}$ are based on a time frame prior to the emergence of the wholesale cost shock. 

As an alternative to the indirect bite variable in equation \ref{eq:6}, one could use producers' geographic proximity as an instrument for retailers' exposure to pass-through to wholesale prices. This assumes that retailers purchase more from producers located nearby than those further away. However, I find little evidence supporting this assumption. Instead, a large share of retailers' wholesale purchases are from producers located in other parts of the state (see appendix \ref{appendix:g}), which suggests that a distance-based instrument is unlikely to capture retailers' exposure to pass-through to wholesale prices.

A key assumption is that equation \ref{eq:5} separately identifies direct and indirect pass-through. One way to test this is to examine whether the direct pass-through estimates $\hat{\beta}_l$ are affected by the inclusion of indirect bite as an additional variable. If estimates for direct pass-through were to change, this would cast doubt on the main identification strategy and, by extension, the results from the previous section. I show in appendix \ref{appendix:d} that direct pass-through estimates are unaffected by the inclusion of indirect bite.

In equation \ref{eq:5}, the indirect pass-through rate flows from the parameter $\psi_l$. For a given minimum wage hike, $\psi_l$ measures the percent change in retailer $r$'s prices resulting from a percentage point increase in indirect minimum wage exposure $l$ months before the minimum wage hike. As with direct pass-through, indirect pass-through is best illustrated in terms of cumulative price level effects. Therefore, I again normalize the effect to zero in a baseline period $m$ months before each hike and report the cumulative treatment effect as the sum of $\psi_l$ at various lags: $E_L = \sum_{l=-m}^{L}\psi_l$. I report the pre-treatment coefficients in a similar manner, with $P_L = -\sum_{l=m}^{-L-1}\psi_{-l}$. 

\begin{figure}[!htbp]
\caption{Indirect pass-through to retail prices and direct pass-through to wholesale prices}
\centering
	\begin{subfigure}{.4\textwidth}
	\centering
	    \includegraphics[width=\linewidth]{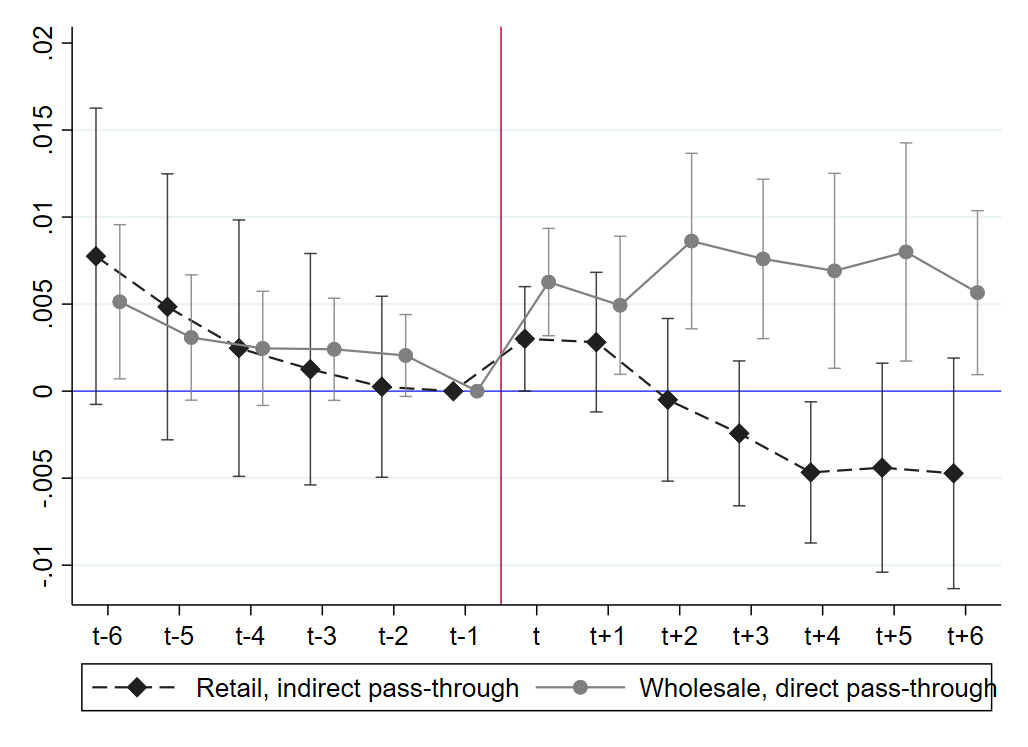}
		\caption{Unadjusted}
		\label{fig:6.2a}		
	\end{subfigure}\hfil
	\begin{subfigure}{.4\textwidth}
	\centering
	    \includegraphics[width=\linewidth]{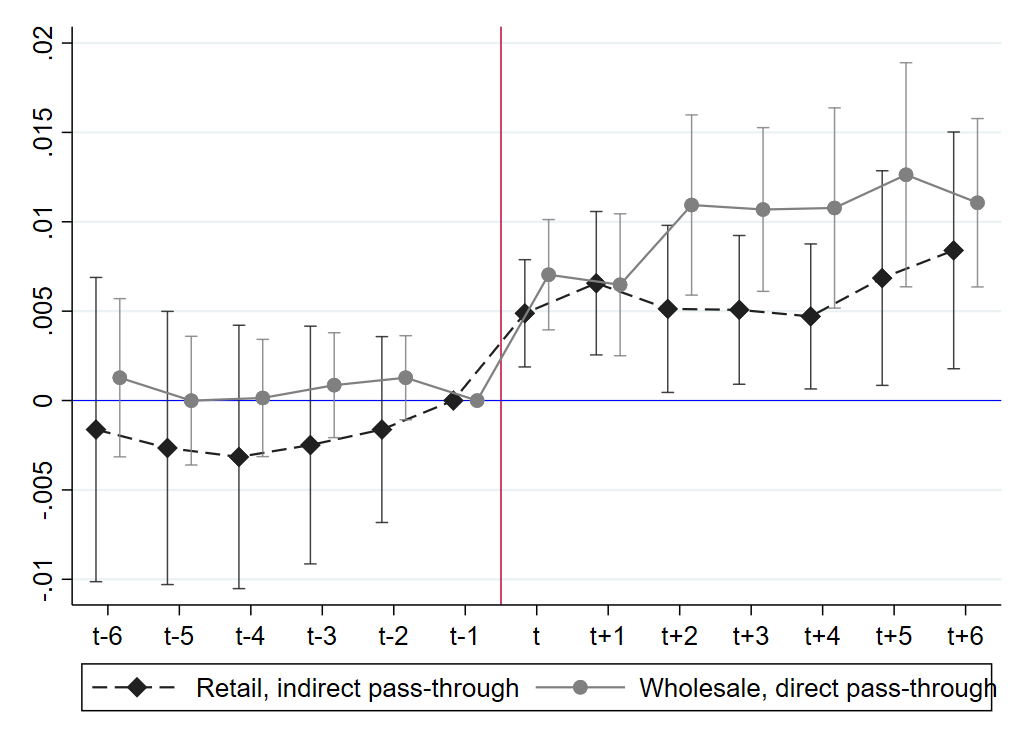}
		\caption{Trend-adjusted}
		\label{fig:6.2b}
	\end{subfigure}\hfil
\label{fig:6.2}
\par \bigskip
\rule{\textwidth}{0.4pt}
\begin{minipage}[h]{\textwidth}
\medskip
\small \emph{Notes:} The figures compare direct pass-through to the wholesale price level and indirect pass-through to the retail price level. Wholesale price effects are estimated from equation \ref{eq:2} with time fixed effects. Indirect retail price effects are estimated from equation \ref{eq:5} with time fixed effects and county-level controls. The estimated coefficients $\beta_l$ and $\psi_l$, respectively, are summed up to cumulative effects $E_L$, relative to the normalized baseline period in $t-1$. Panel (a) shows estimated effects when the dependent variable is not adjusted for a pre-trend. Panel (b) shows estimated effects when the dependent variable is trend-adjusted following the \cite{gb2021} procedure described in section \ref{section:5.1}. Both figures show 90\% confidence intervals of the sums $E_L$ based on SE clustered at the county level. Data source: Top Shelf Data and Washington ESD, July 2018-August 2021.
\end{minipage}
\end{figure}

Figure \ref{fig:6.2a} illustrates that the time path of indirect pass-through to retail prices is remarkably similar to the estimates of direct pass-through to wholesale prices from section \ref{section:4}. The figure reveals a downward-sloping pre-trend interrupted by an inflationary shock in the treatment period, followed by a continuation of the pre-trend into the post-treatment period. As in section \ref{section:5.1}, to quantify the pre-treatment trend I take the average of the distributed lag coefficients for the pre-baseline period, $\bar{\hat{\psi}}_{pre} = 1/4 \sum_{l=2}^{5} \hat{\psi}_{-l}$. I find no statistically significant difference between $\bar{\hat{\psi}}_{pre}$ and the bite-specific trend for pass-through to wholesale prices.\footnote{$\bar{\hat{ \psi}}_{pre} = -.00102$ ($90\%$ confidence interval: $-0.00329$ to $0.00126$), which overlaps with the pre-trend for pass-through to wholesale prices, $\bar{\hat{ \beta}}_{pre} = -0.00197$ ($90\%$ confidence interval: $-0.00367$ to $-0.00026$). See appendix \ref{appendix:c} for details.} Accordingly, I apply the \cite{gb2021} procedure and re-estimate equation \ref{eq:5} with the dependent variable adjusted for the indirect bite-specific trend.\footnote{Unlike in section \ref{section:4}, region-time FE cannot be used to control for the pre-trend since the dependent variable and indirect bite stem from different sets of establishments (retailers and producers, respectively). Moreover, as shown in appendix \ref{appendix:g}, retailers purchase a large share of products from producers located in other regions of the state, meaning region-time FE based on a retailer's region will not capture the indirect bite-specific trend.} 

Note that while it is informative to compare the time paths of indirect pass-through to retail prices and direct pass-through to wholesale prices, the implied pass-through elasticities are not directly comparable. The reason is that retail prices are higher than wholesale prices, meaning a given pass-through elasticity corresponds to a much larger absolute pass-through (in dollar terms) to retail prices than to wholesale prices. In contrast, direct and indirect pass-through to retail prices are directly comparable because they correspond to a single price level (i.e. the retail price level).

In Table \ref{tab:6.1}, I report cumulative effects of indirect pass-through to retail prices relative to the normalized baseline period two months prior to the hike. For the baseline specification, at the average indirect bite (18.14\%), a 10\% minimum wage hike corresponds to a 1.22\% increase in retail prices two months after the hike.

\begin{table}[!htbp] 
\centering
\caption{Indirect pass-through of minimum wage hikes to the retail price level}
\label{tab:6.1}
\renewcommand{\tabcolsep}{1pt}{
\def\sym#1{\ifmmode^{#1}\else\(^{#1}\)\fi}
\begin{tabular*}{\hsize}{@{\hskip\tabcolsep\extracolsep\fill}l*{6}{c}}
\toprule
            &\multicolumn{1}{c}{(1)}&\multicolumn{1}{c}{(2)}&\multicolumn{1}{c}{(3)}&\multicolumn{1}{c}{(4)}&\multicolumn{1}{c}{(5)}&\multicolumn{1}{c}{(6)}\\
            
            &\multicolumn{1}{c}{\parbox{1cm}{Baseline}}&\multicolumn{1}{c}{\parbox{1cm}{No controls}}&\multicolumn{1}{c}{\parbox{1cm}{County FE}}&\multicolumn{1}{c}{\parbox{1cm}{Reg.-time FE}}&\multicolumn{1}{c}{\parbox{1cm}{Winsor-ized}}&\multicolumn{1}{c}{\parbox{1cm}{Outliers}} \\
\midrule
\midrule
\addlinespace
$E_0$       &     0.00650** & 0.00658** & 0.00708*** & 0.00686*** & 0.00787** & 0.00905** \\
            &   (0.00256) & (0.00256) & (0.00270) & (0.00263) & (0.00330) & (0.00383) \\
\addlinespace
$E_2$       &     0.00675 & 0.00685 & 0.00798 & 0.00694 & 0.00912** & 0.0122*** \\
            &   (0.00454) & (0.00452) & (0.00505) & (0.00457) & (0.00438) & (0.00465) \\
\addlinespace
$E_4$       &      0.00633* & 0.00689* & 0.00770* & 0.00663* & 0.00922* & 0.0134** \\
            &   (0.00376) & (0.00378) & (0.00425) & (0.00382) & (0.00526) & (0.00621) \\
\midrule
$\sum \text{Pre-event}$   
            & -1.94e-07 & -0.000618 & -0.000782 & -0.000469 & 3.81e-08 & -2.90e-08 \\ 
            &   (0.00514) & (0.00514) & (0.00504) & (0.00521) & (0.00540) & (0.00582) \\ 
\midrule
\(N\)       &       13,559 & 13,559 & 13,559 & 13,559 & 13,689 & 13,689 \\
Time FE     &        YES          &   YES               &   YES                     &   YES                 &         YES               &   YES  \\
Controls    &        YES          &   NO               &   YES                      &   YES                  &         YES                &   YES \\
County FE   &        NO           &   NO                &   YES                      &   NO                 &         NO                &   NO  \\
Region-date FE &  NO           &   NO                &   NO                      &   YES                 &         NO                &   NO  \\
Trimmed     &        YES          &   YES               &   YES                     &   YES                 &         NO                &   NO  \\
Winsorized &        NO           &   NO                &   NO                      &   NO                  &   YES                     &   NO  \\
Trend-adjusted &  YES          &   YES               &   YES                     &   YES                 &         YES               &   YES  \\

\bottomrule
\end{tabular*}
\begin{minipage}[h]{\textwidth}
\medskip
\small \emph{Notes:} The dependent variable is the establishment-level inflation rate for cannabis retailers, adjusted for a bite-specific trend as detailed in section \ref{section:5.1}. The listed coefficients are the sum of the distributed lag coefficients $E_L$, $L$ months after the minimum wage hikes, relative to the normalized baseline period in $t-2$. The distributed lag coefficients are estimated from equation \ref{eq:2}. The baseline specification in (1) includes as controls the monthly unemployment rate and monthly average wage, both at the county level. (2) excludes controls. (3) controls for county-level price trends. (4) includes region-time FE but not county FE. (5) uses a winsorized outcome (99\% winsorization). (6) does not trim or winsorize the outcome. Standard errors are clustered at the county level and are shown in parentheses. \sym{*} \(p<0.10\), \sym{**} \(p<0.05\), \sym{***} \(p<0.01\). Data from Washington ESD and Top Shelf Data, July 2018-August 2021.
\end{minipage}

}

\end{table}

\subsection*{Comparing indirect and direct pass-through to retail prices: a discussion}

\begin{table}[!htbp] 
\centering
\caption{Minimum wage elasticity of the price level}
\label{tab:6.2}
\renewcommand{\tabcolsep}{1pt}{
\def\sym#1{\ifmmode^{#1}\else\(^{#1}\)\fi}
\begin{tabular*}{\hsize}{@{\hskip\tabcolsep\extracolsep\fill}l*{5}{c}}
\toprule
& \multicolumn{3}{c}{Wholesale} & \multicolumn{2}{c}{Retail} \\
\addlinespace
\cline{2-4} \cline{5-6} 
\addlinespace
&\multicolumn{1}{c}{(1)} &\multicolumn{1}{c}{(2)} &\multicolumn{1}{c}{(3)} &\multicolumn{1}{c}{(4)} &\multicolumn{1}{c}{(5)} \\

& \multicolumn{1}{c}{\parbox{1cm}{\centering Trend-adjusted}} &\multicolumn{1}{c}{\parbox{1cm}{\centering Reg.-time FE}} & \multicolumn{1}{c}{\parbox{1cm}{\centering Un-adjusted}} & \multicolumn{1}{c}{\parbox{1cm}{\centering Direct}} & \multicolumn{1}{c}{\parbox{1cm}{\centering Indirect}} \\
\midrule
$E_2$     &  0.010 & 0.009 & 0.007 & 0.004  & 0.007 \\

\addlinespace
Avg. bite       &    17.20 & 17.20 & 17.20 & 19.37 & 18.14 \\

\addlinespace
MW pass-through elasticity   &  0.172 & 0.155 & 0.120 & 0.077 & 0.127 \\

\addlinespace

\bottomrule
\end{tabular*}
\begin{minipage}[h]{\textwidth}
\medskip
\small \emph{Notes:} This table reports the minimum wage elasticity of prices, two periods after the hike, computed at the bite average bite: $E_2 \times \overline{Bite} \times 0.01$. $E_2$ are taken from tables \ref{tab:5.1}, \ref{tab:5.2}, and \ref{tab:6.1}. Data: Top Shelf Data and Washington ESD, (August 2018 - July 2021).

\end{minipage}

}

\end{table}

To facilitate comparison, Table \ref{tab:6.2} summarizes the main pass-through effects for wholesale and retail prices. Several facts stand out. First, the direct pass-through elasticity for retail prices of 0.077 is in line with existing findings from other retail sectors. \cite{leung2021}, for example, finds a minimum wage pass-through elasticity of 0.06-0.08 for grocery store prices in the U.S.\footnote{Using different scanner data from the U.S., \cite{renkin2020} find an elasticity of 0.036 for grocery store prices.} 

Second, the total minimum wage elasticity of the retail price level, which is obtained by summing the direct and indirect elasticities, is .204\%---much larger than that from direct pass-through alone.\footnote{Note that the direct and indirect pass-through elasticities are computed with different average bites. One could instead compute both elasticities from a single average bite value, but this would be an out of sample prediction for at least one of the elasticities. Since the average bite for indirect pass-through (18.14\%) is similar to that for direct pass-through (19.43\%), I compute the elasticities using their own average bite values. The total elasticity is then the sum of the direct and indirect elasticities.} Thus, failing to account for indirect pass-through dramatically underestimates the effect of minimum wage hikes on retail cannabis prices.

Third, indirect pass-through accounts for over 60\% of the total retail pass-through elasticity, whereas direct pass-through accounts for less than 40\%. To the extent that the cost of goods sold (COGS) makes up a larger share of retailers’ variable costs compared to labor costs (as in other retail sectors), it is reasonable for indirect pass-through to dominate direct pass-through.\footnote{For example, \cite{renkin2020} find that COGS account for more than 80\% of retailers' variable costs.} Indeed, I show in appendix Table \ref{tab:e1} that COGS makes up 60-80\% of variable costs for cannabis retailers, which confirms that the relative magnitudes of direct and indirect pass-through to retail prices are in line with retailers' variable cost structure.\footnote{Note that labor costs and COGS comprise the two largest parts of retailers' variable costs \citep{renkin2020}.}






\section{Dissecting the price effects}\label{section:6}

\subsection{Effects on markups over marginal input cost}

In this subsection, I estimate the impact of minimum wage hikes on retail markups over marginal input cost (MIC) with the aim of determining the degree of wholesale cost pass-through to retail prices. I estimate the following equation for retail establishments:
\begin{equation}\label{eq:7}
   \Delta \mu_{r,t} = \sum_{l =-5}^{6} \beta_l \Delta MW_{r,t-l} \times Bite_{k(r),t-l} + \sum_{l =-5}^{6} \psi_l IB_{r,P,t-l} + X_{k(r),q(t)} + \gamma_t + \epsilon_{r,t}.
\end{equation}
The dependent variable, $\Delta \mu_{r,t}$, is the monthly percent change in MIC markup for establishment $r$ (see appendix \ref{appendix:f} for details on the markup index). Table \ref{tab:6.3} displays the estimated markup effects and compares them to the estimates for direct and indirect pass-through to retail prices found in sections \ref{section:4}-\ref{section:5}. Comparing $E_4$ in columns 1 and 2 reveals that the direct effect on the markup is very similar to the direct effect on the price level. This is not surprising, since direct pass-through to retail prices entails a price increase that is---by definition---independent of wholesale costs. In contrast, columns 3 and 4 reveal a large indirect effect on the price level in $E_4$ (significant at the 10\% level) but a small and statistically insignificant indirect effect on markups. This indicates no markup adjustment to the wholesale cost shock on the part of retailers and implies a full pass-through of the wholesale cost shock to retail prices.

\begin{table}[!htbp] 
\centering
\caption{Minimum wage effects on retail prices and markups over marginal input cost}
\label{tab:6.3}
\renewcommand{\tabcolsep}{1pt}{
\def\sym#1{\ifmmode^{#1}\else\(^{#1}\)\fi}
\begin{tabular*}{\hsize}{@{\hskip\tabcolsep\extracolsep\fill}l*{4}{c}}
\toprule
\addlinespace

& \multicolumn{2}{c}{Direct effects} & \multicolumn{2}{c}{Indirect effects} \\
\addlinespace

\cline{2-3} \cline{4-5} \\

& \multicolumn{1}{c}{(1)} & \multicolumn{1}{c}{(2)} & \multicolumn{1}{c}{(3)} & \multicolumn{1}{c}{(4)}\\
\addlinespace
& \multicolumn{1}{c}{\parbox{1cm}{Price level}} & \multicolumn{1}{c}{\parbox{1cm}{Markup}} & \multicolumn{1}{c}{\parbox{1cm}{Price level}} & \multicolumn{1}{c}{\parbox{1cm}{Markup}} \\
\addlinespace
\midrule
\midrule
\addlinespace
$E_0$     & 0.00332** & 0.00277** & 0.00650** & 0.00338 \\
 & (0.00144) & (0.00137) & (0.00256) & (0.00293) \\
\addlinespace

$E_2$     & 0.00362** & 0.00293** & 0.00675 & 0.00210 \\
& (0.00145) & (0.00145) & (0.00454) & (0.00475) \\

\addlinespace

$E_4$     & 0.00497** & 0.00506*** & 0.00633* & -0.000940  \\
& (0.00200) & (0.00176) & (0.00376) &  (0.00432) \\

\addlinespace

\midrule
$\sum \text{Pre-event}$   
           & -0.00059 &  0.000461 & -1.94e-07 & -0.000760 \\ 
           & (0.00108) &   (0.00149) & (0.00514) & (0.00633) \\ 
\midrule
\(N\)      & 14,044 & 14,040 & 13,559 & 13,552  \\
Time FE &        YES    &   YES     &   YES   &  YES \\
Controls &        YES    &   YES     &   YES   &  YES \\
Trimmed & YES    &   YES     &   YES   &  YES \\
Trend-adjusted &  NO    &   NO     &   YES   &  YES \\

\bottomrule

\end{tabular*}
\begin{minipage}[h]{\textwidth}
\medskip
\small \emph{Notes:} This table reports the effects of minimum wage hikes on retail markups over marginal input cost. The dependent variable is the establishment-level percent change in markup over marginal input cost. For ease of comparison, column (1) is copied from column 1 in table \ref{tab:5.2} and shows the direct pass-through effect of minimum wage hikes on the retail price level. Column (2) reports the direct effect on markups over MIC. Column (3) is copied from column 1 in table \ref{tab:6.1} and shows the indirect pass-through effect of minimum wage hikes on the retail price level. Column (4) reports the indirect effect on markups over MIC. The dependent variable in column (4) is adjusted for a bite-specific pre-trend as detailed in section \ref{section:5.1}. All specifications include time fixed effects and county controls (unemployment rate and average wage). The listed coefficients are the sum of the distributed lag coefficients $E_L$, $L$ months after the minimum wage hikes, relative to the normalized baseline period in $t-2$. Standard errors of the sums are clustered at the county level and are shown in parentheses. \sym{*} \(p<0.10\), \sym{**} \(p<0.05\), \sym{***} \(p<0.01\). Data from Washington ESD and Top Shelf Data, August 2018-July 2021.
\end{minipage}

}

\end{table}

\subsection{Price effects by scale of production}

One concern in the cannabis industry is that rules governing production capacity for producers have created an uneven playing field, and a common complaint is that small producers operate on slim margins while large establishments enjoy a higher degree of market power \citep{lcbtier1}. If larger producers have more market power, then they may be able to absorb cost shocks along other margins (e.g. by adjusting profits), in which case pass-through elasticities may decrease with establishment size. Therefore, in this section I test for heterogeneous pass-through across different establishment sizes.

Producer licenses are based on a three-tier system governing the square footage of plant canopy that an establishment is legally permitted to operate. Tier 1 producers can grow up to 2,000 square feet of plant canopy, tier 2 can grow up to 10,000 square feet, while tier 3 can operate up to 30,000 square feet. Tiers were assigned to establishments before the market fist opened in 2014 and once assigned an establishment cannot switch tiers.\footnote{While producers can produce below the threshold for their tier, actual canopy usage has been found to be proportional to those thresholds \citep{lcb2019b}.} Thus, the scale of production is effectively predetermined for cannabis producers during the sample period.


In contrast to producers, there is no tier system governing the scale of production for cannabis retailers. Nevertheless, since a retail firm can operate up to five establishments, one can approximate retail firm size by distinguishing between independent stores and those belonging to a chain.\footnote{Each establishment has it's own cannabis business license. Since the number of licenses in the market is fixed, a firm must purchase an existing license in order to open another retail location.} Retail establishments belonging to a chain may exhibit different cost pass-through compared to independent establishments for a variety of reasons. For example, if chains spread cost shocks across locations then pass-through effects will be attenuated compared to independent establishments. Alternatively, if chains have lower prices to begin with, then they may have more scope to increase prices via cost pass-through. Lower initial prices also imply that a dollar increase in price corresponds to a larger pass-through elasticity than that same dollar increase at a higher price point. Another possibility is that chains may have higher profit margins and hence may be able to offset cost shocks by reducing profits. Finally, a chain's very existence may reflect superior management, better pricing strategies, or other unobservable differences that influence its ability to pass cost shocks through to prices. Ultimately, effect sizes for independent vs. chain stores are difficult to predict a priori.

I estimate equation \ref{eq:5} separately for each subsample and report the results in Table \ref{tab:6.4}. Columns 1-3 show that pass-through to wholesale prices is monotonically decreasing with establishment size: small producers exhibit much larger price level effects than medium-sized producers, while effects for large producers are close to zero and statistically insignificant.\footnote{The difference between effect sizes for small and large producers is statistically significant for $E_2$ but not for $E_0$ or $E_4$.} I interpret this as evidence that small producers may be less able to absorb the minimum wage cost shock via other margins of adjustment (e.g. by adjusting profits) compared to large producers. 

For retailers, the opposite pattern holds, as direct pass-through effects monotonically increase with chain size (though effects are only statistically significant for establishments belonging to 'large' chains). Results are less straightforward for indirect pass-through to retail prices. On the one hand, indirect pass-through monotonically increases with chain size for $E_0$ and $E_2$, thereby mimicking the pattern for direct pass-through. However, at $E_4$ effects resemble an inverse U-shape, with large effects for mid-sized chains and comparatively small effects for both independent stores and large chains. Taken together, retail establishments belonging to chains appear to have higher pass-through elasticities than independent stores.\footnote{In appendix \ref{appendix:f} I show that average retail prices for product subcategories are similar for independent and chain stores. Average wholesale prices are similarly homogeneous across producer tiers. Thus, heterogeneous pass-through is not a result of different initial price levels across the scale of production.} 


\newgeometry{margin = 3.5cm} 
\begin{landscape}
\begin{table}
\centering
\caption{Minimum wage pass-through to prices, by establishment size}
\label{tab:6.4}
\renewcommand{\tabcolsep}{1pt}{
\def\sym#1{\ifmmode^{#1}\else\(^{#1}\)\fi}
\begin{tabular*}{\hsize}{@{\hskip\tabcolsep\extracolsep\fill}l*{9}{c}}
\toprule

&\multicolumn{3}{c}{Wholesale} &\multicolumn{6}{c}{Retail} \\
\addlinespace
\cline{2-4} \cline{5-10} \\
& \multicolumn{3}{c}{Direct} &\multicolumn{3}{c}{Direct} &\multicolumn{3}{c}{Indirect} \\
\addlinespace
\cline{2-4} \cline{5-7} \cline{8-10} \\

&\multicolumn{1}{c}{(1)}&\multicolumn{1}{c}{(2)}&\multicolumn{1}{c}{(3)}&\multicolumn{1}{c}{(4)}&\multicolumn{1}{c}{(5)}&\multicolumn{1}{c}{(6)} &\multicolumn{1}{c}{(7)}&\multicolumn{1}{c}{(8)}&\multicolumn{1}{c}{(9)}\\
\addlinespace
            
&\multicolumn{1}{c}{\parbox{1cm}{\centering Small}}&\multicolumn{1}{c}{\parbox{1cm}{\centering Medium}}&\multicolumn{1}{c}{\parbox{1cm}{\centering Large}} &\multicolumn{1}{c}{\parbox{1cm}{\centering Indep-endent}}&\multicolumn{1}{c}{\parbox{1cm}{\centering 2-3 stores}}&\multicolumn{1}{c}{\parbox{1cm}{\centering 4+ stores}} &\multicolumn{1}{c}{\parbox{1cm}{\centering Indep-endent}}&\multicolumn{1}{c}{\parbox{1cm}{\centering 2-3 stores}}&\multicolumn{1}{c}{\parbox{1cm}{\centering 4+ stores}} \\
\midrule
\addlinespace
$E_0$       &     0.014* & 0.006 & 0.003 & 0.002 & 0.004 & 0.005** & 0.005 & 0.007 & 0.014 \\
            &   (0.007) & (0.003) & (0.003) & (0.002) & (0.002) & (0.002) & (0.004) & (0.004) & (0.009) \\
\addlinespace
$E_2$       &     0.025* & 0.01* & 1.61e-04 & 0.002 & 0.004 & 0.005* & 0.007 & 0.007 & 0.009 \\
            &  (0.010) & (0.004) & (0.004) & (0.002) & (0.002) & (0.002) & (0.006) & (0.006) & (0.010) \\
\addlinespace
$E_4$       &    0.023* & 0.010 & -5.91e-04 & 0.004 & 0.004 & 0.008*** & -1.14e-04 & 0.022*** & 0.007  \\
            &    (0.010) & (0.006) & (0.007) & (0.003) & (0.003) & (0.002) & (0.006) & (0.005) & (0.010) \\
\midrule
$\sum \text{Pre-event}$   
            &  -7.03e-08 & 1.46e-07 & -1.28e-08 & -0.002 & -1.53e-04 & 0.002 & -6.82e-08 & -1.32e-07 & 1.21e-08 \\ 
            &   (0.007) & (0.003) & (0.005) & (0.002) & (0.003) & (0.003) & (0.005) & (0.009) & (0.012) \\ 
\midrule
\(N\)       &    2,673 & 6,968 & 5,136 & 5,566 & 5,211 & 3,267 & 5,461 & 4,943 & 3,155  \\
Time FE     &        YES          &   YES               &   YES                     &   YES                 &         YES               &   YES  &   YES                 &         YES               &   YES \\
Controls    &        NO          &   NO               &   NO                      &   YES                  &         YES                &   YES  &   YES                  &         YES                &   YES \\
Trend-adjusted  &        YES          &   YES               &   YES  &   NO                 &         NO                &   NO  &   YES                 &         YES                &   YES \\

\bottomrule
\end{tabular*}
\begin{minipage}[h]{1.7\textwidth}
\medskip
\small \emph{Notes:} This table shows price level effects when estimating equation \ref{eq:5} for sub-samples based on firm size. For wholesale price effects, small corresponds to tier 1 producer-processors, medium to tier 2, and large to tier 3. For retailers, establishments are categorized by chain size.  The listed coefficients are sums of the distributed lag coefficients $E_L$, $L$ months after the minimum wage hikes, relative to the normalized baseline period in $t-2$.Standard errors are clustered at the county level and are shown in parentheses. \sym{*} \(p<0.10\), \sym{**} \(p<0.05\), \sym{***} \(p<0.01\). Data from Washington ESD and Top Shelf Data, August 2018-July 2021.
\end{minipage}

}

\end{table}
\end{landscape}
\restoregeometry

\section{Other margins of adjustment to minimum wage hikes}\label{section:7}

While the primary focus of this paper is the price level effects of minimum wage hikes, firms may adjust to the cost shock along other margins as well. In this section, I examine several other possible channels for firm adjustment.

\subsection{Employment effects}

In this section, I examine the employment effects of minimum wage hikes during the sample period. Since employment information is not available for cannabis establishments, I use monthly employment data from the QCEW at the 5-digit NAICS industry level. I estimate the following distributed lag equation: 
\begin{equation}
   \Delta \ln Emp_{k,t} = \sum_{l =-5}^{6} \beta_l \Delta MW_{t-l} \times Bite_{k,t-l} + X_{k,q(t)} + \gamma_t + \epsilon_{k,t}.\label{eq:8}
\end{equation}
The dependent variable is the first difference of (log) county employment at the 5-digit industry level.\footnote{Cannabis retailers belong to NAICS 45399 ("all other miscellaneous store retailers"). Cannabis producer-processors that grow indoors belong to NAICS 11141 ("food crops grown under cover"). I do not estimate employment effects for the NAICS industry containing outdoor growers because the majority of producer-processors in Washington state grow indoors.} The treatment intensity $\sum_{l =-5}^{6} \beta_l \Delta MW_{t-l} \times Bite_{k,t-l}$ is the same as in the previous sections but for one difference: Since equation \ref{eq:8} is at the county level, $\Delta MW$ does not include citywide minimum wage hikes. Appendix Table \ref{tab:j.1} shows that employment effects are mostly insignificant. Overall, these results suggest that the minimum wage has no effect on employment for cannabis establishments. However, I caution against over-interpreting these results. Since cannabis workers are a subset of employees at the 5-digit NAICS level, one cannot definitively rule out employment effects at cannabis establishments.

\subsection{Demand feedback}

This paper treats minimum wage hikes as a cost shock to cannabis establishments. However, it is conceivable that by raising the incomes of low-wage workers, minimum wage hikes affect demand for cannabis products which in turn could contribute to the retail price effects found in the main part of the paper.\footnote{\cite{leung2021} finds more than full minimum wage pass-through to grocery store prices in the U.S. and attributes part of the price effect to demand feedback.} In this subsection, I test for such demand effects. If the retail price elasticity of demand for cannabis products is non-zero, then a regression of quantity sold on treatment intensity will suffer from simultaneity. Nevertheless, the resulting bias will lead to conservative estimates and hence provide a lower bound---and useful test---of possible demand effects.\footnote{Treatment intensity is endogenous because it simultaneously affects prices and quantity demanded. Since $cov(p_{r,t}, \Delta MW_{r,t-l} \times Bite_{k(r),t-l}) > 0$, $cov(q_{r,t}, \Delta MW_{r,t-l} \times Bite_{k(r),t-l} > 0)$, and $cov(q_{r,t}, p_{j,t}) < 0$, OLS estimates are negatively biased.} To test for demand effects I estimate the following variant of equation \ref{eq:2} for retailers:
\begin{equation}\label{eq:9}
   \Delta q_{r,t} = \sum_{l =-5}^{6} \beta_l \Delta MW_{r,t-l} \times Bite_{k(r),t-l} + \gamma_t + \epsilon_{r,t}.
\end{equation}
The dependent variable is the natural logarithm of the monthly establishment-level quantity index.\footnote{The quantity index is constructed similarly to the price index used in the previous sections. Since a quantity index requires products to have a common unit of measurement, equation \ref{eq:9} is restricted to the product category 'Usable Marijuana' and product quantities are converted to grams.} My results show that minimum wage hikes have no effect on cannabis consumption in the nine month period spanning $t-2$ through $t+6$ (see Appendix Table \ref{tab:j.2}). However, a positive pre-trend is found for the periods $t-6$ through $t-2$. This pre-trend coupled with the flat treatment effects after $t-2$ preclude a strong conclusion on the effects of minimum wage hikes on cannabis demand.

\subsection{Productivity: a discussion}

Productivity is further channel through which the minimum wage may affect firms and workers, and several mechanisms have been proposed in the literature. Since the minimum wage reduces the price of physical capital relative to labor, firms may substitute machinery for workers. \cite{mayernis2018} find evidence of this in the context of the 2004 minimum wage reform in China. However, I view this scenario as unlikely in the cannabis context, since cannabis production leaves little scope for technological adjustments. Most producers grow cannabis indoors in a setting averse to mechanization, and most cannabis is harvested, dried, and trimmed by hand to produce aesthetically pleasing, higher-priced buds \citep{miller2022}.

\cite{ku2022} proposes a different mechanism through which the minimum wage might affect productivity: if the minimum wage causes workers to anticipate potential layoffs, workers may increase their individual effort to avoid being laid off. Alternatively, firms may substitute out of low-skilled labor and into high-skilled labor. I do not observe employment at the establishment level so I cannot test either of these mechanisms. 

The cost shock could also induce firms to adopt better management or organizational practices, which has been shown to increase productivity without the need for physical capital investment \citep{bloom2013,atkin2017}. To the extent that productivity-enhancing changes to business processes are implemented, they are likely long-term adjustments that carry a considerable lag before any productivity gains are realized. In that case, pass-through effects would decrease with successive minimum wage hikes. Yet I find the opposite to be true: of the three minimum wage events in my sample period, event 1 has the smallest pass-through effects while event 3 has the largest. Ultimately, however, I do not observe worker or firm productivity and therefore cannot rule out this channel.

\section{Discussion and conclusion}\label{section:8}

\subsection{Discussion}
One issue with this type of analysis is the degree to which results from one industry can be used to infer price dynamics in other industries. Some of the characteristics that make Washington's cannabis industry an ideal laboratory for studying minimum wage pass-through also set the industry apart. The prevalence of small-scale indoor cultivation, along with rules governing the scale of cultivation, mean that cannabis production is likely to be  more labor intensive than other agricultural industries (I discuss this topic in appendix \ref{appendix:g}). Accordingly, the labor share of variable cost for cannabis producers is expected to be higher, and hence, the cost shock imposed by the minimum wage may be larger compared to other agricultural industries. At the retail level, however, the variable cost structure for cannabis establishments is remarkably similar to conventional retail settings (see appendix \ref{appendix:e}), meaning the relative importance of the wholesale and labor cost shocks should be similar.

Another consideration when comparing industries is the price elasticity of demand, a key determinant of cost pass-through. Conditional on the degree of competition, a higher price elasticity reduces the ability of firms to pass cost shocks onto consumers \citep{weyl2013}. This is an important consideration in the context of cannabis given its potentially addictive nature.\footnote{Epidemiological evidence on cannabis addiction is mixed. Recent evidence suggests that 9\% of regular cannabis users become dependent, compared to 67.5\% for nicotine, 22.7\% for alcohol, and 20.9\% for cocaine \citep{lopez2015}.} Using a similar cannabis scanner dataset as this paper, \cite{hollenbeck2021} estimate an overall retail price elasticity of 1.04 (relative to the outside good), which is higher than elasticities for cigarettes and beer but lower than that for spirits.\footnote{Estimates of aggregate price elasticities range from .16 \citep{gordon2015} to 0.8 \citep{becker1994} for cigarettes; 0.69 to 0.72 for beer \citep{miller2017b}; and 2.8 for spirits \citep{miravete2018}.} Interestingly, this elasticity is not substantially different from those found in non-addictive product markets.\footnote{\cite{miller2017a}, for example, estimate an elasticity of 0.92 for the Portland cement industry.} This suggests that based on the price elasticity alone, retail cost pass-through in cannabis is not expected to differ from other industries studied in the literature.

A final distinction is that the cannabis market operates under statewide autarky. This implies that cannabis retailers may be constrained in their response to the wholesale cost shock since the set of substitutable wholesale products is partly determined by geography. In contrast, retailers in other industries can leverage interstate trade networks to substitute out of products with high pass-through to wholesale prices. Therefore, I view my indirect retail pass-through estimates as more applicable for industries with home bias (e.g. grocery stores), but potentially less applicable for industries with a high degree of geographic substitutability along the supply chain (e.g. drugstores and general merchandise stores).\footnote{Using the 2007 Commodity Flow Survey, \cite{renkin2020} provide evidence of substantial home bias in US grocery consumption.}

\subsection{Conclusion}
In this paper, I study the effects of minimum wage increases on wholesale and retail prices in Washington state's legal recreational cannabis industry. I use scanner-level data to estimate pass-through elasticities across a set of predetermined minimum wage hikes from 2018 to 2021. When ignoring pass-through to wholesale prices, I find that a 10\% increase in the minimum wage raises retail prices by 0.77\%. Yet, I also find substantial pass-through to wholesale prices: a 10\% increase in the minimum wage raises wholesale prices by 1.72\%. The existence of pass-through to wholesale prices implies that retailers face a wholesale cost shock in addition to the labor cost shock. When the empirical model is augmented to account for the wholesale cost shock, retail pass-through more than doubles to 2.04\%. I find that retailers do not adjust markups to wholesale pass-through, indicating a full pass-through of the wholesale cost shock to retail prices. Moreover, pass-through to wholesale prices decreases with the scale of production, which suggests that large producers may adjust to the labor cost shock along other margins. The findings in this paper highlight the importance of examining the entire supply chain---beyond the final point of sale---when investigating the product market effects of minimum wage hikes. 

\newpage
\bibliography{export.bib} 

\begin{thebibliography}{}

\bibitem [\protect \citeauthoryear {%
Aaronson%
}{%
Aaronson%
}{%
{\protect \APACyear {2001}}%
}]{%
aaronson2001}
\APACinsertmetastar {%
aaronson2001}%
\begin{APACrefauthors}%
Aaronson, D.%
\end{APACrefauthors}%
\unskip\
\newblock
\APACrefYearMonthDay{2001}{2}{}.
\newblock
{\BBOQ}\APACrefatitle {Price Pass-Through and the Minimum Wage} {Price
  pass-through and the minimum wage}.{\BBCQ}
\newblock
\APACjournalVolNumPages{Review of Economics and Statistics}{83}{}{158-169}.
\newblock
\begin{APACrefDOI} \doi{10.1162/003465301750160126} \end{APACrefDOI}
\PrintBackRefs{\CurrentBib}

\bibitem [\protect \citeauthoryear {%
Aizpurua-Olaizola%
\ \protect \BOthers {.}}{%
Aizpurua-Olaizola%
\ \protect \BOthers {.}}{%
{\protect \APACyear {2016}}%
}]{%
aiz2016}
\APACinsertmetastar {%
aiz2016}%
\begin{APACrefauthors}%
Aizpurua-Olaizola, O.%
, Soydaner, U.%
, Öztürk, E.%
, Schibano, D.%
, Simsir, Y.%
, Navarro, P.%
\BDBL {}Usobiaga, A.%
\end{APACrefauthors}%
\unskip\
\newblock
\APACrefYearMonthDay{2016}{2}{}.
\newblock
{\BBOQ}\APACrefatitle {Evolution of the Cannabinoid and Terpene Content during
  the Growth of Cannabis sativa Plants from Different Chemotypes} {Evolution of
  the cannabinoid and terpene content during the growth of cannabis sativa
  plants from different chemotypes}.{\BBCQ}
\newblock
\APACjournalVolNumPages{Journal of Natural Products}{79}{}{324-331}.
\newblock
\begin{APACrefDOI} \doi{10.1021/acs.jnatprod.5b00949} \end{APACrefDOI}
\PrintBackRefs{\CurrentBib}

\bibitem [\protect \citeauthoryear {%
Allegretto%
\ \BBA {} Reich%
}{%
Allegretto%
\ \BBA {} Reich%
}{%
{\protect \APACyear {2018}}%
}]{%
reich2018}
\APACinsertmetastar {%
reich2018}%
\begin{APACrefauthors}%
Allegretto, S.%
\BCBT {}\ \BBA {} Reich, M.%
\end{APACrefauthors}%
\unskip\
\newblock
\APACrefYearMonthDay{2018}{1}{}.
\newblock
{\BBOQ}\APACrefatitle {Are Local Minimum Wages Absorbed by Price Increases?
  Estimates from Internet-Based Restaurant Menus} {Are local minimum wages
  absorbed by price increases? estimates from internet-based restaurant
  menus}.{\BBCQ}
\newblock
\APACjournalVolNumPages{ILR Review}{71}{}{35-63}.
\newblock
\begin{APACrefDOI} \doi{10.1177/0019793917713735} \end{APACrefDOI}
\PrintBackRefs{\CurrentBib}

\bibitem [\protect \citeauthoryear {%
Ashenfelter%
\ \BBA {} Štepán Jurajda%
}{%
Ashenfelter%
\ \BBA {} Štepán Jurajda%
}{%
{\protect \APACyear {2022}}%
}]{%
ash2022}
\APACinsertmetastar {%
ash2022}%
\begin{APACrefauthors}%
Ashenfelter, O.%
\BCBT {}\ \BBA {} Štepán Jurajda.%
\end{APACrefauthors}%
\unskip\
\newblock
\APACrefYearMonthDay{2022}{}{}.
\newblock
{\BBOQ}\APACrefatitle {Wages, Minimum Wages, and Price Pass-through: The Case
  of McDonald's Restaurants} {Wages, minimum wages, and price pass-through: The
  case of mcdonald's restaurants}.{\BBCQ}
\newblock
\APACjournalVolNumPages{Journal of Labor Economics}{40}{}{179-201}.
\PrintBackRefs{\CurrentBib}

\bibitem [\protect \citeauthoryear {%
Atkin%
, Chaudhry%
, Chaudry%
, Khandelwal%
\BCBL {}\ \BBA {} Verhoogen%
}{%
Atkin%
\ \protect \BOthers {.}}{%
{\protect \APACyear {2017}}%
}]{%
atkin2017}
\APACinsertmetastar {%
atkin2017}%
\begin{APACrefauthors}%
Atkin, D.%
, Chaudhry, A.%
, Chaudry, S.%
, Khandelwal, A\BPBI K.%
\BCBL {}\ \BBA {} Verhoogen, E.%
\end{APACrefauthors}%
\unskip\
\newblock
\APACrefYearMonthDay{2017}{8}{}.
\newblock
{\BBOQ}\APACrefatitle {Organizational barriers to technology adoption: Evidence
  from soccer-ball producers in Pakistan} {Organizational barriers to
  technology adoption: Evidence from soccer-ball producers in pakistan}.{\BBCQ}
\newblock
\APACjournalVolNumPages{Quarterly Journal of Economics}{132}{}{1101-1164}.
\newblock
\begin{APACrefDOI} \doi{10.1093/qje/qjx010} \end{APACrefDOI}
\PrintBackRefs{\CurrentBib}

\bibitem [\protect \citeauthoryear {%
Barbagallo%
}{%
Barbagallo%
}{%
{\protect \APACyear {2021}}%
}]{%
cbt2021}
\APACinsertmetastar {%
cbt2021}%
\begin{APACrefauthors}%
Barbagallo, P.%
\end{APACrefauthors}%
\unskip\
\newblock
\APACrefYearMonthDay{2021}{6}{}.
\newblock
{\BBOQ}\APACrefatitle {Pioneer State} {Pioneer state}.{\BBCQ}
\newblock
\APACjournalVolNumPages{Cannabis Business Times}{}{}{}.
\newblock
\begin{APACrefURL}
  \url{https://www.cannabisbusinesstimes.com/article/state-of-washington-state-cannabis/}
  \end{APACrefURL}
\PrintBackRefs{\CurrentBib}

\bibitem [\protect \citeauthoryear {%
Barcott%
, With%
, Levenson%
\BCBL {}\ \BBA {} Kudialis%
}{%
Barcott%
\ \protect \BOthers {.}}{%
{\protect \APACyear {2022}}%
}]{%
leafly2022}
\APACinsertmetastar {%
leafly2022}%
\begin{APACrefauthors}%
Barcott, B.%
, With, B\BPBI W.%
, Levenson, M\BPBI S.%
\BCBL {}\ \BBA {} Kudialis, C.%
\end{APACrefauthors}%
\unskip\
\newblock
\APACrefYearMonthDay{2022}{}{}.
\newblock
\APACrefbtitle {Jobs report 2022} {Jobs report 2022}\ \APACbVolEdTR {}{Report}.
\newblock
\APACaddressInstitution{}{Leafly}.
\PrintBackRefs{\CurrentBib}

\bibitem [\protect \citeauthoryear {%
Bartik%
}{%
Bartik%
}{%
{\protect \APACyear {1991}}%
}]{%
bartik1991}
\APACinsertmetastar {%
bartik1991}%
\begin{APACrefauthors}%
Bartik, T\BPBI J.%
\end{APACrefauthors}%
\unskip\
\newblock
\APACrefYear{1991}.
\newblock
\APACrefbtitle {Who Benefits from State and Local Economic Development
  Policies?} {Who benefits from state and local economic development policies?}
\newblock
\APACaddressPublisher{}{W.E. Upjohn Institute}.
\newblock
\begin{APACrefDOI} \doi{10.17848/9780585223940} \end{APACrefDOI}
\PrintBackRefs{\CurrentBib}

\bibitem [\protect \citeauthoryear {%
Becker%
, Grossman%
\BCBL {}\ \BBA {} Murphy%
}{%
Becker%
\ \protect \BOthers {.}}{%
{\protect \APACyear {1994}}%
}]{%
becker1994}
\APACinsertmetastar {%
becker1994}%
\begin{APACrefauthors}%
Becker, G\BPBI S.%
, Grossman, M.%
\BCBL {}\ \BBA {} Murphy, K.%
\end{APACrefauthors}%
\unskip\
\newblock
\APACrefYearMonthDay{1994}{}{}.
\newblock
{\BBOQ}\APACrefatitle {An empirical analysis of cigarette addiction} {An
  empirical analysis of cigarette addiction}.{\BBCQ}
\newblock
\APACjournalVolNumPages{The American Economic Review}{84}{}{396-418}.
\PrintBackRefs{\CurrentBib}

\bibitem [\protect \citeauthoryear {%
Bertrand%
, Duflo%
\BCBL {}\ \BBA {} Mullainathan%
}{%
Bertrand%
\ \protect \BOthers {.}}{%
{\protect \APACyear {2004}}%
}]{%
bertrand2004}
\APACinsertmetastar {%
bertrand2004}%
\begin{APACrefauthors}%
Bertrand, M.%
, Duflo, E.%
\BCBL {}\ \BBA {} Mullainathan, S.%
\end{APACrefauthors}%
\unskip\
\newblock
\APACrefYearMonthDay{2004}{2}{}.
\newblock
{\BBOQ}\APACrefatitle {How Much Should We Trust Differences-In-Differences
  Estimates?} {How much should we trust differences-in-differences
  estimates?}{\BBCQ}
\newblock
\APACjournalVolNumPages{The Quarterly Journal of Economics}{119}{}{249-275}.
\newblock
\begin{APACrefDOI} \doi{10.1162/003355304772839588} \end{APACrefDOI}
\PrintBackRefs{\CurrentBib}

\bibitem [\protect \citeauthoryear {%
Bloom%
, Eifert%
, Mahajan%
, McKenzie%
\BCBL {}\ \BBA {} Roberts%
}{%
Bloom%
\ \protect \BOthers {.}}{%
{\protect \APACyear {2013}}%
}]{%
bloom2013}
\APACinsertmetastar {%
bloom2013}%
\begin{APACrefauthors}%
Bloom, N.%
, Eifert, B.%
, Mahajan, A.%
, McKenzie, D.%
\BCBL {}\ \BBA {} Roberts, J.%
\end{APACrefauthors}%
\unskip\
\newblock
\APACrefYearMonthDay{2013}{2}{}.
\newblock
{\BBOQ}\APACrefatitle {Does management matter? Evidence from india} {Does
  management matter? evidence from india}.{\BBCQ}
\newblock
\APACjournalVolNumPages{Quarterly Journal of Economics}{128}{}{1-51}.
\newblock
\begin{APACrefDOI} \doi{10.1093/qje/qjs044} \end{APACrefDOI}
\PrintBackRefs{\CurrentBib}

\bibitem [\protect \citeauthoryear {%
Bonnet%
, Dubois%
, Boas%
\BCBL {}\ \BBA {} Klapper%
}{%
Bonnet%
\ \protect \BOthers {.}}{%
{\protect \APACyear {2013}}%
}]{%
bonnet2013}
\APACinsertmetastar {%
bonnet2013}%
\begin{APACrefauthors}%
Bonnet, C.%
, Dubois, P.%
, Boas, S\BPBI B\BPBI V.%
\BCBL {}\ \BBA {} Klapper, D.%
\end{APACrefauthors}%
\unskip\
\newblock
\APACrefYearMonthDay{2013}{5}{}.
\newblock
{\BBOQ}\APACrefatitle {Empirical Evidence on the Role of Nonlinear Wholesale
  Pricing and Vertical Restraints on Cost Pass-Through} {Empirical evidence on
  the role of nonlinear wholesale pricing and vertical restraints on cost
  pass-through}.{\BBCQ}
\newblock
\APACjournalVolNumPages{Review of Economics and Statistics}{95}{}{500-515}.
\newblock
\begin{APACrefDOI} \doi{10.1162/REST_a_00267} \end{APACrefDOI}
\PrintBackRefs{\CurrentBib}

\bibitem [\protect \citeauthoryear {%
Borusyak%
, Hull%
\BCBL {}\ \BBA {} Jaravel%
}{%
Borusyak%
\ \protect \BOthers {.}}{%
{\protect \APACyear {2022}}%
}]{%
Borusyak2022}
\APACinsertmetastar {%
Borusyak2022}%
\begin{APACrefauthors}%
Borusyak, K.%
, Hull, P.%
\BCBL {}\ \BBA {} Jaravel, X.%
\end{APACrefauthors}%
\unskip\
\newblock
\APACrefYearMonthDay{2022}{1}{}.
\newblock
{\BBOQ}\APACrefatitle {Quasi-Experimental Shift-Share Research Designs}
  {Quasi-experimental shift-share research designs}.{\BBCQ}
\newblock
\APACjournalVolNumPages{The Review of Economic Studies}{89}{}{181-213}.
\newblock
\begin{APACrefDOI} \doi{10.1093/restud/rdab030} \end{APACrefDOI}
\PrintBackRefs{\CurrentBib}

\bibitem [\protect \citeauthoryear {%
Bossler%
\ \BBA {} Schank%
}{%
Bossler%
\ \BBA {} Schank%
}{%
{\protect \APACyear {2022}}%
}]{%
schank2022}
\APACinsertmetastar {%
schank2022}%
\begin{APACrefauthors}%
Bossler, M.%
\BCBT {}\ \BBA {} Schank, T.%
\end{APACrefauthors}%
\unskip\
\newblock
\APACrefYearMonthDay{2022}{4}{}.
\newblock
{\BBOQ}\APACrefatitle {Wage Inequality in Germany after the Minimum Wage
  Introduction} {Wage inequality in germany after the minimum wage
  introduction}.{\BBCQ}
\newblock
\APACjournalVolNumPages{Journal of Labor Economics}{}{}{}.
\newblock
\begin{APACrefDOI} \doi{10.1086/720391} \end{APACrefDOI}
\PrintBackRefs{\CurrentBib}

\bibitem [\protect \citeauthoryear {%
Burstein%
\ \BBA {} Gopinath%
}{%
Burstein%
\ \BBA {} Gopinath%
}{%
{\protect \APACyear {2014}}%
}]{%
burstein2014}
\APACinsertmetastar {%
burstein2014}%
\begin{APACrefauthors}%
Burstein, A.%
\BCBT {}\ \BBA {} Gopinath, G.%
\end{APACrefauthors}%
\unskip\
\newblock
\APACrefYearMonthDay{2014}{}{}.
\newblock
{\BBOQ}\APACrefatitle {International Prices and Exchange Rates} {International
  prices and exchange rates}.{\BBCQ}
\newblock
\BIn{} \APACrefbtitle {Handbook of International Economics, 4th ed.} {Handbook
  of international economics, 4th ed.}\ (\BVOL~4, \BPG~391-451).
\newblock
\APACaddressPublisher{}{Elsevier}.
\newblock
\begin{APACrefDOI} \doi{https://doi.org/10.1016/B978-0-444-54314-1.00007-0}
  \end{APACrefDOI}
\PrintBackRefs{\CurrentBib}

\bibitem [\protect \citeauthoryear {%
Callaway%
, Goodman-Bacon%
\BCBL {}\ \BBA {} Sant'anna%
}{%
Callaway%
\ \protect \BOthers {.}}{%
{\protect \APACyear {2021}}%
}]{%
callaway2021}
\APACinsertmetastar {%
callaway2021}%
\begin{APACrefauthors}%
Callaway, B.%
, Goodman-Bacon, A.%
\BCBL {}\ \BBA {} Sant'anna, P\BPBI H\BPBI C.%
\end{APACrefauthors}%
\unskip\
\newblock
\APACrefYearMonthDay{2021}{}{}.
\newblock
\APACrefbtitle {Difference-in-Differences with a Continuous Treatment}
  {Difference-in-differences with a continuous treatment}\ \APACbVolEdTR
  {}{Working paper}.
\newblock
\begin{APACrefDOI} \doi{https://doi.org/10.48550/arXiv.2107.02637}
  \end{APACrefDOI}
\PrintBackRefs{\CurrentBib}

\bibitem [\protect \citeauthoryear {%
Card%
}{%
Card%
}{%
{\protect \APACyear {1992}}%
}]{%
card1992}
\APACinsertmetastar {%
card1992}%
\begin{APACrefauthors}%
Card, D.%
\end{APACrefauthors}%
\unskip\
\newblock
\APACrefYearMonthDay{1992}{10}{}.
\newblock
{\BBOQ}\APACrefatitle {Using Regional Variation in Wages to Measure the Effects
  of the Federal Minimum Wage} {Using regional variation in wages to measure
  the effects of the federal minimum wage}.{\BBCQ}
\newblock
\APACjournalVolNumPages{Industrial and Labor Relations Review}{46}{}{22}.
\newblock
\begin{APACrefDOI} \doi{10.2307/2524736} \end{APACrefDOI}
\PrintBackRefs{\CurrentBib}

\bibitem [\protect \citeauthoryear {%
Caulkins%
\ \BBA {} Stever%
}{%
Caulkins%
\ \BBA {} Stever%
}{%
{\protect \APACyear {2010}}%
}]{%
caulkins2010}
\APACinsertmetastar {%
caulkins2010}%
\begin{APACrefauthors}%
Caulkins, J\BPBI P.%
\BCBT {}\ \BBA {} Stever, H\BPBI G.%
\end{APACrefauthors}%
\unskip\
\newblock
\APACrefYearMonthDay{2010}{}{}.
\newblock
\APACrefbtitle {Estimated Cost of Production for Legalized Cannabis} {Estimated
  cost of production for legalized cannabis}\ \APACbVolEdTR {}{Working paper}.
\newblock
\APACaddressInstitution{}{RAND Corporation}.
\PrintBackRefs{\CurrentBib}

\bibitem [\protect \citeauthoryear {%
Cervantes%
}{%
Cervantes%
}{%
{\protect \APACyear {2006}}%
}]{%
cervantes2006}
\APACinsertmetastar {%
cervantes2006}%
\begin{APACrefauthors}%
Cervantes, J.%
\end{APACrefauthors}%
\unskip\
\newblock
\APACrefYear{2006}.
\newblock
\APACrefbtitle {Marijuana Horticulture: The Indoor/Outdoor Medical Grower's
  Bible} {Marijuana horticulture: The indoor/outdoor medical grower's bible}.
\newblock
\APACaddressPublisher{}{Van Patten Publishing}.
\PrintBackRefs{\CurrentBib}

\bibitem [\protect \citeauthoryear {%
Draca%
, Machin%
\BCBL {}\ \BBA {} Reenen%
}{%
Draca%
\ \protect \BOthers {.}}{%
{\protect \APACyear {2011}}%
}]{%
draca2011}
\APACinsertmetastar {%
draca2011}%
\begin{APACrefauthors}%
Draca, M.%
, Machin, S.%
\BCBL {}\ \BBA {} Reenen, J\BPBI V.%
\end{APACrefauthors}%
\unskip\
\newblock
\APACrefYearMonthDay{2011}{}{}.
\newblock
{\BBOQ}\APACrefatitle {Minimum wages and firm profitability} {Minimum wages and
  firm profitability}.{\BBCQ}
\newblock
\APACjournalVolNumPages{American Economic Journal: Applied
  Economics}{3}{}{129-151}.
\newblock
\begin{APACrefDOI} \doi{10.1257/app.3.1.129} \end{APACrefDOI}
\PrintBackRefs{\CurrentBib}

\bibitem [\protect \citeauthoryear {%
Dube%
, Lester%
\BCBL {}\ \BBA {} Reich%
}{%
Dube%
\ \protect \BOthers {.}}{%
{\protect \APACyear {2016}}%
}]{%
dube2016}
\APACinsertmetastar {%
dube2016}%
\begin{APACrefauthors}%
Dube, A.%
, Lester, T\BPBI W.%
\BCBL {}\ \BBA {} Reich, M.%
\end{APACrefauthors}%
\unskip\
\newblock
\APACrefYearMonthDay{2016}{7}{}.
\newblock
{\BBOQ}\APACrefatitle {Minimum Wage Shocks, Employment Flows, and Labor Market
  Frictions} {Minimum wage shocks, employment flows, and labor market
  frictions}.{\BBCQ}
\newblock
\APACjournalVolNumPages{Journal of Labor Economics}{34}{}{663-704}.
\newblock
\begin{APACrefDOI} \doi{10.1086/685449} \end{APACrefDOI}
\PrintBackRefs{\CurrentBib}

\bibitem [\protect \citeauthoryear {%
Dustmann%
, Lindner%
, Schönberg%
, Umkehrer%
\BCBL {}\ \BBA {} Berge%
}{%
Dustmann%
\ \protect \BOthers {.}}{%
{\protect \APACyear {2022}}%
}]{%
lindner2022}
\APACinsertmetastar {%
lindner2022}%
\begin{APACrefauthors}%
Dustmann, C.%
, Lindner, A.%
, Schönberg, U.%
, Umkehrer, M.%
\BCBL {}\ \BBA {} Berge, P\BPBI V.%
\end{APACrefauthors}%
\unskip\
\newblock
\APACrefYearMonthDay{2022}{2}{}.
\newblock
{\BBOQ}\APACrefatitle {Reallocation Effects of the Minimum Wage} {Reallocation
  effects of the minimum wage}.{\BBCQ}
\newblock
\APACjournalVolNumPages{Quarterly Journal of Economics}{137}{}{267-328}.
\newblock
\begin{APACrefDOI} \doi{10.1093/qje/qjab028} \end{APACrefDOI}
\PrintBackRefs{\CurrentBib}

\bibitem [\protect \citeauthoryear {%
Eichenbaum%
, Jaimovich%
\BCBL {}\ \BBA {} Rebelo%
}{%
Eichenbaum%
\ \protect \BOthers {.}}{%
{\protect \APACyear {2011}}%
}]{%
eichenbaum2011}
\APACinsertmetastar {%
eichenbaum2011}%
\begin{APACrefauthors}%
Eichenbaum, M.%
, Jaimovich, N.%
\BCBL {}\ \BBA {} Rebelo, S.%
\end{APACrefauthors}%
\unskip\
\newblock
\APACrefYearMonthDay{2011}{}{}.
\newblock
{\BBOQ}\APACrefatitle {Reference Prices, Costs, and Nominal Rigidities}
  {Reference prices, costs, and nominal rigidities}.{\BBCQ}
\newblock
\APACjournalVolNumPages{The American Economic Review}{101}{}{234-262}.
\newblock
\begin{APACrefDOI} \doi{10.1257/aer.101.1.234} \end{APACrefDOI}
\PrintBackRefs{\CurrentBib}

\bibitem [\protect \citeauthoryear {%
Fougere%
, Gautier%
\BCBL {}\ \BBA {} Bihan%
}{%
Fougere%
\ \protect \BOthers {.}}{%
{\protect \APACyear {2010}}%
}]{%
fougere2010}
\APACinsertmetastar {%
fougere2010}%
\begin{APACrefauthors}%
Fougere, D.%
, Gautier, E.%
\BCBL {}\ \BBA {} Bihan, H\BPBI L.%
\end{APACrefauthors}%
\unskip\
\newblock
\APACrefYearMonthDay{2010}{10}{}.
\newblock
{\BBOQ}\APACrefatitle {Restaurant Prices and the Minimum Wage} {Restaurant
  prices and the minimum wage}.{\BBCQ}
\newblock
\APACjournalVolNumPages{Journal of Money, Credit and Banking}{42}{}{1199-1234}.
\newblock
\begin{APACrefDOI} \doi{10.1111/j.1538-4616.2010.00339.x} \end{APACrefDOI}
\PrintBackRefs{\CurrentBib}

\bibitem [\protect \citeauthoryear {%
Goldsmith-Pinkham%
, Sorkin%
\BCBL {}\ \BBA {} Swift%
}{%
Goldsmith-Pinkham%
\ \protect \BOthers {.}}{%
{\protect \APACyear {2020}}%
}]{%
goldsmith2020}
\APACinsertmetastar {%
goldsmith2020}%
\begin{APACrefauthors}%
Goldsmith-Pinkham, P.%
, Sorkin, I.%
\BCBL {}\ \BBA {} Swift, H.%
\end{APACrefauthors}%
\unskip\
\newblock
\APACrefYearMonthDay{2020}{}{}.
\newblock
{\BBOQ}\APACrefatitle {Bartik Instruments: What, When, Why, and How} {Bartik
  instruments: What, when, why, and how}.{\BBCQ}
\newblock
\APACjournalVolNumPages{American Economic Review}{110}{}{2586-2624}.
\newblock
\begin{APACrefDOI} \doi{10.1257/aer.20181047} \end{APACrefDOI}
\PrintBackRefs{\CurrentBib}

\bibitem [\protect \citeauthoryear {%
Goodman-Bacon%
}{%
Goodman-Bacon%
}{%
{\protect \APACyear {2021}}%
}]{%
gb2021}
\APACinsertmetastar {%
gb2021}%
\begin{APACrefauthors}%
Goodman-Bacon, A.%
\end{APACrefauthors}%
\unskip\
\newblock
\APACrefYearMonthDay{2021}{}{}.
\newblock
{\BBOQ}\APACrefatitle {Difference-in-differences with variation in treatment
  timing} {Difference-in-differences with variation in treatment
  timing}.{\BBCQ}
\newblock
\APACjournalVolNumPages{Journal of Econometrics}{225}{}{254-277}.
\newblock
\begin{APACrefDOI} \doi{10.1016/j.jeconom.2021.03.014} \end{APACrefDOI}
\PrintBackRefs{\CurrentBib}

\bibitem [\protect \citeauthoryear {%
Gopalan%
, Hamilton%
, Kalda%
\BCBL {}\ \BBA {} Sovich%
}{%
Gopalan%
\ \protect \BOthers {.}}{%
{\protect \APACyear {2021}}%
}]{%
gopalan2021}
\APACinsertmetastar {%
gopalan2021}%
\begin{APACrefauthors}%
Gopalan, R.%
, Hamilton, B\BPBI H.%
, Kalda, A.%
\BCBL {}\ \BBA {} Sovich, D.%
\end{APACrefauthors}%
\unskip\
\newblock
\APACrefYearMonthDay{2021}{7}{}.
\newblock
{\BBOQ}\APACrefatitle {State minimum wages, employment, and wage spillovers:
  Evidence from administrative payroll data} {State minimum wages, employment,
  and wage spillovers: Evidence from administrative payroll data}.{\BBCQ}
\newblock
\APACjournalVolNumPages{Journal of Labor Economics}{39}{}{673-707}.
\newblock
\begin{APACrefDOI} \doi{10.1086/711355} \end{APACrefDOI}
\PrintBackRefs{\CurrentBib}

\bibitem [\protect \citeauthoryear {%
Gordon%
\ \BBA {} Sun%
}{%
Gordon%
\ \BBA {} Sun%
}{%
{\protect \APACyear {2015}}%
}]{%
gordon2015}
\APACinsertmetastar {%
gordon2015}%
\begin{APACrefauthors}%
Gordon, B\BPBI R.%
\BCBT {}\ \BBA {} Sun, B.%
\end{APACrefauthors}%
\unskip\
\newblock
\APACrefYearMonthDay{2015}{}{}.
\newblock
{\BBOQ}\APACrefatitle {A dynamic model of rational addiction: Evaluating
  cigarette taxes} {A dynamic model of rational addiction: Evaluating cigarette
  taxes}.{\BBCQ}
\newblock
\APACjournalVolNumPages{Marketing Science}{34}{}{452-470}.
\newblock
\begin{APACrefDOI} \doi{10.1287/mksc.2014.0885} \end{APACrefDOI}
\PrintBackRefs{\CurrentBib}

\bibitem [\protect \citeauthoryear {%
Hallenberg%
}{%
Hallenberg%
}{%
{\protect \APACyear {2017}}%
}]{%
hallenberg2017}
\APACinsertmetastar {%
hallenberg2017}%
\begin{APACrefauthors}%
Hallenberg, P.%
\end{APACrefauthors}%
\unskip\
\newblock
\APACrefYearMonthDay{2017}{2}{}.
\newblock
{\BBOQ}\APACrefatitle {A 51st state, called Liberty, would have political clout
  and an ag-based economy} {A 51st state, called liberty, would have political
  clout and an ag-based economy}.{\BBCQ}
\newblock
\APACjournalVolNumPages{The Spokesman Review}{}{}{}.
\newblock
\begin{APACrefURL}
  \url{https://www.spokesman.com/stories/2017/feb/12/a-51st-state-called-liberty-would-have-political-c/}
  \end{APACrefURL}
\PrintBackRefs{\CurrentBib}

\bibitem [\protect \citeauthoryear {%
Hansen%
, Miller%
\BCBL {}\ \BBA {} Weber%
}{%
Hansen%
\ \protect \BOthers {.}}{%
{\protect \APACyear {2022}}%
}]{%
hansen2021}
\APACinsertmetastar {%
hansen2021}%
\begin{APACrefauthors}%
Hansen, B.%
, Miller, K.%
\BCBL {}\ \BBA {} Weber, C.%
\end{APACrefauthors}%
\unskip\
\newblock
\APACrefYearMonthDay{2022}{8}{}.
\newblock
{\BBOQ}\APACrefatitle {Vertical integration and production inefficiency in the
  presence of a gross receipts tax} {Vertical integration and production
  inefficiency in the presence of a gross receipts tax}.{\BBCQ}
\newblock
\APACjournalVolNumPages{Journal of Public Economics}{212}{}{}.
\newblock
\begin{APACrefDOI} \doi{10.1016/j.jpubeco.2022.104693} \end{APACrefDOI}
\PrintBackRefs{\CurrentBib}

\bibitem [\protect \citeauthoryear {%
Harasztosi%
\ \BBA {} Lindner%
}{%
Harasztosi%
\ \BBA {} Lindner%
}{%
{\protect \APACyear {2019}}%
}]{%
lindner2019}
\APACinsertmetastar {%
lindner2019}%
\begin{APACrefauthors}%
Harasztosi, P.%
\BCBT {}\ \BBA {} Lindner, A.%
\end{APACrefauthors}%
\unskip\
\newblock
\APACrefYearMonthDay{2019}{}{}.
\newblock
{\BBOQ}\APACrefatitle {Who Pays for the Minimum Wage?} {Who pays for the
  minimum wage?}{\BBCQ}
\newblock
\APACjournalVolNumPages{American Economic Review}{109}{}{2693-2727}.
\newblock
\begin{APACrefDOI} \doi{10.1257/aer.20171445} \end{APACrefDOI}
\PrintBackRefs{\CurrentBib}

\bibitem [\protect \citeauthoryear {%
Hollenbeck%
\ \BBA {} Uetake%
}{%
Hollenbeck%
\ \BBA {} Uetake%
}{%
{\protect \APACyear {2021}}%
}]{%
hollenbeck2021}
\APACinsertmetastar {%
hollenbeck2021}%
\begin{APACrefauthors}%
Hollenbeck, B.%
\BCBT {}\ \BBA {} Uetake, K.%
\end{APACrefauthors}%
\unskip\
\newblock
\APACrefYearMonthDay{2021}{8}{}.
\newblock
{\BBOQ}\APACrefatitle {Taxation and Market Power in the Legal Marijuana
  Industry} {Taxation and market power in the legal marijuana industry}.{\BBCQ}
\newblock
\APACjournalVolNumPages{The RAND Journal of Economics}{52}{}{559-595}.
\newblock
\begin{APACrefDOI} \doi{https://doi.org/10.1111/1756-2171.12384}
  \end{APACrefDOI}
\PrintBackRefs{\CurrentBib}

\bibitem [\protect \citeauthoryear {%
Hong%
\ \BBA {} Li%
}{%
Hong%
\ \BBA {} Li%
}{%
{\protect \APACyear {2017}}%
}]{%
hong2017}
\APACinsertmetastar {%
hong2017}%
\begin{APACrefauthors}%
Hong, G\BPBI H.%
\BCBT {}\ \BBA {} Li, N.%
\end{APACrefauthors}%
\unskip\
\newblock
\APACrefYearMonthDay{2017}{3}{}.
\newblock
{\BBOQ}\APACrefatitle {Market structure and cost pass-through in retail}
  {Market structure and cost pass-through in retail}.{\BBCQ}
\newblock
\APACjournalVolNumPages{Review of Economics and Statistics}{99}{}{151-166}.
\newblock
\begin{APACrefDOI} \doi{10.1162/REST_a_00560} \end{APACrefDOI}
\PrintBackRefs{\CurrentBib}

\bibitem [\protect \citeauthoryear {%
Hummels%
, Jørgensen%
, Munch%
\BCBL {}\ \BBA {} Xiang%
}{%
Hummels%
\ \protect \BOthers {.}}{%
{\protect \APACyear {2014}}%
}]{%
hummels2014}
\APACinsertmetastar {%
hummels2014}%
\begin{APACrefauthors}%
Hummels, D.%
, Jørgensen, R.%
, Munch, J.%
\BCBL {}\ \BBA {} Xiang, C.%
\end{APACrefauthors}%
\unskip\
\newblock
\APACrefYearMonthDay{2014}{}{}.
\newblock
{\BBOQ}\APACrefatitle {The wage effects of offshoring: Evidence from danish
  matched worker-firm data} {The wage effects of offshoring: Evidence from
  danish matched worker-firm data}.{\BBCQ}
\newblock
\APACjournalVolNumPages{American Economic Review}{104}{}{1597-1629}.
\newblock
\begin{APACrefDOI} \doi{10.1257/aer.104.6.1597} \end{APACrefDOI}
\PrintBackRefs{\CurrentBib}

\bibitem [\protect \citeauthoryear {%
{International Labour Organization}%
}{%
{International Labour Organization}%
}{%
{\protect \APACyear {2021}}%
}]{%
ilo2021}
\APACinsertmetastar {%
ilo2021}%
\begin{APACrefauthors}%
{International Labour Organization}.%
\end{APACrefauthors}%
\unskip\
\newblock
\APACrefYearMonthDay{2021}{}{}.
\newblock
\APACrefbtitle {Minimum Wage Policy Guide} {Minimum wage policy guide}\
  \APACbVolEdTR {}{Report}.
\newblock
\APACaddressInstitution{}{International Labour Organization}.
\newblock
\begin{APACrefURL}
  \url{https://www.ilo.org/wcmsp5/groups/public/---ed_protect/---protrav/---travail/documents/publication/wcms_508566.pdf}
  \end{APACrefURL}
\PrintBackRefs{\CurrentBib}

\bibitem [\protect \citeauthoryear {%
Jaravel%
}{%
Jaravel%
}{%
{\protect \APACyear {2019}}%
}]{%
Jaravel2019}
\APACinsertmetastar {%
Jaravel2019}%
\begin{APACrefauthors}%
Jaravel, X.%
\end{APACrefauthors}%
\unskip\
\newblock
\APACrefYearMonthDay{2019}{5}{}.
\newblock
{\BBOQ}\APACrefatitle {The Unequal Gains from Product Innovations: Evidence
  from the U.S. Retail Sector} {The unequal gains from product innovations:
  Evidence from the u.s. retail sector}.{\BBCQ}
\newblock
\APACjournalVolNumPages{The Quarterly Journal of Economics}{134}{}{715-783}.
\newblock
\begin{APACrefDOI} \doi{10.1093/qje/qjy031} \end{APACrefDOI}
\PrintBackRefs{\CurrentBib}

\bibitem [\protect \citeauthoryear {%
Jiang%
\ \BBA {} Miller%
}{%
Jiang%
\ \BBA {} Miller%
}{%
{\protect \APACyear {2022}}%
}]{%
miller2022}
\APACinsertmetastar {%
miller2022}%
\begin{APACrefauthors}%
Jiang, S.%
\BCBT {}\ \BBA {} Miller, K.%
\end{APACrefauthors}%
\unskip\
\newblock
\APACrefYearMonthDay{2022}{}{}.
\newblock
{\BBOQ}\APACrefatitle {Watching the grass grow: does recreational cannabis
  legalization affect retail and agricultural wages?} {Watching the grass grow:
  does recreational cannabis legalization affect retail and agricultural
  wages?}{\BBCQ}
\newblock
\APACjournalVolNumPages{Journal of Cannabis Research}{4}{}{42}.
\newblock
\begin{APACrefDOI} \doi{10.1186/s42238-022-00149-6} \end{APACrefDOI}
\PrintBackRefs{\CurrentBib}

\bibitem [\protect \citeauthoryear {%
Ku%
}{%
Ku%
}{%
{\protect \APACyear {2022}}%
}]{%
ku2022}
\APACinsertmetastar {%
ku2022}%
\begin{APACrefauthors}%
Ku, H.%
\end{APACrefauthors}%
\unskip\
\newblock
\APACrefYearMonthDay{2022}{}{}.
\newblock
{\BBOQ}\APACrefatitle {Does Minimum Wage Increase Labor Productivity? Evidence
  from Piece Rate Workers} {Does minimum wage increase labor productivity?
  evidence from piece rate workers}.{\BBCQ}
\newblock
\APACjournalVolNumPages{Journal of Labor Economics}{}{}{}.
\newblock
\begin{APACrefDOI} \doi{10.1086/716347} \end{APACrefDOI}
\PrintBackRefs{\CurrentBib}

\bibitem [\protect \citeauthoryear {%
Leung%
}{%
Leung%
}{%
{\protect \APACyear {2021}}%
}]{%
leung2021}
\APACinsertmetastar {%
leung2021}%
\begin{APACrefauthors}%
Leung, J\BPBI H.%
\end{APACrefauthors}%
\unskip\
\newblock
\APACrefYearMonthDay{2021}{}{}.
\newblock
{\BBOQ}\APACrefatitle {Minimum Wage and Real Wage Inequality: Evidence from
  Pass-Through to Retail Prices} {Minimum wage and real wage inequality:
  Evidence from pass-through to retail prices}.{\BBCQ}
\newblock
\APACjournalVolNumPages{The Review of Economics and Statistics}{103}{}{1-16}.
\newblock
\begin{APACrefDOI} \doi{10.1162/rest_a_00915} \end{APACrefDOI}
\PrintBackRefs{\CurrentBib}

\bibitem [\protect \citeauthoryear {%
Lindo%
, Myers%
, Schlosser%
\BCBL {}\ \BBA {} Cunningham%
}{%
Lindo%
\ \protect \BOthers {.}}{%
{\protect \APACyear {2020}}%
}]{%
cunningham2020}
\APACinsertmetastar {%
cunningham2020}%
\begin{APACrefauthors}%
Lindo, J\BPBI M.%
, Myers, C\BPBI K.%
, Schlosser, A.%
\BCBL {}\ \BBA {} Cunningham, S.%
\end{APACrefauthors}%
\unskip\
\newblock
\APACrefYearMonthDay{2020}{}{}.
\newblock
{\BBOQ}\APACrefatitle {How Far Is Too Far? New Evidence on Abortion Clinic
  Closures, Access, and Abortions} {How far is too far? new evidence on
  abortion clinic closures, access, and abortions}.{\BBCQ}
\newblock
\APACjournalVolNumPages{Journal of Human Resources}{55}{}{1137-1160}.
\newblock
\begin{APACrefURL}
  \url{http://jhr.uwpress.org/lookup/doi/10.3368/jhr.55.4.1217-9254R3}
  \end{APACrefURL}
\newblock
\begin{APACrefDOI} \doi{10.3368/jhr.55.4.1217-9254R3} \end{APACrefDOI}
\PrintBackRefs{\CurrentBib}

\bibitem [\protect \citeauthoryear {%
Lopez-Quintero%
\ \protect \BOthers {.}}{%
Lopez-Quintero%
\ \protect \BOthers {.}}{%
{\protect \APACyear {2011}}%
}]{%
lopez2015}
\APACinsertmetastar {%
lopez2015}%
\begin{APACrefauthors}%
Lopez-Quintero, C.%
, de~los Cobos, J\BPBI P.%
, Hasin, D\BPBI S.%
, Okuda, M.%
, Wang, S.%
, Grant, B\BPBI F.%
\BCBL {}\ \BBA {} Blanco, C.%
\end{APACrefauthors}%
\unskip\
\newblock
\APACrefYearMonthDay{2011}{5}{}.
\newblock
{\BBOQ}\APACrefatitle {Probability and predictors of transition from first use
  to dependence on nicotine, alcohol, cannabis, and cocaine: Results of the
  National Epidemiologic Survey on Alcohol and Related Conditions (NESARC)}
  {Probability and predictors of transition from first use to dependence on
  nicotine, alcohol, cannabis, and cocaine: Results of the national
  epidemiologic survey on alcohol and related conditions (nesarc)}.{\BBCQ}
\newblock
\APACjournalVolNumPages{Drug and Alcohol Dependence}{115}{}{120-130}.
\newblock
\begin{APACrefDOI} \doi{10.1016/j.drugalcdep.2010.11.004} \end{APACrefDOI}
\PrintBackRefs{\CurrentBib}

\bibitem [\protect \citeauthoryear {%
Lucca%
, Nadauld%
\BCBL {}\ \BBA {} Shen%
}{%
Lucca%
\ \protect \BOthers {.}}{%
{\protect \APACyear {2019}}%
}]{%
lucca2019}
\APACinsertmetastar {%
lucca2019}%
\begin{APACrefauthors}%
Lucca, D\BPBI O.%
, Nadauld, T.%
\BCBL {}\ \BBA {} Shen, K.%
\end{APACrefauthors}%
\unskip\
\newblock
\APACrefYearMonthDay{2019}{}{}.
\newblock
{\BBOQ}\APACrefatitle {Credit Supply and the Rise in College Tuition: Evidence
  from the Expansion in Federal Student Aid Programs} {Credit supply and the
  rise in college tuition: Evidence from the expansion in federal student aid
  programs}.{\BBCQ}
\newblock
\APACjournalVolNumPages{Review of Financial Studies}{32}{}{423-466}.
\newblock
\begin{APACrefDOI} \doi{10.1093/rfs/hhy069} \end{APACrefDOI}
\PrintBackRefs{\CurrentBib}

\bibitem [\protect \citeauthoryear {%
Mayneris%
, Poncet%
\BCBL {}\ \BBA {} Zhang%
}{%
Mayneris%
\ \protect \BOthers {.}}{%
{\protect \APACyear {2018}}%
}]{%
mayernis2018}
\APACinsertmetastar {%
mayernis2018}%
\begin{APACrefauthors}%
Mayneris, F.%
, Poncet, S.%
\BCBL {}\ \BBA {} Zhang, T.%
\end{APACrefauthors}%
\unskip\
\newblock
\APACrefYearMonthDay{2018}{}{}.
\newblock
{\BBOQ}\APACrefatitle {Improving or disappearing: Firm-level adjustments to
  minimum wages in China} {Improving or disappearing: Firm-level adjustments to
  minimum wages in china}.{\BBCQ}
\newblock
\APACjournalVolNumPages{Journal of Development Economics}{135}{}{20-42}.
\newblock
\begin{APACrefDOI} \doi{10.1016/j.jdeveco.2018.06.010} \end{APACrefDOI}
\PrintBackRefs{\CurrentBib}

\bibitem [\protect \citeauthoryear {%
K.~Miller%
\ \BBA {} Seo%
}{%
K.~Miller%
\ \BBA {} Seo%
}{%
{\protect \APACyear {2021}}%
}]{%
miller2021}
\APACinsertmetastar {%
miller2021}%
\begin{APACrefauthors}%
Miller, K.%
\BCBT {}\ \BBA {} Seo, B.%
\end{APACrefauthors}%
\unskip\
\newblock
\APACrefYearMonthDay{2021}{3}{}.
\newblock
{\BBOQ}\APACrefatitle {The effect of cannabis legalization on substance demand
  and tax revenues} {The effect of cannabis legalization on substance demand
  and tax revenues}.{\BBCQ}
\newblock
\APACjournalVolNumPages{National Tax Journal}{74}{}{109-145}.
\newblock
\begin{APACrefDOI} \doi{10.1086/712915} \end{APACrefDOI}
\PrintBackRefs{\CurrentBib}

\bibitem [\protect \citeauthoryear {%
N\BPBI H.~Miller%
, Osborne%
\BCBL {}\ \BBA {} Sheu%
}{%
N\BPBI H.~Miller%
\ \protect \BOthers {.}}{%
{\protect \APACyear {2017}}%
}]{%
miller2017a}
\APACinsertmetastar {%
miller2017a}%
\begin{APACrefauthors}%
Miller, N\BPBI H.%
, Osborne, M.%
\BCBL {}\ \BBA {} Sheu, G.%
\end{APACrefauthors}%
\unskip\
\newblock
\APACrefYearMonthDay{2017}{3}{}.
\newblock
{\BBOQ}\APACrefatitle {Pass-through in a concentrated industry: empirical
  evidence and regulatory implications} {Pass-through in a concentrated
  industry: empirical evidence and regulatory implications}.{\BBCQ}
\newblock
\APACjournalVolNumPages{RAND Journal of Economics}{48}{}{69-93}.
\newblock
\begin{APACrefDOI} \doi{10.1111/1756-2171.12168} \end{APACrefDOI}
\PrintBackRefs{\CurrentBib}

\bibitem [\protect \citeauthoryear {%
N\BPBI H.~Miller%
\ \BBA {} Weinberg%
}{%
N\BPBI H.~Miller%
\ \BBA {} Weinberg%
}{%
{\protect \APACyear {2017}}%
}]{%
miller2017b}
\APACinsertmetastar {%
miller2017b}%
\begin{APACrefauthors}%
Miller, N\BPBI H.%
\BCBT {}\ \BBA {} Weinberg, M\BPBI C.%
\end{APACrefauthors}%
\unskip\
\newblock
\APACrefYearMonthDay{2017}{}{}.
\newblock
{\BBOQ}\APACrefatitle {Understanding the Price Effects of the MillerCoors Joint
  Venture} {Understanding the price effects of the millercoors joint
  venture}.{\BBCQ}
\newblock
\APACjournalVolNumPages{Econometrica}{85}{}{1763-1791}.
\newblock
\begin{APACrefDOI} \doi{10.3982/ecta13333} \end{APACrefDOI}
\PrintBackRefs{\CurrentBib}

\bibitem [\protect \citeauthoryear {%
Miravete%
, Seim%
\BCBL {}\ \BBA {} Thurk%
}{%
Miravete%
\ \protect \BOthers {.}}{%
{\protect \APACyear {2018}}%
}]{%
miravete2018}
\APACinsertmetastar {%
miravete2018}%
\begin{APACrefauthors}%
Miravete, E\BPBI J.%
, Seim, K.%
\BCBL {}\ \BBA {} Thurk, J.%
\end{APACrefauthors}%
\unskip\
\newblock
\APACrefYearMonthDay{2018}{}{}.
\newblock
{\BBOQ}\APACrefatitle {Market Power and the Laffer Curve} {Market power and the
  laffer curve}.{\BBCQ}
\newblock
\APACjournalVolNumPages{Econometrica}{86}{}{1651-1687}.
\newblock
\begin{APACrefDOI} \doi{10.3982/ecta12307} \end{APACrefDOI}
\PrintBackRefs{\CurrentBib}

\bibitem [\protect \citeauthoryear {%
Nadreau%
, Fortenbery%
\BCBL {}\ \BBA {} Mick%
}{%
Nadreau%
\ \protect \BOthers {.}}{%
{\protect \APACyear {2020}}%
}]{%
wsu2020}
\APACinsertmetastar {%
wsu2020}%
\begin{APACrefauthors}%
Nadreau, T\BPBI P.%
, Fortenbery, T\BPBI R.%
\BCBL {}\ \BBA {} Mick, T\BPBI B.%
\end{APACrefauthors}%
\unskip\
\newblock
\APACrefYearMonthDay{2020}{}{}.
\newblock
\APACrefbtitle {2020 Contributions of the Washington Cannabis Sector} {2020
  contributions of the washington cannabis sector}\ \APACbVolEdTR {}{Report}.
\newblock
\APACaddressInstitution{}{Washington State University Impact Center}.
\PrintBackRefs{\CurrentBib}

\bibitem [\protect \citeauthoryear {%
Nakamura%
\ \BBA {} Zerom%
}{%
Nakamura%
\ \BBA {} Zerom%
}{%
{\protect \APACyear {2010}}%
}]{%
nakamura2010}
\APACinsertmetastar {%
nakamura2010}%
\begin{APACrefauthors}%
Nakamura, E.%
\BCBT {}\ \BBA {} Zerom, D.%
\end{APACrefauthors}%
\unskip\
\newblock
\APACrefYearMonthDay{2010}{}{}.
\newblock
{\BBOQ}\APACrefatitle {Accounting for Incomplete Pass-Through} {Accounting for
  incomplete pass-through}.{\BBCQ}
\newblock
\APACjournalVolNumPages{Review of Economic Studies}{77}{}{1192-1230}.
\newblock
\begin{APACrefDOI} \doi{10.1111/j.1467-937X.2009.00589.x} \end{APACrefDOI}
\PrintBackRefs{\CurrentBib}

\bibitem [\protect \citeauthoryear {%
Neumark%
, Salas%
\BCBL {}\ \BBA {} Wascher%
}{%
Neumark%
\ \protect \BOthers {.}}{%
{\protect \APACyear {2014}}%
}]{%
neumark2014}
\APACinsertmetastar {%
neumark2014}%
\begin{APACrefauthors}%
Neumark, D.%
, Salas, J\BPBI M\BPBI I.%
\BCBL {}\ \BBA {} Wascher, W.%
\end{APACrefauthors}%
\unskip\
\newblock
\APACrefYearMonthDay{2014}{5}{}.
\newblock
{\BBOQ}\APACrefatitle {Revisiting the Minimum Wage—Employment Debate:
  Throwing Out the Baby with the Bathwater?} {Revisiting the minimum
  wage—employment debate: Throwing out the baby with the bathwater?}{\BBCQ}
\newblock
\APACjournalVolNumPages{ILR Review}{67}{}{608-648}.
\newblock
\begin{APACrefDOI} \doi{10.1177/00197939140670S307} \end{APACrefDOI}
\PrintBackRefs{\CurrentBib}

\bibitem [\protect \citeauthoryear {%
Patrick%
\ \protect \BOthers {.}}{%
Patrick%
\ \protect \BOthers {.}}{%
{\protect \APACyear {2022}}%
}]{%
mtf2022}
\APACinsertmetastar {%
mtf2022}%
\begin{APACrefauthors}%
Patrick, M\BPBI E.%
, Schulenberg, J\BPBI E.%
, Miech, R\BPBI A.%
, Johnston, L\BPBI D.%
, O'malley, P\BPBI M.%
\BCBL {}\ \BBA {} Bachman, J\BPBI G.%
\end{APACrefauthors}%
\unskip\
\newblock
\APACrefYearMonthDay{2022}{9}{}.
\newblock
\APACrefbtitle {Monitoring the Future Panel Study Annual Report.} {Monitoring
  the future panel study annual report.}
\newblock
\begin{APACrefDOI} \doi{10.7826/ISR} \end{APACrefDOI}
\PrintBackRefs{\CurrentBib}

\bibitem [\protect \citeauthoryear {%
Rambachan%
\ \BBA {} Roth%
}{%
Rambachan%
\ \BBA {} Roth%
}{%
{\protect \APACyear {2023}}%
}]{%
roth2022a}
\APACinsertmetastar {%
roth2022a}%
\begin{APACrefauthors}%
Rambachan, A.%
\BCBT {}\ \BBA {} Roth, J.%
\end{APACrefauthors}%
\unskip\
\newblock
\APACrefYearMonthDay{2023}{2}{}.
\newblock
{\BBOQ}\APACrefatitle {A More Credible Approach to Parallel Trends} {A more
  credible approach to parallel trends}.{\BBCQ}
\newblock
\APACjournalVolNumPages{The Review of Economic Studies}{}{}{}.
\newblock
\begin{APACrefDOI} \doi{10.1093/restud/rdad018} \end{APACrefDOI}
\PrintBackRefs{\CurrentBib}

\bibitem [\protect \citeauthoryear {%
Renkin%
, Montialoux%
\BCBL {}\ \BBA {} Siegenthaler%
}{%
Renkin%
\ \protect \BOthers {.}}{%
{\protect \APACyear {2022}}%
}]{%
renkin2020}
\APACinsertmetastar {%
renkin2020}%
\begin{APACrefauthors}%
Renkin, T.%
, Montialoux, C.%
\BCBL {}\ \BBA {} Siegenthaler, M.%
\end{APACrefauthors}%
\unskip\
\newblock
\APACrefYearMonthDay{2022}{9}{}.
\newblock
{\BBOQ}\APACrefatitle {The Pass-Through of Minimum Wages into U.S. Retail
  Prices: Evidence from Supermarket Scanner Data} {The pass-through of minimum
  wages into u.s. retail prices: Evidence from supermarket scanner
  data}.{\BBCQ}
\newblock
\APACjournalVolNumPages{The Review of Economics and
  Statistics}{104}{}{890-908}.
\newblock
\begin{APACrefDOI} \doi{10.1162/rest_a_00981} \end{APACrefDOI}
\PrintBackRefs{\CurrentBib}

\bibitem [\protect \citeauthoryear {%
Schaneman%
}{%
Schaneman%
}{%
{\protect \APACyear {2021}}%
}]{%
schaneman2021}
\APACinsertmetastar {%
schaneman2021}%
\begin{APACrefauthors}%
Schaneman, B.%
\end{APACrefauthors}%
\unskip\
\newblock
\APACrefYearMonthDay{2021}{12}{}.
\newblock
{\BBOQ}\APACrefatitle {Washington state cannabis oversupply spurs calls for
  change} {Washington state cannabis oversupply spurs calls for change}.{\BBCQ}
\newblock
\APACjournalVolNumPages{MJBizDaily}{}{}{}.
\newblock
\begin{APACrefURL}
  \url{https://mjbizdaily.com/washington-state-cannabis-supply-hits-new-low-spurs-calls-change/}
  \end{APACrefURL}
\PrintBackRefs{\CurrentBib}

\bibitem [\protect \citeauthoryear {%
Schmidheiny%
\ \BBA {} Siegloch%
}{%
Schmidheiny%
\ \BBA {} Siegloch%
}{%
{\protect \APACyear {2023}}%
}]{%
siegloth2020}
\APACinsertmetastar {%
siegloth2020}%
\begin{APACrefauthors}%
Schmidheiny, K.%
\BCBT {}\ \BBA {} Siegloch, S.%
\end{APACrefauthors}%
\unskip\
\newblock
\APACrefYearMonthDay{2023}{}{}.
\newblock
{\BBOQ}\APACrefatitle {On Event Studies and Distributed-Lags in Two-Way Fixed
  Effects Models: Identification, Equivalence, and Generalization} {On event
  studies and distributed-lags in two-way fixed effects models: Identification,
  equivalence, and generalization}.{\BBCQ}
\newblock
\APACjournalVolNumPages{Journal of Applied Econometrics}{}{}{}.
\PrintBackRefs{\CurrentBib}

\bibitem [\protect \citeauthoryear {%
{US Census Bureau}%
}{%
{US Census Bureau}%
}{%
{\protect \APACyear {2017}}%
}]{%
naics2007}
\APACinsertmetastar {%
naics2007}%
\begin{APACrefauthors}%
{US Census Bureau}.%
\end{APACrefauthors}%
\unskip\
\newblock
\APACrefYearMonthDay{2017}{}{}.
\newblock
\APACrefbtitle {North American Industrial Classification System.} {North
  american industrial classification system.}
\newblock
\APACaddressPublisher{}{Executive Office of the President of the United
  States}.
\newblock
\begin{APACrefURL} \url{census.gov/naics} \end{APACrefURL}
\PrintBackRefs{\CurrentBib}

\bibitem [\protect \citeauthoryear {%
Wallach%
}{%
Wallach%
}{%
{\protect \APACyear {2014}}%
}]{%
wallach2014}
\APACinsertmetastar {%
wallach2014}%
\begin{APACrefauthors}%
Wallach, P.%
\end{APACrefauthors}%
\unskip\
\newblock
\APACrefYearMonthDay{2014}{8}{}.
\newblock
\APACrefbtitle {Washington's Marijuana Legalization Grows Knowledge, Not Just
  Pot: A Report on the State's Strategy to Assess Reform} {Washington's
  marijuana legalization grows knowledge, not just pot: A report on the state's
  strategy to assess reform}\ \APACbVolEdTR {}{Report}.
\newblock
\APACaddressInstitution{}{Brookings Institution}.
\PrintBackRefs{\CurrentBib}

\bibitem [\protect \citeauthoryear {%
{Washington State Legislature}%
}{%
{Washington State Legislature}%
}{%
{\protect \APACyear {1960}}%
}]{%
wamw}
\APACinsertmetastar {%
wamw}%
\begin{APACrefauthors}%
{Washington State Legislature}.%
\end{APACrefauthors}%
\unskip\
\newblock
\APACrefYearMonthDay{1960}{}{}.
\newblock
\APACrefbtitle {Minimum wages, {WAC} 296-128-050.} {Minimum wages, {WAC}
  296-128-050.}
\newblock
\begin{APACrefURL}
  \url{https://app.leg.wa.gov/wac/default.aspx?cite=296-128&full=true#296-128-050}
  \end{APACrefURL}
\PrintBackRefs{\CurrentBib}

\bibitem [\protect \citeauthoryear {%
{Washington State Liquor and Cannabis Board}%
}{%
{Washington State Liquor and Cannabis Board}%
}{%
{\protect \APACyear {2019}}%
}]{%
lcb2019b}
\APACinsertmetastar {%
lcb2019b}%
\begin{APACrefauthors}%
{Washington State Liquor and Cannabis Board}.%
\end{APACrefauthors}%
\unskip\
\newblock
\APACrefYearMonthDay{2019}{3}{}.
\newblock
\APACrefbtitle {Year One Canopy Report} {Year one canopy report}\ \APACbVolEdTR
  {}{Report}.
\PrintBackRefs{\CurrentBib}

\bibitem [\protect \citeauthoryear {%
{Washington State Liquor and Cannabis Board}%
}{%
{Washington State Liquor and Cannabis Board}%
}{%
{\protect \APACyear {2021}}%
}]{%
lcbtier1}
\APACinsertmetastar {%
lcbtier1}%
\begin{APACrefauthors}%
{Washington State Liquor and Cannabis Board}.%
\end{APACrefauthors}%
\unskip\
\newblock
\APACrefYearMonthDay{2021}{4}{}.
\newblock
\APACrefbtitle {Tier 1 Cannabis Producer Licensee Survey Report} {Tier 1
  cannabis producer licensee survey report}\ \APACbVolEdTR {}{Report}.
\newblock
\begin{APACrefURL}
  \url{https://lcb.wa.gov/sites/default/files/publications/rules/2021%20Proposed%20Rules/Tier1_Report%20_FINAL_Rev%203%20Data%20Attached.pdf}
  \end{APACrefURL}
\PrintBackRefs{\CurrentBib}

\bibitem [\protect \citeauthoryear {%
Weyl%
\ \BBA {} Fabinger%
}{%
Weyl%
\ \BBA {} Fabinger%
}{%
{\protect \APACyear {2013}}%
}]{%
weyl2013}
\APACinsertmetastar {%
weyl2013}%
\begin{APACrefauthors}%
Weyl, E\BPBI G.%
\BCBT {}\ \BBA {} Fabinger, M.%
\end{APACrefauthors}%
\unskip\
\newblock
\APACrefYearMonthDay{2013}{}{}.
\newblock
{\BBOQ}\APACrefatitle {Pass-Through as an Economic Tool: Principles of
  Incidence under Imperfect Competition} {Pass-through as an economic tool:
  Principles of incidence under imperfect competition}.{\BBCQ}
\newblock
\APACjournalVolNumPages{Journal of Political Economy}{121}{}{}.
\PrintBackRefs{\CurrentBib}

\bibitem [\protect \citeauthoryear {%
Xu%
}{%
Xu%
}{%
{\protect \APACyear {2022}}%
}]{%
xu2022}
\APACinsertmetastar {%
xu2022}%
\begin{APACrefauthors}%
Xu, C.%
\end{APACrefauthors}%
\unskip\
\newblock
\APACrefYearMonthDay{2022}{11}{}.
\newblock
{\BBOQ}\APACrefatitle {RESHAPING GLOBAL TRADE: THE IMMEDIATE AND LONG-RUN
  EFFECTS OF BANK FAILURES} {Reshaping global trade: The immediate and long-run
  effects of bank failures}.{\BBCQ}
\newblock
\APACjournalVolNumPages{Quarterly Journal of Economics}{137}{}{2107-2161}.
\newblock
\begin{APACrefDOI} \doi{10.1093/qje/qjac016} \end{APACrefDOI}
\PrintBackRefs{\CurrentBib}

\end{thebibliography}

\newpage

\appendix
\setcounter{table}{0}
\setcounter{figure}{0}
\renewcommand{\thetable}{\Alph{section}\arabic{table}}
\renewcommand{\thefigure}{\Alph{section}\arabic{figure}}

\newpage
\section{Firm costs in the cannabis industry}
\label{appendix:g}

\subsection{Cannabis labor}

\subsubsection*{The labor intensive nature of cannabis production}

Most cannabis plants are dioecious, meaning there are separate male and female plants, and buds with high concentrations of psychoactive compounds are exclusively produced by unpollinated female plants. Since pollination leads to seed production and inferior buds, producers are typically forced to discard the entire crop if it becomes cross-pollinated.  Therefore, unlike other dioecious crops like fruits or nuts, where males and females must co-mingle, cannabis producers must carefully identify and remove any male cannabis plants from the growing area, since even a single male plant can pollinate---and thereby ruin---an entire crop. This laborious process is compounded by a heavy reliance on indoor cultivation, which is generally considered to be more labor intensive than outdoor production.\footnote{Indoor cultivation offers stable growing conditions, year-round harvests, and enables more potent buds \cite{aiz2016}.} When cannabis plant buds have matured, they are harvested and trimmed by hand, a process which takes up to six hours per pound \citep{cervantes2006}. Trimming is particularly labor intensive, as workers use hand trimmers to manually shape the harvested buds.\footnote{As noted by \citep{miller2021}, growers have shied away from mechanized trimmers since hand-trimming allows producers to extract higher quality buds and fetch higher prices from consumers.} Other tasks like filling pre-roll shakers are also largely done by hand.\par


\subsubsection*{Wages in cannabis}

Wages in cannabis are significantly lower than in other industries in Washington state. This should come as no surprise: at the retail level, budtending is a low-skill job that requires no formal education, while the same holds for most jobs at the producer level.\footnote{Producers typically employ a small number of "master growers" who are trained in cultivation, along with a much larger number of low-wage employees engaged in garden labor (e.g. harvesting, drying, trimming), filling pre-rolls, packaging, delivery and other manual labor tasks.} Table \ref{tab:g.1} shows the average annual wage for cannabis establishments for the years 2018-2020 and compares it to the statewide average for the corresponding NAICS industry and all industries combined. For producers, the annual gross wage gap to NAICS 111 is less than 3\%; for retailers, the gap to NAICS 453 ranges from 8\% to 11\%. When converted to hourly wages (assuming 2,080 hours per year), the wage gap between cannabis producers and NAICS 111 ranges from \$0.22 to \$0.37 per hour. For cannabis retailers, the gap is slightly larger: on average, cannabis employees earned between \$0.95 and \$1.58 less per hour than than NAICS 453, which amounts to a difference of 8\% to 11\%. 

\begin{table}[!htbp] 
\centering
\caption{Annual gross wages in the Washington state cannabis industry}
\label{tab:g.1}
\renewcommand{\tabcolsep}{.5pt}{
\def\sym#1{\ifmmode^{#1}\else\(^{#1}\)\fi}
\begin{tabular*}{\hsize}{@{\hskip\tabcolsep\extracolsep\fill}l*{8}{c}}
\toprule

& \multicolumn{3}{c}{Wholesale} & \multicolumn{3}{c}{Retail} & & \\
\addlinespace
\cline{2-4} \cline{5-7} 
\addlinespace
\multicolumn{1}{c}{Year} & \multicolumn{1}{c}{\parbox{1cm}{Cannabis wholesale}} & \multicolumn{1}{c}{\parbox{1cm}{NAICS 111}} & \multicolumn{1}{c}{\parbox{1cm}{NAICS 111419}} & \multicolumn{1}{c}{\parbox{1cm}{Cannabis retail}} & \multicolumn{1}{c}{\parbox{1cm}{NAICS 453}} 
& \multicolumn{1}{c}{\parbox{1cm}{NAICS 453998}}  & \multicolumn{1}{c}{\parbox{1cm}{All private inds.}} & \multicolumn{1}{c}{\parbox{1cm}{Min. wage}} \\
\addlinespace

\midrule

2018  &  \$27,906 & \$28,804 & \$28,371 & \$26,126 & \$28,116 & \$31,848 & \$66,156 & \$23,920 \\

\addlinespace
2019 & \$29,713 & \$30,499 & \$30,417 & \$27,468 & \$29,798 & \$32,922 & \$57,185 & \$24,960 \\

\addlinespace

2020 & \$32,315 & \$33,026 & \$33,459 & \$29,534 & \$32,847 & \$34,847 & \$76,801 & \$28,080 \\

\bottomrule

\end{tabular*}
\begin{minipage}[h]{\textwidth}
\medskip
\small \emph{Notes:} This table compares average annual gross wage for workers at cannabis establishments for the years 2018-2020. Average annual gross wage is obtained by dividing total wages by average covered employment. Minimum wage is based on 2,080 hours per year. Data for 2021 is not available. Data from Washington state ESD and High Peak Strategy.
\end{minipage}

}

\end{table}

\newpage
\subsection{Variable cost structure for cannabis retailers}

\label{appendix:e}
Table \ref{tab:e1} illustrates that cannabis retailers have a similar variable cost structure as other retail industries studied in the literature. \cite{renkin2020}, for example, find that for U.S. grocery stores, COGS accounts for 83\% of variable costs. Note that in most retail settings, cost of goods sold (COGS) and labor cost together account for 99\% of variable cost \citep{renkin2020}. Other expenditures like packaging and transport costs typically make up less than 1\% of variable cost. 
 
\begin{table}[!htbp] 
\centering
\caption{COGS and the labor share of costs for cannabis retailers}
\label{tab:e1}
\renewcommand{\tabcolsep}{.5pt}{
\def\sym#1{\ifmmode^{#1}\else\(^{#1}\)\fi}
\begin{tabular*}{\hsize}{@{\hskip\tabcolsep\extracolsep\fill}l*{5}{c}}
\toprule 

& \multicolumn{2}{c}{Average expenditure} & & \multicolumn{2}{c}{Variable cost share} \\
\cline{2-3} \cline{5-6} \\

\multicolumn{1}{c}{Year} & \multicolumn{1}{c}{Labor} & \multicolumn{1}{c}{COGS} & & \multicolumn{1}{c}{Labor} & \multicolumn{1}{c}{COGS} \\

\midrule

2018  &  \$324,582 & \$702,358 & & 0.32 & 0.68 \\
\addlinespace

2019 & \$370,897 & \$1,187,462 & & 0.24 & 0.76 \\
\addlinespace

2020 & \$407,273 & \$1,584,301 & & 0.20 & 0.80 \\
\bottomrule

\end{tabular*}
\begin{minipage}[h]{\textwidth}
\medskip
\small \emph{Notes:} This table compares average annual labor expenditure and COGS expenditure for cannabis retail establishments in Washington state for the years 2018-2020. Aggregate payroll data on cannabis retailers is from the Washington state ESD and High Peak Strategy (2018-2020). Labor expenditure equals total wages divided by the number of active retail establishments. Establishments with missing UI data are excluded from total wages and establishment counts. COGS is the average annual wholesale expenditure for cannabis retailers in the estimation sample. Wholesale purchases from processor-only licenses are included. Wholesale expenditure data from Top Shelf Data (2018-2020). 
\end{minipage}

}
\end{table}

\newpage
\subsection{The geography of wholesale costs}

Table \ref{tab:h.1} shows the percentage of retailers' wholesale costs in relation to a producer's geographic location. Column 1 shows that only 5.22\% of retailers' wholesale expenditures go to producers located in the same city as the retailer. Column 2 shows that less than 15\% goes to producers in the same county as the retailer. For column 3, I sort counties into their respective 3-digit zip codes (retailers are located in 14 3-digit zip codes compared to 37 counties). Column 3 shows that less than 16\% of wholesale cost goes to producers located in the same 3-digit zip code. Next, I sort counties into three regions (west, central, east), defined by well-established topographic and economic boundaries. Column 4 shows that 62\% of wholesale sales go to retailers in a different region than the producer. Column 5 looks at the subset of establishments located in the west and east regions of the state, thus dropping producers in the central region. The east and west regions are non-contiguous and are located on opposite sides of the state. For establishments located in these two regions, 23.9\% of wholesale sales go to retailers located in the other region, that is to say, retailers on the opposite side of the state. Because the majority of establishments are located in the west and east regions, this share amounts to 21.4\% of all of wholesale expenditures in the industry. Taken together, the results from Table \ref{tab:h.1} illustrate that there is no home bias in wholesale cannabis purchases.

\begin{table}[!htbp] 
\centering
\caption{Share of retailers' wholesale costs by geographic proximity}
\label{tab:h.1}
\renewcommand{\tabcolsep}{1pt}{
\def\sym#1{\ifmmode^{#1}\else\(^{#1}\)\fi}
\begin{small}\begin{tabular*}{\hsize}{@{\hskip\tabcolsep\extracolsep\fill}l*{6}{c}}
 \toprule
 \addlinespace
            &\multicolumn{1}{c}{(1)} &\multicolumn{1}{c}{(2)} & \multicolumn{1}{c}{(3)} &\multicolumn{1}{c}{(4)} &\multicolumn{1}{c}{(5)}  &\multicolumn{1}{c}{(6)}\\
\addlinespace
            &\multicolumn{1}{c}{\parbox{1cm}{Same city}}&\multicolumn{1}{c}{\parbox{1cm}{Same county}} &\multicolumn{1}{c}{\parbox{1cm}{Same 3-digit zip code}} &\multicolumn{1}{c}{\parbox{1cm}{Same region}} &\multicolumn{1}{c}{\parbox{1cm}{Non-contiguous region}}  &\multicolumn{1}{c}{\parbox{1cm}{Same state}}\\
\addlinespace
\midrule
\multicolumn{1}{l}{\parbox{2cm}{Percent of \\ wholesale expenditure}}   & 5.22\% & 14.67\% & 15.59\% & 62.08\% & 23.90\%  & 100\%\\
\bottomrule
\end{tabular*}\end{small}
\begin{minipage}[h]{\textwidth}
\medskip
\small \emph{Notes:} This table shows the share of retailers' wholesale expenditure according to wholesalers' geographic proximity. The shares are based on 5.92 million unique wholesaler-retailer-product-month observations from August 2018 through July 2021. Retailers are located in 14 3-digit zip codes and 35 counties. Region groups counties into three categories: west, central, or east. Data from Top Shelf Data.
\end{minipage}
}

\end{table}

\newpage
\subsection{The relationship between producer and retail bite}

This section illustrates that the relationship between producer and retail bite within a county is weak. The raw correlation between bite for NAICS 111 and NAICS 453 is 0.18, but when controlling for county controls and FE, the relationship shrinks to -0.03. This indicates a very weak relationship between bite within counties.

\begin{table}[!htbp] 
\centering
\caption{Within-county correlation between producer and retail bite}
\label{tab:h.2}
\renewcommand{\tabcolsep}{1pt}{
\def\sym#1{\ifmmode^{#1}\else\(^{#1}\)\fi}
\begin{small}\begin{tabular*}{.3\hsize}{@{\hskip\tabcolsep\extracolsep\fill}l*{3}{c}}
 \toprule
 \addlinespace
           & \multicolumn{1}{c}{(1)} &\multicolumn{1}{c}{(2)} & \multicolumn{1}{c}{(3)} \\
\addlinespace
&\multicolumn{1}{c}{\parbox{1cm}{\centering Raw correlation}} &\multicolumn{1}{c}{\parbox{1cm}{\centering FE}} &\multicolumn{1}{c}{\parbox{1cm}{\centering No FE}} \\
\addlinespace
\midrule
     & 0.18  & -0.03	& -0.05 \\
      &  & (0.09) & (0.09) \\
$N$ & 96 & 96 & 96 \\        
\bottomrule
\end{tabular*}\end{small}
\begin{minipage}[h]{\textwidth}
\medskip
\small \emph{Notes:} Column 1 shows the unconditional within-county correlation for bite. Columns 2 and 3 show OLS estimates from a county-level regression of bite for NAICS 111 on bite for NAICS 453, with county controls (log average wage and the unemployment rate, both for Q3). SE are in parentheses. Data from WA ESD, 2018-2020.
\end{minipage}
}

\end{table}

            


\newpage
\section{Details on prices and price indexes}
\label{appendix:a}
\label{appendix:f}
\setcounter{table}{0}
\setcounter{figure}{0}

\subsection{Price descriptive statistics}

Producer licenses are based on a three-tier system governing the square footage of plant canopy that an establishment is legally permitted to operate. Tier 1 producers can grow up to 2,000 square feet of plant canopy, tier 2 can grow up to 10,000 square feet, while tier 3 can operate up to 30,000 square feet.\footnote{Tiers were assigned to establishments before the market fist opened in 2014 and once assigned an establishment cannot switch tiers. While producers can produce below the threshold for their tier, actual canopy usage has been found to be proportional to those thresholds \citep{lcb2019b}.} A common complaint among cannabis producers is that small producers operate on slim margins while large producers enjoy a higher degree of market power \citep{lcbtier1}. If large producers have more market power, then one would expect higher markups (all else equal). While producer markups are not observable in the data, wholesale prices serve as a proxy for markups under the assumption of homogeneous marginal input costs across producers and product subcategories. Table \ref{tab:a.00} shows a U-shaped relationship between wholesale prices and the scale of production for nearly all product subcategories. For 3.5g and 7g of usable marijuana, average wholesale prices differ by less than 1\% between small and large producers. The price gap grows to 5\% for 14g products and 21\% for 28g products, reflecting a flatter quantity discount for large producers compared to small producers.\footnote{For producers, the LCB reports the unit weight for some product types (e.g. flower lots) in 1g units regardless of how the product is actually bundled. As a result, the average wholesale price listed under 1g is partly based on per-gram prices of larger package sizes. Therefore, caution is warranted when interpreting differences in wholesale prices for 1g products across the production tiers.} However, note that 3.5g and 7g products generate over 2.5 times as much revenue as the higher unit weights, which indicates that for the majority of their sales, large producers are not able to charge higher prices compared to small producers.\footnote{One cannot rule out unobserved quality differences in large packages between small and large producers. If small producers sell lower quality buds in large package sizes (compared to large producers), then this may explain the steeper quantity discounts for small producers.} This implies that the heterogeneous pass-through across the scale of production shown in Table \ref{tab:6.4} in the main part of the paper is not the result of different initial price levels between smaller and larger producers.

\begin{table}[!htbp] 
\centering
\caption{Average wholesale prices for major product subcategories}
\label{tab:a.00}
\renewcommand{\tabcolsep}{1pt}{
\def\sym#1{\ifmmode^{#1}\else\(^{#1}\)\fi}
\begin{tabular*}{\hsize}{@{\hskip\tabcolsep\extracolsep\fill}l*{4}{c}}
\toprule

&\multicolumn{1}{c}{(1)}&\multicolumn{1}{c}{(2)}&\multicolumn{1}{c}{(3)} & \\
\addlinespace
\multicolumn{1}{c}{\parbox{2cm}{Unit weight (grams)}} &\multicolumn{1}{c}{\parbox{2cm}{\centering Small producers}} &\multicolumn{1}{c}{\parbox{2cm}{\centering Medium-sized producers}} & \multicolumn{1}{c}{\parbox{2cm}{\centering Large producers}} & \multicolumn{1}{c}{\parbox{2cm}{Sales (millions of \$)}} \\

\addlinespace
\midrule
\addlinespace

&\multicolumn{4}{c}{Concentrate for inhalation} \\
\cline{2-5}
\addlinespace
$0.5$g    & 9.60 & 12.85 & 11.09 & 32.03 \\

\addlinespace
$1$g   & 11.22 & 9.25 & 11.57 & 300.32 \\

\addlinespace
&\multicolumn{4}{c}{Usable marijuana} \\
\cline{2-5}
\addlinespace

$1$g$^{+}$    & 4.00 & 3.25 & 2.97 & 99.83 \\

\addlinespace
$3.5$g   & 11.70 & 10.25 & 11.85 & 257.83 \\

\addlinespace
$7$g   & 22.33 & 20.33 & 22.66 & 74.03 \\

\addlinespace
$14$g   & 37.53 & 32.63 & 39.47  & 49.81 \\

\addlinespace
$28$g  & 55.75 & 54.56 & 67.58 & 75.65 \\

\addlinespace

\bottomrule
\end{tabular*}
\begin{minipage}[h]{\textwidth}
\medskip
\small \emph{Notes:} This table reports average wholesale prices for the main cannabis product subcategories. Column 1 shows average prices for tier 1 producers (less than 2,000 sq. ft. of plant canopy); Column 2 corresponds to tier 2 producers (2,000 to 10,000 sq. ft. of plant canopy); Column 3 corresponds to tier 3 producers (10,000 to 30,000 sq. ft. of plant canopy). Data: Top Shelf Data, (August 2018 - July 2021). \\
$^{+}$ For producers, the LCB reports the unit weight for some product types (e.g. flower lots) in 1g units regardless of how the product is actually bundled. As a result, for usable marijuana the average wholesale price listed under 1g is partly based on per-gram prices of larger package sizes.
\end{minipage}

}

\end{table}

Table \ref{tab:a.01} reports average retail prices across chain size and product subcategories. Unlike the case with wholesale prices, retail prices exhibit an inverse U-shaped relationship between price and chain size, as independent retailers and large chains have higher average prices compared to mid-sized chains (2-3 stores). Importantly, however, for a given product subcategory, average prices are quite homogeneous across chain size. This again indicates that the heterogeneous pass-through shown in Table \ref{tab:6.4} in the main part of the paper does not reflect differences in initial prices between independent and chain stores. 

\begin{table}[!htbp] 
\centering
\caption{Average retail prices for major product subcategories}
\label{tab:a.01}
\renewcommand{\tabcolsep}{1pt}{
\def\sym#1{\ifmmode^{#1}\else\(^{#1}\)\fi}
\begin{tabular*}{\hsize}{@{\hskip\tabcolsep\extracolsep\fill}l*{4}{c}}
\toprule

&\multicolumn{1}{c}{(1)}&\multicolumn{1}{c}{(2)}&\multicolumn{1}{c}{(3)} & \\
\multicolumn{1}{c}{Unit weight (grams)} &\multicolumn{1}{c}{\parbox{1cm}{\centering Independent stores}} &\multicolumn{1}{c}{\parbox{1cm}{\centering 2-3 stores}} & \multicolumn{1}{c}{\parbox{1cm}{\centering 4+ stores}} &\multicolumn{1}{c}{\parbox{2cm}{\centering Sales (millions of \$)}}\\

\addlinespace
\midrule
\addlinespace

 &\multicolumn{4}{c}{Concentrate for inhalation} \\
\cline{2-5}
\addlinespace
$0.5$g    & 28.57 & 29.41 & 28.82 & 116.27 \\

\addlinespace
$1$g   & 29.51 & 29.55 & 29.45 & 969.54 \\

\addlinespace
&\multicolumn{4}{c}{Usable marijuana} \\
\cline{2-5}
\addlinespace

$1$g    & 8.47 & 8.55 & 8.49 & 276.14 \\

\addlinespace
$3.5$g   & 28.29 & 29.06 & 28.17 & 677.17 \\

\addlinespace
$7$g   & 51.37 & 53.49 & 51.95 & 193.59 \\

\addlinespace
$14$g   & 82.62 & 83.38 & 80.18 & 137.49 \\

\addlinespace
$28$g  & 125.41 & 130.42 & 128.77 & 215.89 \\

\addlinespace

\bottomrule
\end{tabular*}
\begin{minipage}[h]{\textwidth}
\medskip
\small \emph{Notes:} This table reports average retail prices for the main cannabis product subcategories. Column 1 shows average prices for establishments that do not belong to a retail chain. Column 2 corresponds to establishments belonging to a retail chain comprising 2-3 stores. Column 3 corresponds to establishments belonging to a chain with 4-5 stores. Data: Top Shelf Data, (August 2018 - July 2021).

\end{minipage}

}

\end{table}

\newpage
\subsection{Establishment-level price indexes}

My empirical analysis uses traceability data provided by the data analytic firm Top Shelf Data (TSD), which ingests the raw tracking data from the Liquor and Cannabis Board (LCB) and matches it with additional product information. Note that the raw tracking data from the LCB includes each product's SKU, but TSD does not report this. Instead, each product is identified by a unique combination of five elements: retailer-producer-category-unit weight-product name. For products with no unit weight (such as liquid edibles), the first four elements identify the product. TSD then calculates the average price of product $i$ at retail establishment $j$ in month $t$ as
\begin{equation}
    P_{i,j,t} = \frac{TR_{i,j,t}}{TQ_{i,j,t}}.
\end{equation}
 where $TR_{i,j,t}$ is the revenue from product $i$ at retailer $j$ in month $t$, and $TQ_{i,j,t}$ is total quantity.
 
 To construct establishment-level price indexes, I employ a two step process similar to that used by \cite{renkin2020}. In the first step, I use $P_{i,j,t}$ to construct a geometric mean of month-over-month changes for product subcategory $c$ at establishment $j$:
\begin{equation}\label{eq:a.1}
    I_{c,j,t} = \prod_i \left( \frac{P_{i,j,t}}{P_{i,j,t-1}}\right)^{\omega_{i,c,y(t)}}
\end{equation}
where each subcategory is a unique category-unit weight combination.\footnote{Since unit weight is a major component of cannabis product differentiation (akin to volume in beverage sales), the majority of sales contain information on unit weight. Therefore, in the first step of the establishment index, I choose to aggregate at category-unit weight level rather than the category level.} For example, 1.0g usable marijuana and 2.0 gram usable marijuana are separate subcategories. Following \cite{renkin2020}, the weight $\omega_{i,c,y(t)}$ is the share of product $i$ in total revenue of subcategory $c$ in establishment $j$ during the calendar year of month $t$.\footnote{As pointed out by\cite{renkin2020}, price indexes are often constructed using lagged quantity weights. Since product turnover is high in cannabis retail, lagged weights would limit the number of products used in constructing the price indexes. Thus, contemporaneous weights are used.}

In the second step, I aggregate across subcategories to get the price index for establishment $j$ in month $t$:
\begin{equation}\label{eq:a.2}
    I_{j,t} = \prod_c I_{c,j,t}^{\omega_{c,j,y(t)}}.
\end{equation}
Similar to the last step, the weight $\omega_{c,j,y(t)}$ is the share of subcategory $c$ in total revenue in establishment $j$ during the calendar year of month $t$.

Establishment-level price indexes for producers are constructed in a very similar manner as with retailers, but for two exceptions. First, at the wholesale level a product is identified by a unique combination of four  elements (not five as with retailers): producer-category-unit weight-product name. While a retailer may sell similar products produced by different producers, a producer creates the product and sells it to many retailers, which makes it unnecessary to identify a product at the five-element level. Note that this still allows for wholesale price discrimination, since the wholesale price of a single product may differ among retailers. Second, the wholesale price data exhibits much larger variation in prices compared to the retail data. As a result, the product-level index $\frac{P_{i,j,t}}{P_{i,j,t-1}}$ in eq. \ref{eq:a.1} leads to a few inconceivable outliers such as a 562-factor increase in prices from one month to the next. To prevent outliers from driving results and to reduce standard errors in my estimation, I trim the top and bottom 0.1\% of the product indexes before calculating the subcategory index in equation \ref{eq:a.1}. As Table \ref{tab:a.1} illustrates, trimming does not meaningfully change the location or shape of the distribution but lowers the standard deviation considerably.

\begin{table}[!htbp] 
\centering
\caption{Product-level price indexes}
\label{tab:a.1}
{
\def\sym#1{\ifmmode^{#1}\else\(^{#1}\)\fi}
\begin{tabular*}{\hsize}{@{\hskip\tabcolsep\extracolsep\fill}l*{3}{c}}

\toprule

&\multicolumn{2}{c}{Wholesale} & \multicolumn{1}{c}{Retail}\\
\cline{2-3} \cline{4-4}
\addlinespace

&\multicolumn{1}{c}{No trim} & \multicolumn{1}{c}{0.2\% trim} & \multicolumn{1}{c}{No trim}\\

\midrule
\addlinespace
Mean      &     1.004333 &     1.000440 & 1.000028 \\

\addlinespace

St. dev.      &   0.816940  & 0.026360 & 0.015641\\
\addlinespace
Min    &   0.000667   &     0.652272 & 0.009345 \\
\addlinespace
1\% & 0.940171 & 0.946112 & 0.985232 \\
\addlinespace
25\%       &   0.999989  &  0.999989  &  0.999848 \\
\addlinespace
Median     & 1.000000    & 1.000000 &  1.000000 \\
\addlinespace
75\%       &   1.000000 & 1.000000 & 1.000139\\
\addlinespace
99\% & 1.067935 & 1.060525 &  1.014273 \\
\addlinespace
Max       &   562.785120 & 1.646053 & 15.273730\\

\addlinespace
$N$       &   1,658,554 & 1,657,326 & 7,590,876 \\

\bottomrule
\end{tabular*}
\begin{minipage}[h]{\textwidth}
\medskip
\small \emph{Notes:} This table shows descriptive statistics for product-level price indexes, $\frac{P_{i,j,t}}{P_{i,j,t-1}}$. The price index forms the basis for the subcategory index (i.e. the first step of the establishment index). Product-level price indexes are not trimmed for retailers because they exhibit much less variation than for wholesalers. Data source: Top Shelf Data, August 2018-July 2021.
\end{minipage}

}
\end{table}

\subsection{Establishment-level markup index}

This section describes the construction of the establishment-level markup indexes used in section \ref{section:6}. For each individual product, the markup over marginal input cost (MIC) is defined as:
\begin{equation}
    \mu_{i,r,t} = \frac{P_{i,r,t}}{MC_{i,r,t}}
\end{equation}
where $P_{i,r,t}$ is the price of product $i$ sold by retail establishment $r$ in month $t$, and $MC_{i,r,t}$ is the wholesale price that retailer $r$ pays for that very same product in period $t$. Note that this formulation implicitly assumes that the markup of interest is contemporaneous (period $t$ retail price and period $t$ wholesale price) rather than lagged (period $t$ retail price and period $t-1$ wholesale price).

To construct establishment-level markup indexes, I employ a two step procedure similar to that used for the price indexes throughout the paper. In the first step, I use $\mu_{i,r,t}$ to construct a geometric mean of month-over-month changes in markups for product subcategory $c$ at establishment $r$:
\begin{equation}
    I^{\mu}_{c,r,t} = \prod_i \left( \frac{\mu_{i,r,t}}{\mu_{i,r,t-1}}\right)^{\omega_{i,c,y(t)}}
\end{equation}
where each subcategory is a unique category-unit weight combination. The weight $\omega_{i,c,y(t)}$ is the share of product $i$ in total revenue of subcategory $c$ in establishment $r$ during the calendar year of month $t$.

In the second step, I aggregate across subcategories to get the markup index for establishment $r$ in month $t$:
\begin{equation}
    I^{\mu}_{r,t} = \prod_c I_{c,r,t}^{\omega_{c,r,y(t)}}.
\end{equation}
Here, the weight $\omega_{c,r,y(t)}$ is the share of subcategory $c$ in total revenue in establishment $r$ during the calendar year of month $t$. The dependent variable is then obtained by taking the natural logarithm of markup index: $\Delta \mu_{r,t} = \ln I^{\mu}_{r,t}$.

\newpage
\section{Wage data}
\label{appendix:b}
\setcounter{table}{0}
\setcounter{figure}{0}

\subsection*{NAICS classification for cannabis establishments}

Defining the minimum wage bite variable at the industry-by-county level requires careful consideration of which industry codes to use since establishments in the cannabis industry may fall under more than one North American Industrial Classification System (NAICS) industry code. The underlying principle of the NAICS system---that establishments with similar production processes be grouped together---greatly facilitates this, since the NAICS codes align well with the vertically disintegrated structure of the cannabis industry. For example, NAICS 453 captures all cannabis retailers, since NAICS 453998 (a component of NAICS 453) includes "All Other Miscellaneous Store Retailers (except Tobacco Stores), including Marijuana Stores, Medicinal and Recreational" \citep{naics2007}. At the producer level, NAICS 111 captures all cannabis growers, since NAICS 111998 includes "All Other Miscellaneous Crop Farming, including Marijuana Grown in an Open Field" and NAICS 111419 includes "Other Food Crops Grown Under Cover, including Marijuana Grown Under Cover" \citep{naics2007}. Slightly complicating things is the fact that in addition to growing cannabis, most producers are also processors (i.e. producer-processors). Processing falls under NAICS 424 which includes as a subcomponent "Other Farm Product Raw Material Merchant Wholesalers, including Marijuana Merchant wholesalers" (NAICS 424590).\footnote{A third industry, NAICS 115, may also apply to producer-processors, as it includes support activities for agriculture involving soil preparation, planting, and cultivating. However, to be in NAICS 115 an establishment must primarily perform these activities independent of the agriculture producing establishment, e.g. on a contractual basis. It is very unlikely that an establishment with a coveted producer-processor license would solely operate on a contractual basis without engaging in any production of its own. Therefore, I do not consider NAICS 115 in my analysis.} Importantly, though, NAICS classifies an establishment based on its primary activity, meaning that a producer-processor only belongs to NAICS 424 if the sales and revenue from processing activities exceed those of its own crop production \citep{naics2007}. I view it as more likely that a producer-processor belongs to NAICS 111 for two reasons. First, while it is not possible to directly compare the revenue share of crop production versus processing activities at the establishment level, at the industry level unprocessed "Usable Marijuana" accounts for over 61\% of producer-processors' revenue in my sample period. Second, \cite{miller2022} show that when cannabis was first legalized, the establishment count for NAICS 1114 in Washington increased by a similar count as the number of producer cannabis licenses. Moreover, the state saw a proportional increase in the number of workers and the total wages paid in NAICS 1114 \citep{miller2022}. Therefore, I classify all establishments with a joint producer-processor license as NAICS 111 under the assumption that crop production activities exceed processing activities for these establishments. Establishments with only a processor license (i.e. those allowed to process---but not grow---cannabis) would then be assigned NAICS 424, which is their proper classification. However, the very small number of processor licenses makes it difficult to identify treatment effects, so I drop processor-only licenses from my sample altogether. 

Table \ref{tab:b.1} provides an overview of the representativeness of cannabis employment in the respective 3-digit NAICS industries. The employment share for cannabis retailers is larger than that for producers, but the shares remain relatively constant over time for both producers and retailers. The fact that NAICS 111 is less representative does not imply that measurement error for the producer regressions is greater than that for the retail regressions, since it could be the case that the industries contained in NAICS 453 are more homogeneous than those in NAICS 111. A better indication of measurement error is the relation between cannabis wages and wages in the corresponding 3-digit NAICS industry. Table \ref{tab:g.1} in appendix \ref{appendix:g} shows that mean annual wages for cannabis establishments are remarkably similar to their corresponding NAICS industries and very close to the wage floor imposed by the minimum wage.

\begin{table}[!htbp] 
\centering
\caption{Employment in cannabis relative to 3-digit NAICS industry}
\label{tab:b.1}
\renewcommand{\tabcolsep}{.5pt}{
\def\sym#1{\ifmmode^{#1}\else\(^{#1}\)\fi}
\begin{tabular*}{\hsize}{@{\hskip\tabcolsep\extracolsep\fill}l*{6}{c}}
\toprule

& \multicolumn{3}{c}{Wholesale} & \multicolumn{3}{c}{Retail} \\
\addlinespace
\cline{2-4} \cline{5-7} 
\addlinespace
\multicolumn{1}{c}{Year} & \multicolumn{1}{c}{\parbox{1cm}{Cannabis Wholesale}} & \multicolumn{1}{c}{\parbox{1cm}{NAICS 111}} & \multicolumn{1}{c}{\parbox{1cm}{Emp. share}} & \multicolumn{1}{c}{\parbox{1cm}{Cannabis Retail}} & \multicolumn{1}{c}{\parbox{1cm}{NAICS 453}} & \multicolumn{1}{c}{\parbox{1cm}{Emp. share}} \\
\addlinespace

\midrule

2018  &  4,634 & 68,443 & .07 & 3,988 & 25,411 & .16 \\

\addlinespace
2019 & 4,727 & 64,112 & .07 & 4,618 & 25,908 & .18 \\

\addlinespace

2020 & 5,265 & 61,408 & .09 & 5,047 & 22,517 & .22 \\

\bottomrule

\end{tabular*}
\begin{minipage}[h]{\textwidth}
\medskip
\small \emph{Notes:} This table compares annual average employment at cannabis establishments and the respective NAICS subsectors for the years 2018-2020. Only UI covered employment is included (95\% of US jobs). NAICS 111 and 453 correspond to crop production and miscellaneous store retailers, respectively. Data for 2021 is not available. Data from Washington state ESD.
\end{minipage}

}

\end{table}

A final consideration is the granularity of industrial classification to use for the bite variable. Measuring bite at the three-digit industry level carries several advantages that make it preferable for the main analysis. First, cannabis producer-processors belong to different four-digit NAICS industries depending on whether they grow indoors or outdoors. Since I do not observe whether a given producer-processor grows indoors or outdoors, I would have to assume that all establishments are either indoor or outdoor growers, which increases measurement error. In contrast, the 3-digit NAICS code captures both indoor and outdoor producer-processors and thereby avoids such measurement error. Second, with more detailed NAICS codes, the bite variable does not clear the Census Bureau's data privacy filters for several counties, resulting in a reduced sample size.

\subsection*{Measurement error in NAICS 111}

The nature of agricultural labor in the United States means that one must consider whether the bite variable for NAICS 111 is subject to non-random measurement error. Non-random measurement error could arise for several reasons, and each is discussed in the following subsections.

\subsubsection*{Undocumented workers in NAICS 111}
First, if a significant amount of labor in NAICS 111 is performed by low-wage, undocumented migrants who are not eligible for unemployment insurance (and hence do not factor into the bite variable), then the bite variable may overestimate minimum wage exposure. Counties with more undocumented workers will have a smaller true (unobserved) bite, which amounts to classical errors-in-variables. Several facts speak against this being problematic. First, the prevalence of undocumented agricultural labor likely correlates over time within a county. As such, county fixed effects should sweep away cross-county differences in this measurement error. Second, to the extent that measurement error remains after demeaning, the bias leads to conservative treatment effects by attenuating the OLS estimates.

\subsubsection*{Seasonal labor in NAICS 111}
A second issue is that Washington's crop production is highly seasonal and the major crop types are primarily harvested in Q3. Since the minimum wage hikes in my sample occur on January 1st of each year, the bite variable---calculated two quarters prior to the hike---is based on Q3 wages. As a result, the bite variable may overestimate true minimum wage exposure due to seasonal fluctuations in agricultural labor. If counties with higher observed bite employ more low-wage seasonal labor (e.g. low wage rural counties), then the measurement error is non-random and OLS is biased. Unlike in the previous subsection, this is not classical errors-in-variables. Nevertheless, an easy way to overcome this would be to use Q4 bite instead, since Q4 does not coincide with any major harvest activity and hence should be free of seasonal wage fluctuations. As shown in appendix \ref{appendix:d}, estimates are robust to using Q4 bite, suggesting the main results are not affected by measurement error from seasonal wage fluctuations.

\subsubsection*{Measurement error and treatment effect timing}
Finally, setting aside the reasoning laid out in the previous two subsections, the fact remains that any bias from non-random time-varying measurement error would need to coincide with the timing of the minimum wage hike. In other words, for the main results to be driven by measurement error, the bias would have to cause a sharp inflationary shock at precisely the same time as the hike---not before and not after. I view such a scenario as unlikely.

\newpage
\section{Conditions for valid identification}

\label{appendix:l}
\setcounter{table}{0}
\setcounter{figure}{0}

\subsection{Direct pass-through: Identifying the Average Causal Response (ACR)}

Difference-in-differences with continuous treatment is widely applied in empirical research, yet it is only recently that identification and interpretation of such DiD designs have been given formal treatment. \cite{callaway2021} show that a causal interpretation with continuous treatment requires different assumptions than with a binary treatment. This section highlights key insights from \cite{callaway2021} and relates them to my research setting.

First, recall that the treatment intensity in my main specification (equation \ref{eq:2}) equals the percent increase in the minimum wage times the industry-by-county bite, $\Delta MW_{j} \times Bite_{k(j)}$ (I omit leads and lags here for ease of exposition). The key identifying variation comes from $Bite_{k(j)}$ while the $\Delta MW_{j}$ term acts as a scale parameter that enables interpreting the estimated coefficients as elasticities. In other words, omitting $\Delta MW_{j}$ does not change the results but simply alters the scale and interpretation of the estimated parameters (see appendix Figure \ref{fig:d.4}). Therefore, in the rest of this subsection I refer to $Bite_{k(j)}$ as the relevant treatment intensity when discussing the causal parameter of interest and use the following simplified variant of my main empirical equation
\begin{equation}
   \pi_{j,t} = \beta Bite_{k(j),t} + \gamma_t + \epsilon_{j,t} \nonumber
\end{equation}
where leads and lags have been removed for ease of exposition.

\cite{callaway2021} show that the assumptions required for causal interpretation of $\beta$ depend on the causal parameter of interest and the particular research setting. In settings where all groups are assigned some positive treatment intensity, comparisons between groups with different treatment intensities do not allow inference about the Average Treatment effect on the Treated (ATT) since no group can be used to estimate an untreated counterfactual. This applies to my research setting since there are very few untreated units (i.e. establishments in counties with bite equal to zero) which precludes identifying ATT-type parameters. Instead, the parameter of interest is the Average Causal Response (ACR) which captures the overall causal response of a small change in treatment intensity. The ACR is a weighted average of local comparisons of paths of outcomes known as the Average Causal Response on the Treated (ACRT). Specifically, the ACRT is the average difference between potential outcomes under treatment intensity $d$ compared to potential outcomes under a marginal change in the treatment intensity for the group of units that actually experience treatment intensity $d$.
Using the same notation as \cite{callaway2021},\footnote{Note that for ease of exposition I use the multi-valued treatment (rather than continuous treatment) notation from \cite{callaway2021}.}
\begin{equation}\label{eq:l.1}
    \mathbb{E}[\Delta Y_t | D = d_j] - \mathbb{E}[\Delta Y_t | D = d_{j-1}] = ACRT(d_j | d_j) + \underbrace{ATT(d_{j-1} | d_j) - ATT(d_{j-1} | d_{j-1})}_{\text{"selection bias"}}.
\end{equation}
The left side of equation \ref{eq:l.1} compares the change in the outcome $Y$ between higher- and lower-treatment intensity units, where $d_j$ and $d_{j-1}$ are "neighboring" treatment intensities, with $d_j > d_{j-1}$. The right side illustrates that these comparisons are a combination of (i) the average causal response on the treated of treatment intensity $d$ for units that receive treatment intensity $d$, $ACRT(d|d)$, and (ii) differences in the treatment effects between the two groups at the lower treatment intensity (which \cite{callaway2021} call "selection bias"). Equation \ref{eq:l.1} implies that $ACRT(d|d)$ is identified if the selection bias equals zero among units that received \emph{similar} treatment intensities. Intuitively, zero "local" selection bias implies that treatment effects would be similar across counties with similar bites, had those counties been assigned the same (nearby) bite $d$. While this is an inherently untestable assumption (similar to how parallel trends cannot be tested), I view zero local selection bias as valid in my setting for several reasons. First, "selection on gains" (i.e. treatment effect heterogeneity under a counterfactual treatment) would reflect local differences in the price elasticity of demand, which partly depends on local macroeconomic conditions such as county unemployment and average wages, which I explicitly control for in section \ref{section:5.3}. Second, to the extent that county controls may not capture all of the determinants of cross-county differences in the price elasticity of demand, in section \ref{section:5.3} I also include region-time FE. This involves a much more narrow set of comparisons between culturally, politically, and economically homogeneous counties (i.e. counties where one would expect similar selection on gains). Importantly, pass-through effects remain stable across these specifications, which indicates a lack of selection on gains.

\subsection{Indirect pass-through: Identification with the shift-share instrument}

A growing number of empirical studies use shift-share ("Bartik") instruments, defined as the inner product of growth rates (the "shift" or "shock") and exposure shares. \cite{Borusyak2022} develop an econometric framework in which shift-share (SSIV) identification stems from the (conditional) quasi-random assignment of shocks, thereby allowing for endogeneity in the exposure shares. Central to this framework is "shock orthogonality", a necessary and sufficient condition for instrument validity. This section summarizes the main points from \cite{Borusyak2022} regarding shock orthogonality and relates them to my research setting. 

\subsubsection*{SSIV numerical equivalence}
In what follows, I use the same notation as \cite{Borusyak2022} and abstract from time indexes for ease of exposition. Consider the case where one wants to estimate the causal effect in a reduced form linear model relating outcomes to the instrument and controls
\begin{equation}
    y_l = \beta z_l + \gamma w_l + \varepsilon_l
\end{equation}
In my research setting, $y_l$ equals the monthly inflation rate for retail establishment $l$, and $z_l$ captures retailer $l$'s exposure to the minimum wage-induced wholesale cost shock (note that since the SSIV concerns indirect pass-through and not direct pass-through, I abstract from the latter by including it in the control vector $w_l$). Specifically, $z_l$ is a shift-share instrument from a set of shocks, $g_n$, for $n = 1,...N$, and shares $s_{l,n} > 0$ which define the relative exposure of each observation $l$ to each shock $n$. In my setting, the shocks $g_n$ are the minimum wage bite for cannabis producers located in county $n$, and $s_{l,n}$ is the share of retailer $l$'s wholesale expenditure that goes to producers from county $n$. The SSIV is then the exposure-weighted average of the shocks
\begin{equation}
    z_l = \sum_{n=1}^N s_{l,n}g_n.
\end{equation}
By the Frisch-Waugh-Lovell Theorem the SSIV estimator $\hat{\beta}$ is equivalent to a bivariate IV regression of outcome and treatment residuals
\begin{equation}\label{eq:l.3}
    \hat{\beta} = \frac{\sum_{l=1}^{L}z_l y_l^{\perp}}{\sum_{l=1}^{L}z_l z_l^{\perp}}
\end{equation}
where $y_l^{\perp}$ ($z_l^{\perp}$) is the residual from the projection of $y_l$ ($z_l$) on the control vector $w_l$. The key insight by \cite{Borusyak2022} is that the SSIV estimator $\hat{\beta}$ is numerically equivalent to a shock-level IV regression that uses the shocks $g_n$ as the instrument in estimating the reduced form equation
\begin{equation} \label{eq:l.2}
    \bar{y}_n^{\perp} = \alpha + \beta \bar{z}_n^{\perp} + \bar{\varepsilon}_n^{\perp}
\end{equation}
where $\bar{v}_n = \frac{\sum_{l=1}^{L} s_{ln}v_l}{\sum_{l=1}^{L} s_{ln}}$ is the exposure-weighted average of variable $v_l$ and the IV estimation is weighted by average shock exposure $s_n = \sum_{l=1}^{L} s_{ln}$. Specifically, \cite{Borusyak2022} show that
\begin{equation}\label{eq:l.4}
    \hat{\beta} = \frac{\sum_{n=1}^{N}s_n g_n \bar{y}_{n}^{\perp}}{\sum_{n=1}^{N}s_n g_n \bar{z}_{n}^{\perp}}.
\end{equation}
Note that the IV regression in equation \ref{eq:l.2} contains transformed shock-level aggregates of the variables rather than the original observation-level variables. In my research setting, shock-level refers to the county that one (or more) producers are located in. Thus, $\bar{y}_{n}^{\perp}$ is the average residualized inflation of the retailers most exposed to producers in county $n$ (i.e. the $n$th shock), while $\bar{z}_{n}^{\perp}$ is the average residualized indirect bite for the retailers most exposed to producers from county $n$. Each shock $g_n$ in this regression is weighted by $s_n$, which is the average share of retailers' wholesale expenditures (across all retailers in Washington state) that goes to producers from county $n$.

\subsubsection*{Shock orthogonality}
The equivalence in equation \ref{eq:l.4} highlights that SSIV estimates can be seen as stemming from variation across shocks rather than observation-level exposure shares. \cite{Borusyak2022} build on this equivalence to show that the standard exclusion restriction translates into a "shock orthogonality" condition. In particular, they show that shock orthogonality is satisfied if shocks $g_n$ are (conditionally) quasi-randomly assigned across $n$ regardless of the average exposure $s_n$ or the unobservable average (shock-level) error $\bar{\varepsilon}_n$. In the context of my research setting, this implies that the industry-by-county minimum wage bite for cannabis producers has the same expected value across counties $n$, 1) regardless of the average retail exposure to that county $s_n$, and 2) regardless of the aggregated structural error term. Condition 1) implies that producer counties with high average expenditure shares across retailers do not have systematically different bite than producer counties with low average expenditure shares. As an example, this is violated if the producers in a given county $n$ are particularly popular among retailers statewide (hence, that county has a high $s_n$), and after controlling for county-level observables, the producers in that county have systematically higher (or lower) bite than producers in other counties. The raw correlation between bite and expenditure shares is 0.18, and when conditioning on county observables the relationship shrinks to 0.06 (95\% CI: -.0457195, .1662991). This suggests that there is no systematic relationship between bite and average expenditure shares. Condition 2) implies that there is no relationship between producer bite for county $n$ and the average unexplained variation in retail inflation for the retailers that purchase from producers in county $n$. This cannot be tested because the latter is unobserved. However, I view it as unlikely that after conditioning on county controls and retailer FE, retailers with large unexplained shocks to inflation purchase more from producer-counties with higher (or lower) bite.







\subsubsection*{Instrument relevance}
\begin{figure}[!htbp]
\caption{Effect of indirect bite on retailers' wholesale cost}
\centering
    \includegraphics[width=.5\textwidth]{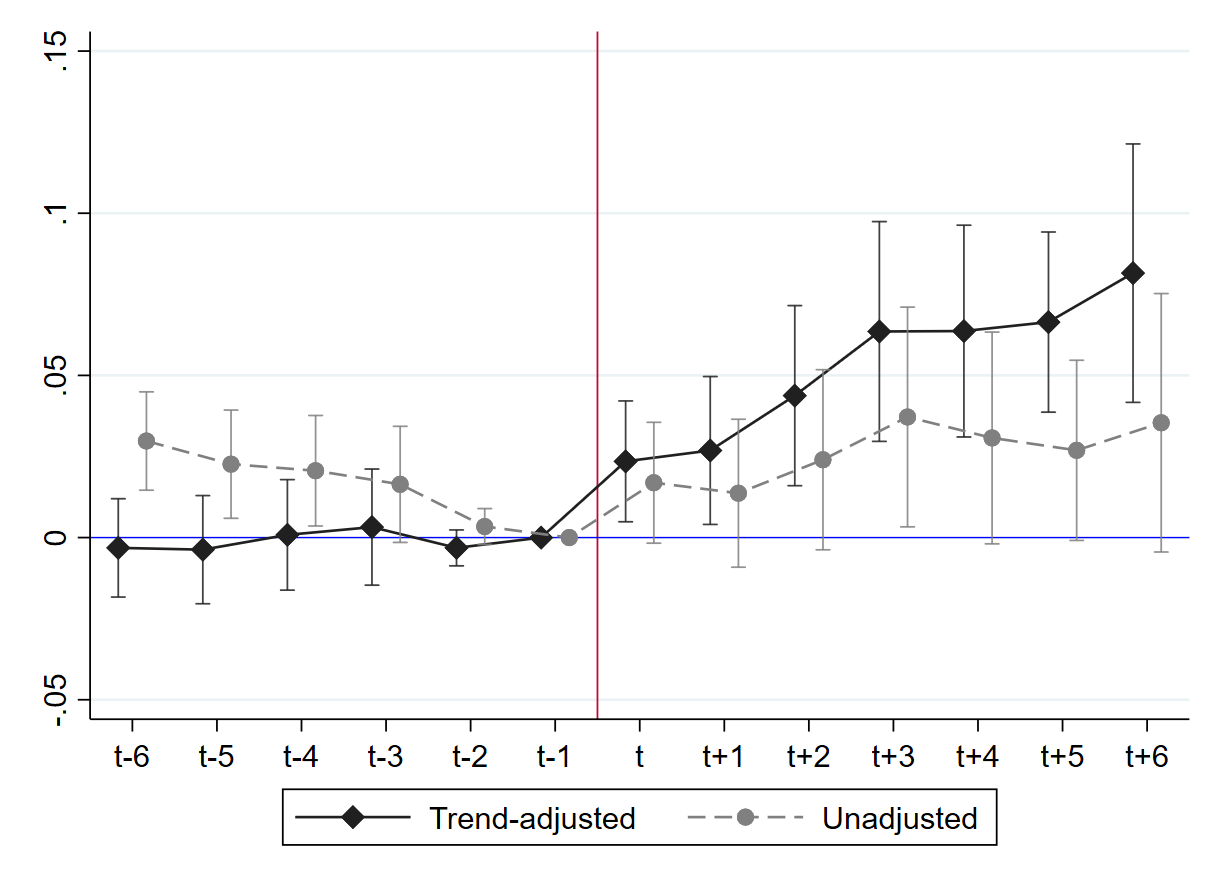}
\label{fig:l.0}
\par \bigskip
\rule{\textwidth}{0.5pt}
\begin{minipage}[h]{\textwidth}
\medskip
\small \emph{Notes:} The figure shows estimates from from equation \ref{eq:l.0} The dependent variable is the establishment-level wholesale cost index for cannabis retailers (in logs). The figure depicts cumulative wholesale cost effects ($E_L$) relative to the baseline period in $t-1$. Cumulative effects $E_L$ are obtained by summing the distributed lag coefficients to lag $L$ as detailed in the main text. The figure shows 90\% confidence intervals of the sums based on SE clustered at the county level.  Data source: Top Shelf Data and Washington ESD, July 2018 to August 2021.
\end{minipage}
\end{figure}

Recall that shock orthogonality is the shock-level equivalent to the exclusion restriction. As in other IV settings, the exclusion restriction is not sufficient for shift-share consistency---instrument relevance is also required. I test for first-stage relevance by estimating a variant of equation \ref{eq:5}
\begin{equation}\label{eq:l.0}
   w_{r,t} = \sum_{l =-5}^{6} \beta_l \Delta MW_{r,t-l} \times Bite_{k(r),t-l} + \sum_{l =-5}^{6} \psi_l IB_{r,P,t-l} + X_{k(r),q(t)} + \gamma_t + \epsilon_{r,t}.
\end{equation}
where $w_{r,t}$ is the (log) wholesale cost index for retail establishment $r$ in month $t$. First-stage relevance is then characterized by the cumulative sum of the distributed lag coefficients $\psi_l$ relative to a normalized baseline period. Figure \ref{fig:l.0} shows that at the average indirect bite (18.14\%), a 10\% increase in the minimum wage corresponds to a 6.13\% (unadjusted) to 12.10\% (trend-adjusted) increase in retailers' wholesale costs relative to the normalized baseline period in $t-2$ (significant at the 10 percent and 1 percent level, respectively).\footnote{The respective F-statistics for the cumulative effects are 3.28 and 11.24.} Note that the first stage estimates are much larger than the reduced form estimates in section \ref{section:5}. This is to be expected since---despite the full pass-through of wholesale costs to retail prices documented in section \ref{section:6}---wholesale prices are lower than retail prices, which means that a dollar increase in wholesale prices amounts to a larger percent change than a dollar increase in retail prices.

\section{Bite-specific trend}
\label{appendix:c}
\setcounter{table}{0}
\setcounter{figure}{0}

\subsection{Direct pass-through to wholesale prices}

\begin{figure}[!htbp]
\caption{Direct pass-through of minimum wage hikes to wholesale prices}
\centering
	\begin{subfigure}{.4\textwidth}
	\centering
		\includegraphics[width=\linewidth]{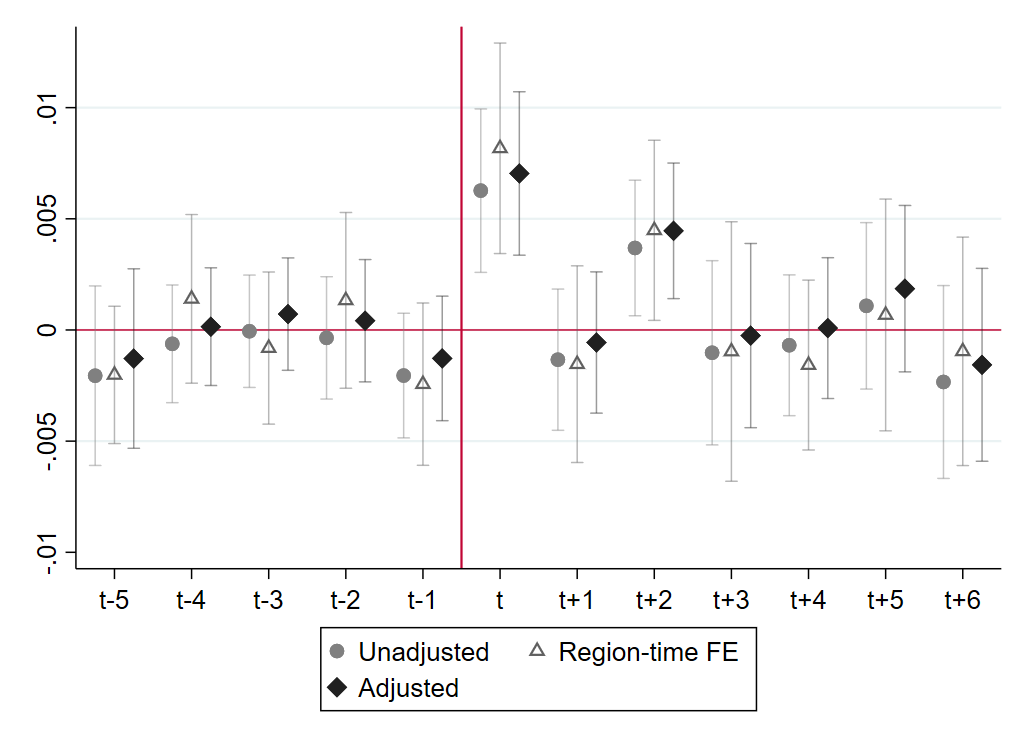}
		\caption{Effect on the wholesale inflation rate (i.e. distributed lag coefficients)}
	\end{subfigure}\hfil
	\begin{subfigure}{.4\textwidth}
    \centering
	\includegraphics[width=\linewidth]{whole_figures_no_10_mo_restriction/config4a1_10_aug_jul_q3bite_no_10_mo_restriction_05_trim_dec_base_trend_adjustment_comparisons_main_3specs_no_county_fe.png}
    \caption{Effect on the wholesale price level (i.e. cumulative sums of distributed lag coefficients)}
    
    \end{subfigure}\hfil
\label{fig:c.0}
\par \bigskip
\rule{\textwidth}{0.4pt}
\begin{minipage}[h]{\textwidth}
\medskip
\small \emph{Notes:} The figures show direct minimum wage pass-through to wholesale prices. Estimates are from equation \ref{eq:2}. The dependent variable is the establishment-level inflation rate for cannabis producers. Panel (a) shows the estimated distributed lag coefficients, $\hat \beta_l$, with 90\% confidence intervals based on SE clustered at the county level. Panel (b) replicates figure \ref{fig:5.1} in the main paper and depicts cumulative price level effects ($E_L$) relative to the baseline period in $t-2$, with 90\% confidence intervals of the sums based on SE clustered at the county level. Data source: Top Shelf Data and Washington ESD, July 2018 to August 2021.
\end{minipage}
\end{figure}

Figure \ref{fig:c.0} illustrates that for all three specifications used in section \ref{section:4} (unadjusted, trend-adjusted, region-time FE), the distributed lag coefficients are not statistically significantly different from zero for $t-5$ through $t-2$, and the period $t$ treatment effects are large and not statistically significantly different from each other. While trend-adjusting the dependent variable does not change the contemporaneous treatment effect, it does affect pass-through estimates over a longer time horizon. Adjusting for the trend results in a permanently higher price level, whereas not adjusting results in the positive effect in period $t$ being undone in subsequent periods. Therefore, it is important to ensure that the trend is robust to a variety of specifications and assumptions. This section provides a detailed exposition of the bite-specific trend and illustrates the empirical validity of the adjustment using two different methods. The first method uses a single, pooled trend for to the entire sample period. This is the method that I use in the main part of the paper (see section \ref{section:4}). The second method estimates a separate trend for each of the three events and only adjusts the dependent variable for event 2, as that is the only event with a significant trend. As I show below, how one adjusts for the trend matters little, as both methods lead to similar results.

\subsubsection*{Pooled trend}
Since my research design pools three minimum wage events, the most obvious way to adjust for the bite-specific trend is to fit a single trend onto the pooled events. This is the strategy I adopt in the main part of the paper. A bite-specific trend in the price \emph{level} occurs if $\bar{\hat{ \beta}}_{pre}$, the average of the distributed lag coefficients in the pre-treatment period (i.e. the average \emph{change} in the price level effect), is statistically significantly different from zero. As Table \ref{tab:c1} illustrates, this is the case when the trend is estimated using all 5 pre-treatment leads. Though the trend extends through 5 leads (i.e. through $t-1$, see Figure \ref{fig:5.1}), I use the estimate from 4 leads (i.e. through $t-2$) when detrending since the base period is set to $t-2$ in my main analysis. Trend-adjustment proceeds as follows: 
\begin{enumerate}
    \item Compute the average of the distributed lag coefficients for the pre-baseline periods:
    \begin{equation*}
        \bar{\hat{ \beta}}_{pre} = 1/4 \sum_{l=2}^{5} \hat{\beta}_{-l}
    \end{equation*}
     where $\bar{\hat{ \beta}}_{pre}$ is the average \emph{change} in the pre-treatment price level effect.
    \item Use $\bar{\hat{ \beta}}_{pre}$ to obtain predicted values for the bite-specific trend over the entire event window: 
    \begin{equation*}
        \hat \pi_{j,t} = \bar{\hat{\beta}}_{pre} \times \Delta MW_{j,t-l} \times Bite_{k(j),t-l}
    \end{equation*}
    \item Estimate equation \ref{eq:2} using adjusted inflation $\tilde \pi_{j,t}$ as the dependent variable:
    \begin{equation*}
        \tilde{\pi}_{j,t} = \pi_{j,t} - \hat \pi_{j,t}
    \end{equation*}
\end{enumerate}

\begin{table}[!htbp] 
\centering
\caption{Bite-specific trend estimates for producers}
\label{tab:c1}
\renewcommand{\tabcolsep}{1pt}{
\def\sym#1{\ifmmode^{#1}\else\(^{#1}\)\fi}
\begin{tabular*}{\hsize}{@{\hskip\tabcolsep\extracolsep\fill}l*{4}{c}}
 \toprule
 \addlinespace
            &\multicolumn{1}{c}{Pooled trend} &\multicolumn{3}{c}{Separate trends} \\
            \addlinespace
            \cline{2-2} \cline{3-5} \\
            \addlinespace
            &\multicolumn{1}{c}{(1)} &\multicolumn{1}{c}{(2)} &\multicolumn{1}{c}{(3)} &\multicolumn{1}{c}{(4)} \\
            & &\multicolumn{1}{c}{Event 1} &\multicolumn{1}{c}{Event 2} &\multicolumn{1}{c}{Event 3} \\
\addlinespace
\midrule
$4$ leads       & -0.00077  & 0.00179 & -0.00090 & -0.00370 \\
            & (0.00069)  &  (0.00261) & (0.00067) & (0.00848) \\
\addlinespace
$5$ leads       & -0.00103*  &  0.00056 & -.00109** & -0.00485    \\
             & (0.00054) &   (0.00209) & (0.00055) & (0.00788)   \\
 \midrule
\(N\)       &  13,033 &  3,996 & 4,646 & 4,391  \\
Time FE     &  YES    &   YES  &  YES  & YES \\
\bottomrule
\end{tabular*}
\begin{minipage}[h]{\textwidth}
\bigskip
\small This table reports the average of the distributed lag coefficients obtained from estimating equation \ref{eq:2}, i.e. the average change in the estimated treatment effect during the pre-treatment period. \sym{*} \(p<0.10\), \sym{**} \(p<0.05\), \sym{***} \(p<0.01\). Data from Washington ESD and Top Shelf Data, August 2018-July 2021.
\end{minipage}
}
\end{table}

\subsubsection*{Separate trends}
A concern with the trend adjustment presented above is that if one or more events exhibit a different trend---or no trend at all---then fitting a single trend onto all events may be misleading. Therefore, in this subsection, I examine each event separately and test for event-specific trends. Event-specific trends are estimated as follows: First, I estimate equation \ref{eq:2} separately for each event using the original (unadjusted) dependent variable. Next, for each event, I compute the average of the distributed lag coefficients for the pre-baseline periods, $\bar{\hat{ \beta}}_{pre,e} = 1/4 \sum_{l=2}^{5} \hat{\beta}_{-l}$, where $\bar{\hat{ \beta}}_{pre}$ is the average pre-treatment \emph{change} in the price level effect for event $e$ (since equation \ref{eq:2} is in first differences). As Table \ref{tab:c1} illustrates, $\bar{\hat{ \beta}}_{pre,e}$ is only statistically significantly different from zero for event 2, meaning there is no bite-specific trend for events 1 and 3. I therefore adjust $ \pi_{j,t}$ for event 2 only and leave the other events unadjusted. Specifically, I use $\bar{\hat{ \beta}}_{pre,e=2}$ to obtain predicted values for the bite-specific trend for the entire event window: $\hat \pi_{j,t(e=2)} = \bar{\hat{\beta}}_{pre(e=2)} \times \Delta MW_{j,t-l(e=2)} \times Bite_{k(j),t-l(e=2)}$. Finally, I estimate equation \ref{eq:2} using adjusted inflation $\tilde \pi_{j,t}$ as the dependent variable 
\begin{equation}\label{}
    \tilde \pi_{j,t(e)} =
    \begin{cases}
        \pi_{j,t(e)} & \text{if}\ e = 1,3\\ 
        \pi_{j,t(e)} - \hat \pi_{j,t(e)} & \text{if}\ e = 2
    \end{cases}
\end{equation} 

\begin{figure}[!htbp]
\caption{Comparing trend adjustment for direct pass-through to wholesale prices}
\centering
	\begin{subfigure}{.4\textwidth}
	\centering
		\includegraphics[width=\linewidth]{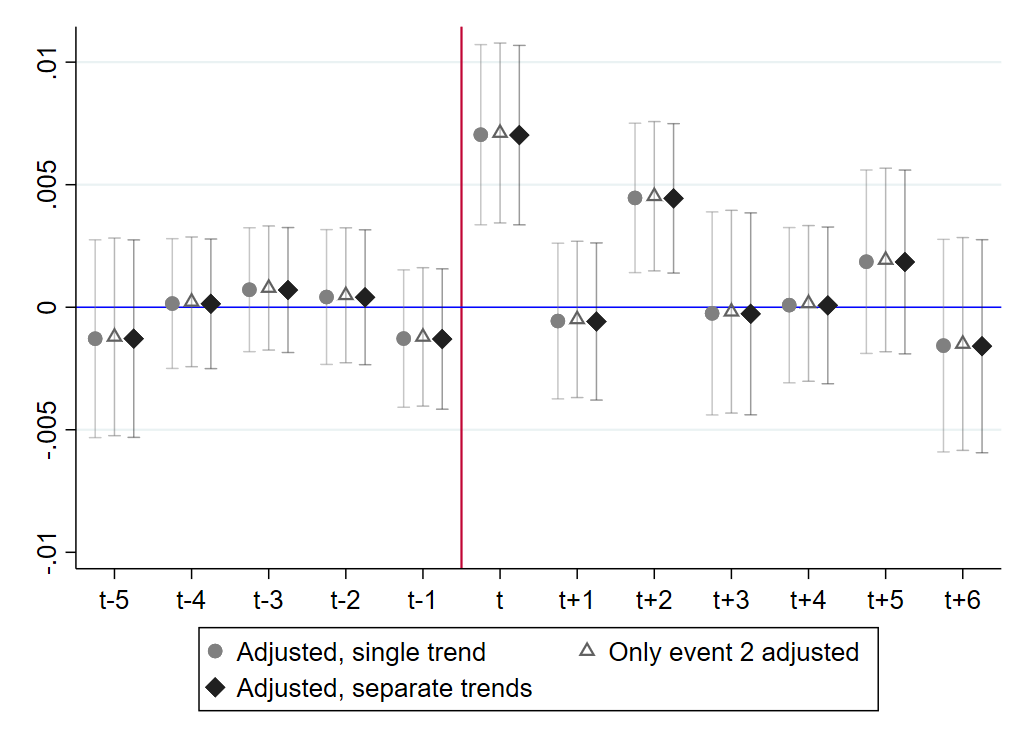} 
		\caption{Effect on the inflation rate}
		\label{fig:c.1a}		
	\end{subfigure}\hfil
	\begin{subfigure}{.4\textwidth}
	\centering
        \includegraphics[width=\linewidth]{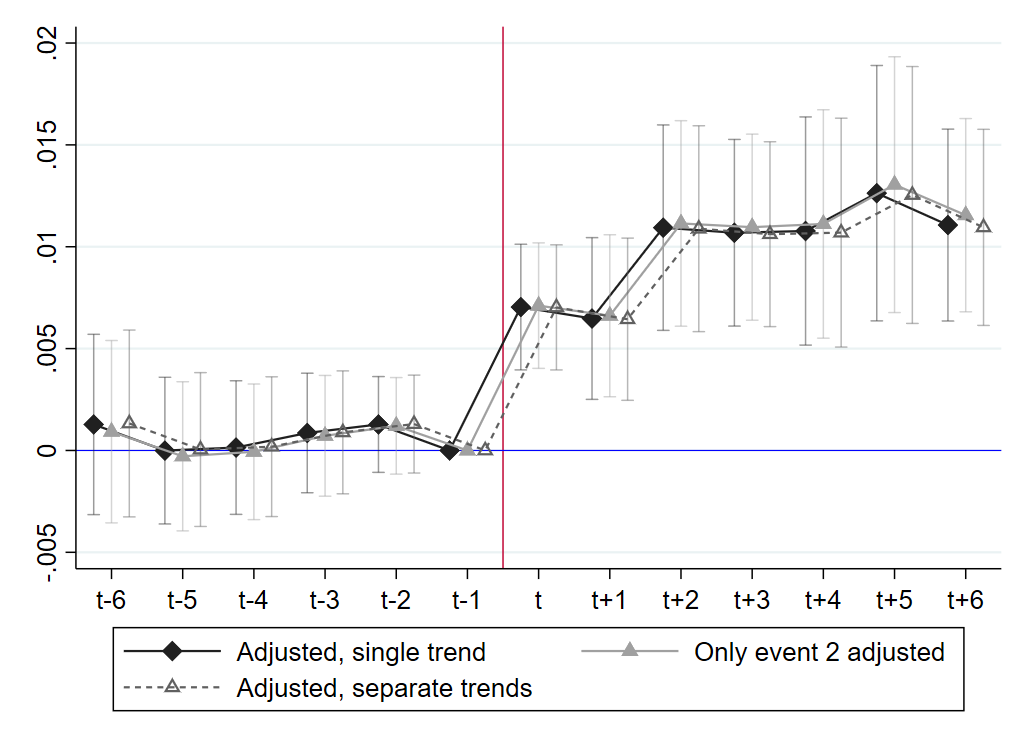} 
		\caption{Effect on the price level}
		\label{fig:c.1b}
	\end{subfigure}\hfil
\label{fig:c.1}
\par \bigskip
\rule{\textwidth}{0.4pt}
\begin{minipage}[h]{\textwidth}
\medskip
\small \emph{Notes:} The figure compares direct pass-through to wholesale prices when the dependent variable is trend-adjusted in different ways. In both panels, estimates are from equation \ref{eq:2} with time fixed effects but no county fixed effects (since the event 2 trend cannot be estimated with county fixed effects). Results are robust to including county fixed effects. Panel (a) shows the estimated distributed lag coefficients, $\hat{\beta}_l$, with 90\% confidence intervals based on SE clustered at the county level. Panel (b) displays cumulative price level effects ($E_L$) relative to the baseline period in $t-1$. Cumulative effects $E_L$ are obtained by summing the distributed lag coefficients to lead or lag $L$ as detailed in the main text. Panel (b) shows the 90\% confidence intervals of the sums based on SE clustered at the county level. In panel b, the normalized base period is set to $t-1$. Data source: Top Shelf Data and Washington ESD, August 2018-July 2021.
\end{minipage}
\end{figure}

Table \ref{tab:c1} shows that events 1 and 3 exhibit no significant pre-trend while event 2 contains a pre-trend that is similar in magnitude to the single trend found in figure \ref{fig:5.1}. Therefore, I adjust the dependent variable for event 2 while leaving events 1 and 3 unadjusted. As Figure \ref{fig:c.1} illustrates, this approach leads to very similar results as the pooled adjustment. As a robustness check, I adjust all 3 events for an event-specific trend irregardless of statistical significance of the trend. Figure \ref{fig:c.1} shows that estimated effects are virtually identical.

\newpage
\subsection{Indirect pass-through to retail prices}

To quantify the pre-treatment trend for indirect pass-through, I again take the average of the distributed lag coefficients for the pre-baseline period, $\bar{\hat{\psi}}_{pre} = 1/4 \sum_{l=2}^{5} \hat{\psi}_{-l}$, where $\bar{\hat{\psi}}_{pre}$ is the average change in the pre-treatment effect. I find that $\bar{\hat{ \psi}}_{pre} = -.0019$ ($90\%$ confidence interval: $-0.0040$ to $0.0002$) which overlaps with the pre-trend for pass-through to wholesale prices, $\bar{\hat{ \beta}}_{pre} = -0.0008$ ($90\%$ confidence interval: $-0.0019$ to $-0.0004$).\footnote{$\bar{\hat{\psi}}_{pre}$ is based on equation \ref{eq:5} with time fixed effects and county-level controls. $\bar{\hat{ \beta}}_{pre}$ is based on equation \ref{eq:2} with time fixed effects. The pre-trend reported here is from pooled events. For both equations, the pre-trend is stable across a variety of specifications.} 

\newpage
\section{Robustness checks}
\label{appendix:d}
\setcounter{table}{0}
\setcounter{figure}{0}

\subsection{Longer event window}

\begin{figure}[!htbp]
\caption{Direct pass-through, 18-month event window}
\centering
	\begin{subfigure}{.4\textwidth}
	\centering
		\includegraphics[width=\linewidth]{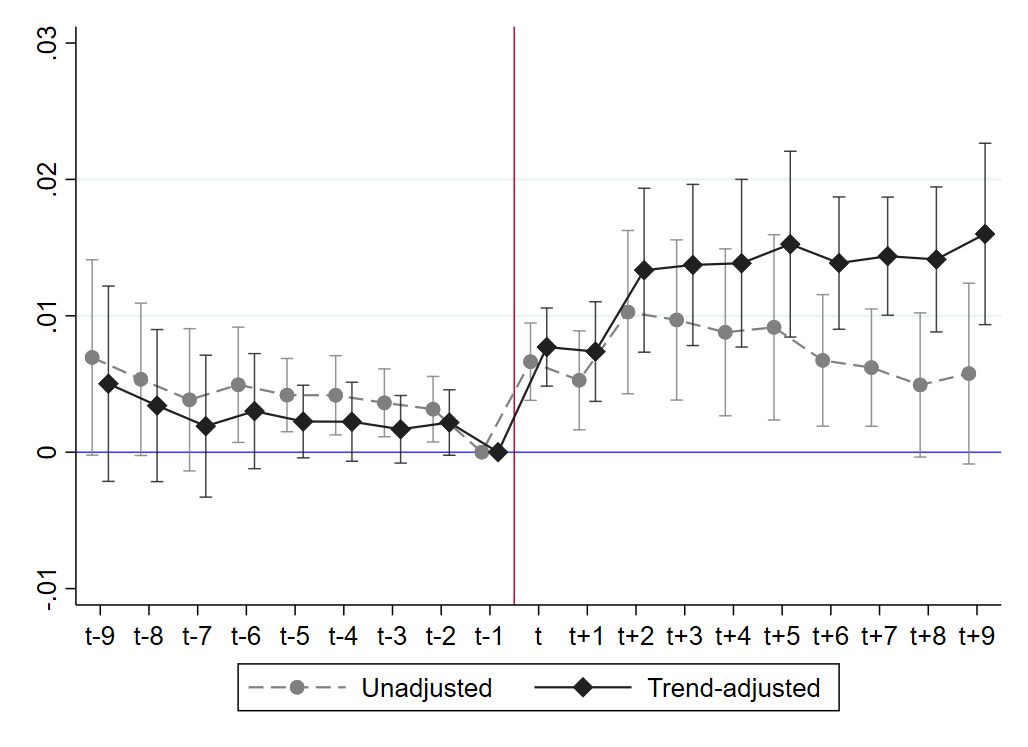} 
		\caption{Wholesale prices}
		\label{fig:k.1a}		
	\end{subfigure}\hfil
	\begin{subfigure}{.4\textwidth}
	\centering
        \includegraphics[width=\linewidth]{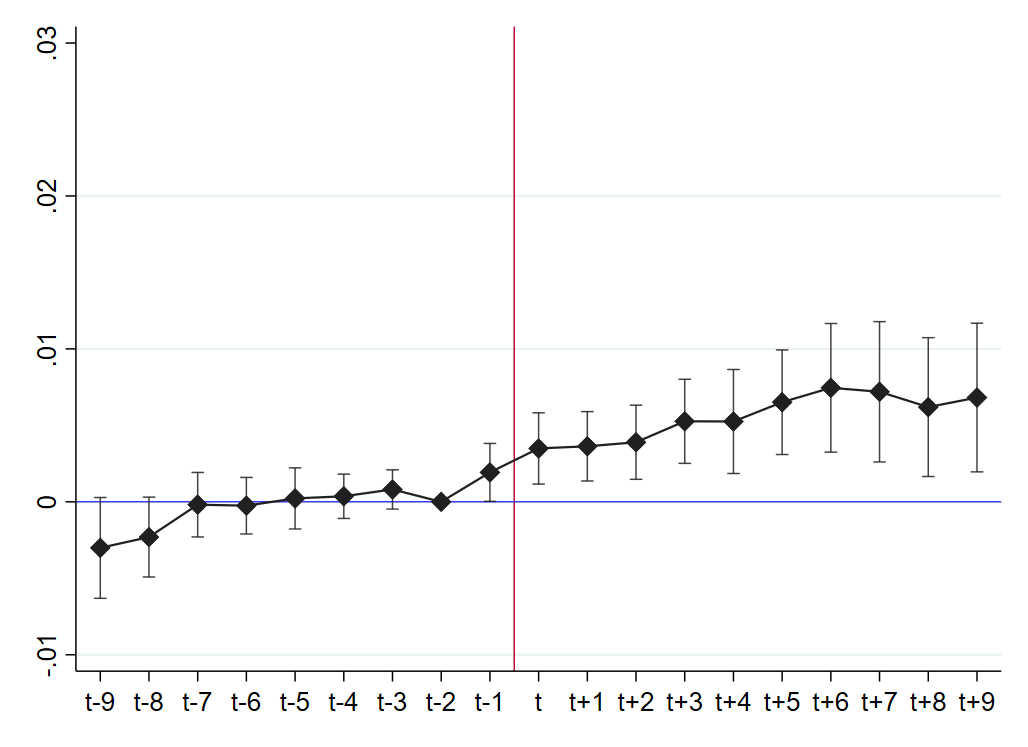}
		\caption{Retail prices}
		\label{fig:k.1b}
	\end{subfigure}\hfil
\label{fig:k.1}
\par \bigskip
\rule{\textwidth}{0.4pt}
\begin{minipage}[h]{\textwidth}
\medskip
\small \emph{Notes:} The figures show direct pass-through to prices over an 18-month event window. Both panels display cumulative price level effects ($E_L$) relative to the normalized baseline period in $t-1$ (wholesale prices) and $t-2$ (retail prices). Cumulative effects $E_L$ are obtained by summing the distributed lag coefficients to lead or lag $L$ as detailed in the main text. Both panels show 90\% confidence intervals of the sums based on SE clustered at the county level. Data source: Top Shelf Data and Washington ESD, March 2018-December 2021.
\end{minipage}
\end{figure}

In this subsection, I show that the results from the main section based on a 12-month event window are unaffected by increasing the size of the event window. As Figures \ref{fig:k.1} and \ref{fig:k.2} illustrate, treatment effects materialize by $t+2$ and remain quite stable thereafter. Therefore, the 12-month event window used in the baseline estimation should adequately capture the short run effects of minimum wage hikes on retail cannabis prices.

\newpage

\subsection{Alternative specifications for direct pass-through}
This subsection reports results from the robustness checks discussed in section \ref{section:5.3}.
\begin{table}[!htbp]
\small
\centering
\caption{Robustness checks for direct pass-through to wholesale prices}
\label{tab:d.1}
\renewcommand{\tabcolsep}{1pt}{
\def\sym#1{\ifmmode^{#1}\else\(^{#1}\)\fi}
\begin{tabular*}{\hsize}{@{\hskip\tabcolsep\extracolsep\fill}l*{6}{c}}
\toprule


& \multicolumn{2}{c}{Alternate bite variable} & \multicolumn{2}{c}{Reverse causality} & \multicolumn{2}{c}{Other} \\
\cline{2-3} \cline{4-5} \cline{6-7}

\addlinespace

&\multicolumn{1}{c}{(1)}&\multicolumn{1}{c}{(2)}&\multicolumn{1}{c}{(3)}&\multicolumn{1}{c}{(4)}&\multicolumn{1}{c}{(5)}&\multicolumn{1}{c}{(6)} \\
\addlinespace

&\multicolumn{1}{c}{\parbox{1.5cm}{\centering Q4 bite}}&\multicolumn{1}{c}{\parbox{1.5cm}{\centering Com-pliance}}&\multicolumn{1}{c}{\parbox{1.5cm}{\centering No Seattle}}&\multicolumn{1}{c}{\parbox{1.5cm}{\centering No King county}} &\multicolumn{1}{c}{\parbox{1.5cm}{\centering Balanced panel}} &\multicolumn{1}{c}{\parbox{1.5cm}{\centering Alt. weights}}\\
\midrule
\midrule
\addlinespace
\multicolumn{1}{l}{\parbox{.5cm}{$E_0$}}       &    0.006** & 0.006*** & 0.006*** & 0.006*** & 0.006*** & 0.008**  \\
    & (0.003) & (0.002) & (0.002) & (0.002) & (0.002) & (0.004) \\

\addlinespace
\multicolumn{1}{l}{\parbox{.5cm}{$E_2$}}       &    0.011** & 0.011*** & 0.010*** & 0.010*** & 0.008*** & 0.013*** \\ 
            &  (0.004) & (0.004) & (0.003) & (0.003) & (0.003) & (0.004)  \\
\addlinespace
\multicolumn{1}{l}{\parbox{.5cm}{$E_4$}}       &   0.009** & 0.011** & 0.010*** & 0.010*** & 0.008** & 0.018*** \\
            &  (0.004) & (0.005) & (0.004) & (0.004) & (0.004) & (0.006) \\
\midrule
\multicolumn{1}{l}{\parbox{.5cm}{$\sum \text{Pre}$\\ $\text{-event}$}}  & -1.4e-07 & 2.1e-08 & -1.2e-07 & -1.7e-07 & 1.7e-07 & -1.0e-07 \\
            &    (0.003) & (0.004) & (0.003) & (0.003) & (0.003) & (0.006) \\
\midrule
\(N\)       &   14,777 & 14,699 & 14,622 & 14,506 & 12,900 & 14,819 \\
Time FE &        YES          &   YES                           &   YES                 &         YES  & YES &   YES \\
Controls &        NO           &   NO                                   &   NO                  &         NO  & NO &   NO              \\
Trend-adjusted &        YES          &   YES                            &   YES                 &         YES  & YES &   YES            \\

\bottomrule
\end{tabular*}
\begin{minipage}[h]{\textwidth}
\medskip
\small \emph{Notes:} The listed coefficients are sums of the distributed lag coefficients $E_L$, $L$ months after the minimum wage hikes, relative to the baseline period in $t-2$. The distributed lag coefficients are estimated from equation \ref{eq:2} with establishment-level inflation rate as the dependent variable. (1) uses Q4 bite in the treatment interaction term $\Delta MW_{j,t-l} \times bite_{k(j),t-l}$, while (2) uses the difference between bite two quarters before and one quarter after the hike. (3)-(4) account for possible endogeneity of Seattle hikes: (3) omits Seattle establishments for event 3 while (4) omits King county establishments for event 3. (5) restricts the panel to establishments that are present at least 10 months for a given event. For (6) the price indexes are constructed with expenditure weights based on the fiscal year starting in July and ending in June of each year. Estimates are unaffected by the inclusion of controls, winsorizing instead of trimming, and not trimming at all (results available on request). Standard errors of the sums $E_L$ are clustered at the county level and are shown in parentheses. \sym{*} \(p<0.10\), \sym{**} \(p<0.05\), \sym{***} \(p<0.01\). Data from Washington ESD and Top Shelf Data, March 2018-December 2021.
\end{minipage}

}
\end{table}

\begin{table}[!htbp]
\small
\centering
\caption{Robustness checks for direct pass-through to retail prices}
\label{tab:d.2}
\renewcommand{\tabcolsep}{1pt}{
\def\sym#1{\ifmmode^{#1}\else\(^{#1}\)\fi}
\begin{tabular*}{\hsize}{@{\hskip\tabcolsep\extracolsep\fill}l*{6}{c}}
\toprule

& \multicolumn{2}{c}{\centering Alternate bite variable} & \multicolumn{2}{c}{\centering Reverse causality} & \multicolumn{2}{c}{\centering Other} \\
\cline{2-3} \cline{4-5} \cline{6-7}

\addlinespace

&\multicolumn{1}{c}{(1)}&\multicolumn{1}{c}{(2)}&\multicolumn{1}{c}{(3)}&\multicolumn{1}{c}{(4)}&\multicolumn{1}{c}{(5)} &\multicolumn{1}{c}{(6)}  \\
\addlinespace

&\multicolumn{1}{c}{\parbox{1.5cm}{\centering Q4 bite}}&\multicolumn{1}{c}{\parbox{1.5cm}{\centering Com-pliance}}&\multicolumn{1}{c}{\parbox{1.5cm}{\centering No Seattle}}&\multicolumn{1}{c}{\parbox{1.5cm}{\centering No King county}}&\multicolumn{1}{c}{\parbox{1.5cm}{\centering Balanced panel}}&\multicolumn{1}{c}{\parbox{1.5cm}{\centering Alt. weights}} \\
\midrule
\addlinespace
\multicolumn{1}{l}{\parbox{.5cm}{$E_0$}}       &     0.003** & 0.004 &  0.003*** & 0.003*** & 0.003** & 0.004*** \\
    & (0.001) & (0.003) & (0.001) & (0.001) & (0.001) & (0.001) \\

\addlinespace
\multicolumn{1}{l}{\parbox{.5cm}{$E_2$}}       &     0.003** & 0.006** & 0.003** & 0.003*** & 0.004** & 0.004** \\ 
            &   (0.002) & (0.003) & (0.001) & (0.001) & (0.002) & (0.002) \\
\addlinespace
\multicolumn{1}{l}{\parbox{.5cm}{$E_4$}}       &    0.004* & 0.008** & 0.004** & 0.005*** & 0.005** & 0.005** \\
            &   (0.002) & (0.003) & (0.002) & (0.002) & (0.002) & (0.002) \\
\midrule
\multicolumn{1}{l}{\parbox{.5cm}{$\sum \text{Pre}$\\ $\text{-event}$}}
            &  -7.6e-04 & 0.002 & -6.0e-05 & -7.7e-04 & -0.001 & 2.8e-04 \\
            &   (0.001) & (0.002) & (0.001) & (0.001) & (0.001) & (0.001) \\
\midrule
\(N\)       &     14,044 & 13,859 & 13,422 & 12,995 & 13,390 & 14,042 \\
Time FE &        YES          &   YES       &   YES       &       YES  &   YES & YES \\
Controls &       YES          &   YES           &   YES       &         YES &   YES     &   YES    \\

\bottomrule
\end{tabular*}
\begin{minipage}[h]{\textwidth}
\medskip
\small \emph{Notes:} The listed coefficients are the sum of the distributed lag coefficients $E_L$, $L$ months after the minimum wage hikes, relative to the baseline period in $t-2$. The distributed lag coefficients are estimated from equation \ref{eq:2} with establishment-level inflation rate as the dependent variable. All specifications include time fixed effects and county level controls (monthly unemployment rate and average monthly wage). (1) uses Q4 bite in the treatment interaction term $\Delta MW_{j,t-l} \times bite_{k(j),t-l}$, while (2) uses the difference between bite two quarters before and one quarter after the hike. (3)-(4) account for possible endogeneity of Seattle hikes: (3) omits Seattle establishments for event 3 while (4) omits King county establishments for event 3. (5) restricts the panel to establishments that are present at least 10 months for a given event. For (6) the price indexes are constructed with expenditure weights based on the fiscal year starting in July and ending in June of each year. Estimates are unaffected by the inclusion of controls, winsorizing instead of trimming, and not trimming at all (results available on request). Standard errors are clustered at the county level and are shown in parentheses. \sym{*} \(p<0.10\), \sym{**} \(p<0.05\), \sym{***} \(p<0.01\). Data from Washington ESD and Top Shelf Data, March 2018-December 2021.
\end{minipage}

}
\end{table}

\newpage
\subsection{Minimum wage compliance and exempt workers}

\begin{figure}[!htbp]
\caption{Average bite one quarter after the minimum wage hike, 2018-2021}

\centering
	\begin{subfigure}{0.5\textwidth}
	\centering
			\includegraphics[width=\linewidth]{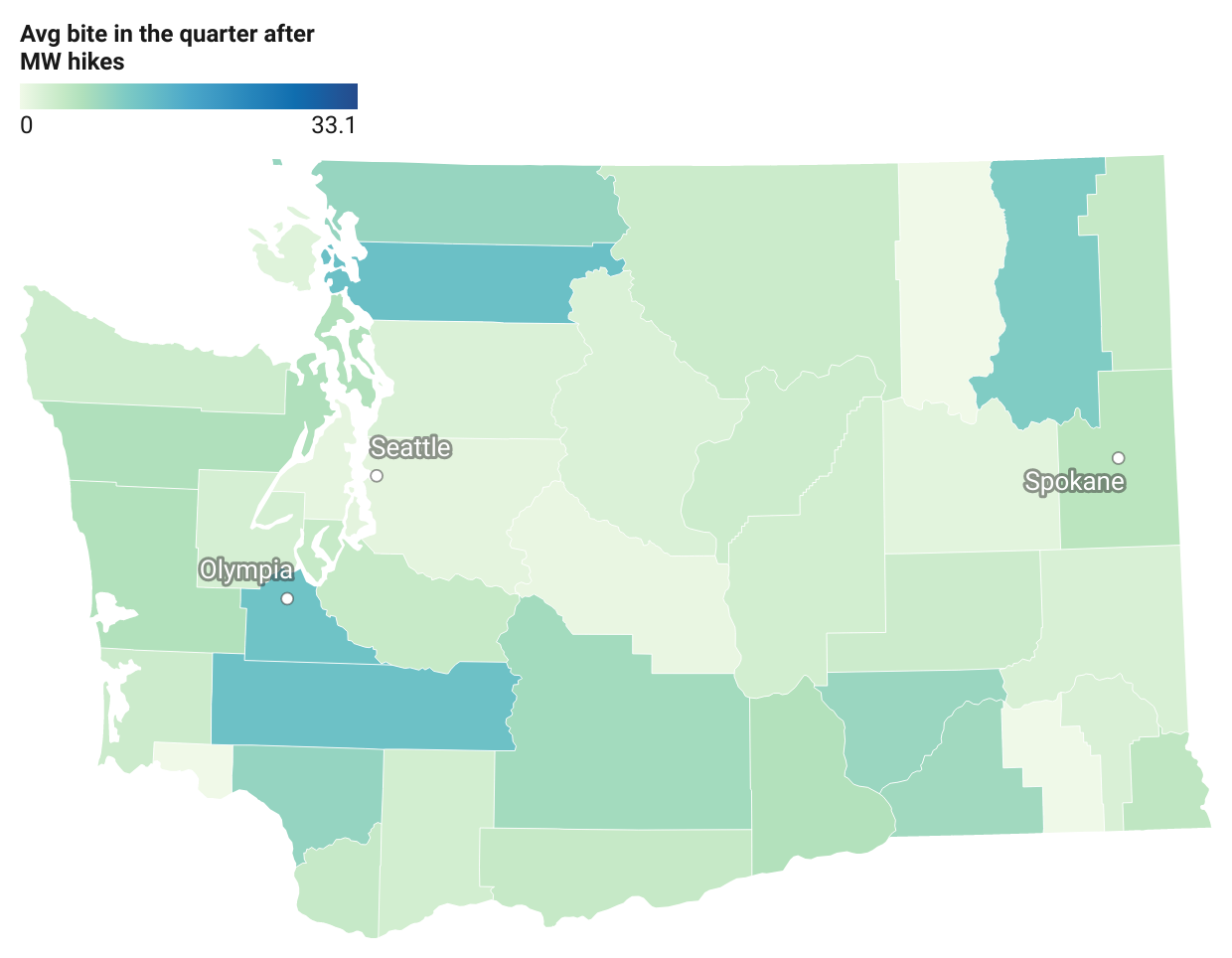}
		\caption{NAICS 111 (Crop production)}
		\label{fig:d.1a}		
	\end{subfigure}\hfil
	\begin{subfigure}{0.5\textwidth}
		\centering
        \includegraphics[width=\linewidth]{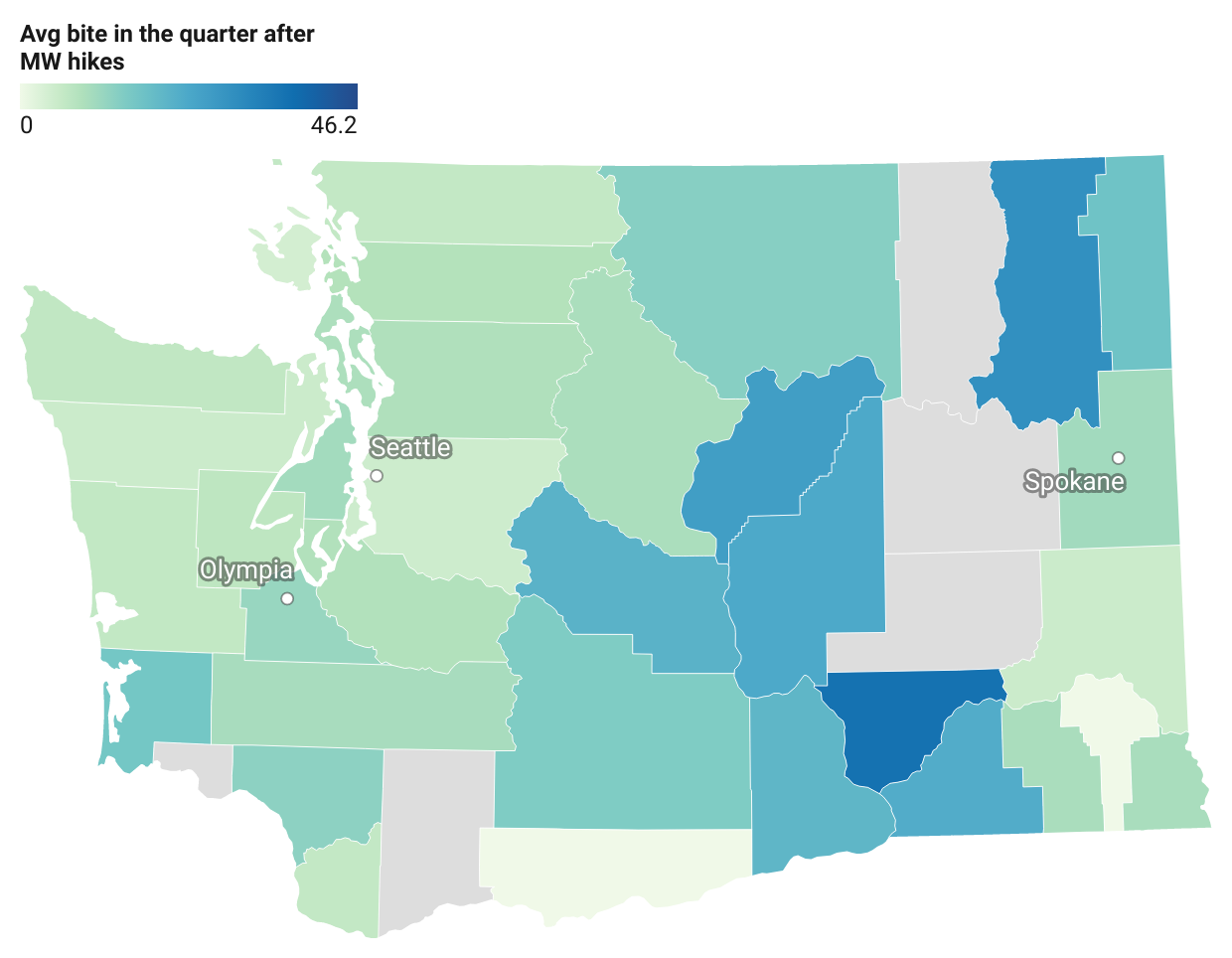}
		\caption{NAICS 453 (Misc. store retailers)}\label{fig:d.1b}
	\end{subfigure}
\label{fig:d.1}
\par \bigskip
\rule{\textwidth}{0.4pt}
\begin{minipage}[h]{\textwidth}
\medskip
\small Notes: The figures show the average bite in the quarter after a minimum wage hike. Cannabis producers belong to NAICS 111 (crop production) while cannabis retailers belong to NAICS 453 (miscellaneous store retailers). Data: Washington ESD, 2019-2021.
\end{minipage}
\end{figure}

When investigating minimum wage effects, it is important to consider the possibility that not all firms or workers comply with minimum wage hikes. If that were the case, then the share of FTE earning below the minimum wage would overestimate the impact of the minimum wage on firm costs, resulting in potentially non-random measurement error in the treatment variable. Luckily, bite lends itself well to measuring minimum wage compliance since bite can also be measured one quarter after the minimum wage hikes. Figure \ref{fig:d.1} shows the average bite one quarter after the minimum wage hikes between 2018-2021. While bite is low for most counties in the crop production industry (panel a), several counties have relatively high bite for miscellaneous store retailers (panel b). At my request, the ESD examined employee-level payroll data at the establishments responsible for these high bite counties and confirmed that the relatively high post-hike bite is a result of minimum wage exemptions rather than non-compliance or data reporting issues.\footnote{The ESD has safeguards in place to flag sub-minimum wages at the employee and firm level. Implausibly low wages are either excluded from the bite variable or the wages are substituted with a previous valid quarter for that employer, adjusting for payroll and inflation.} Under certain circumstances, employers can apply for permission to pay eligible employees less than the state minimum wage.\footnote{Eligibility applies to workers with a disability, employees in job training, student workers in vocational training, student workers employed at an academic institution, and apprentices. Permission must be granted by both the Washington state Department of Labor and Industries and the U.S. Department of Labor.} With the exception of workers with disabilities, however, exempt employees must still be paid 75\% of the state minimum wage (85\% for on-the-job training).\footnote{See \cite{wamw}.} Thus, for exempt employees at the 75\% threshold, the minimum wage hike still corresponds to a wage increase. Moreover, wages slightly above the minimum wage have been shown to be responsive to minimum wage hikes, meaning that the minimum wage hike likely increases wages for exempt employees above the 75\% threshold too.\footnote{For example, \cite{gopalan2021} find that wage increases extend up to \$2.50 above the minimum wage.} 

To summarize, high post-hike bite values in some counties reflect sub-minimum wages paid to exempt employees. Since these employees likely experience a wage increase due to the minimum wage hike, the bite variable in the main analysis (computed two quarters prior to the hike) likely captures true minimum wage exposure. Nevertheless, I test whether removing exempt employees changes the results from the main part of the paper. To do this, I create a new bite variable that is equal to the difference between bite two quarters prior and one quarter after the hike:
\begin{equation}
    \Delta Bite_{k(j)} = Bite_{k(j),Q3,y}-Bite_{k(j),Q1,y+1}
\end{equation}
This effectively nets out non-compliance and exempt employees at the county level. Tables \ref{tab:d.1} and \ref{tab:d.2} show that results are robust to this alternative bite variable.

\newpage
\subsection{Reverse causality}\label{appendix:d.2}

While the overwhelming majority of cities and counties in the sample are subject to exogenous statewide minimum wage hikes, there is one exception: the city of Seattle, located in King county, has a citywide minimum wage that may, under certain circumstances, result in an endogenous bite variable. This section lists the assumptions under which Seattle's minimum wage may be endogenous and reports results that take this potential endogeneity into account.

\begin{figure}[!htbp]
\caption{Seattle citywide minimum wage schedule, 2018-2022}
\centering
	\includegraphics[width=0.4\textwidth]{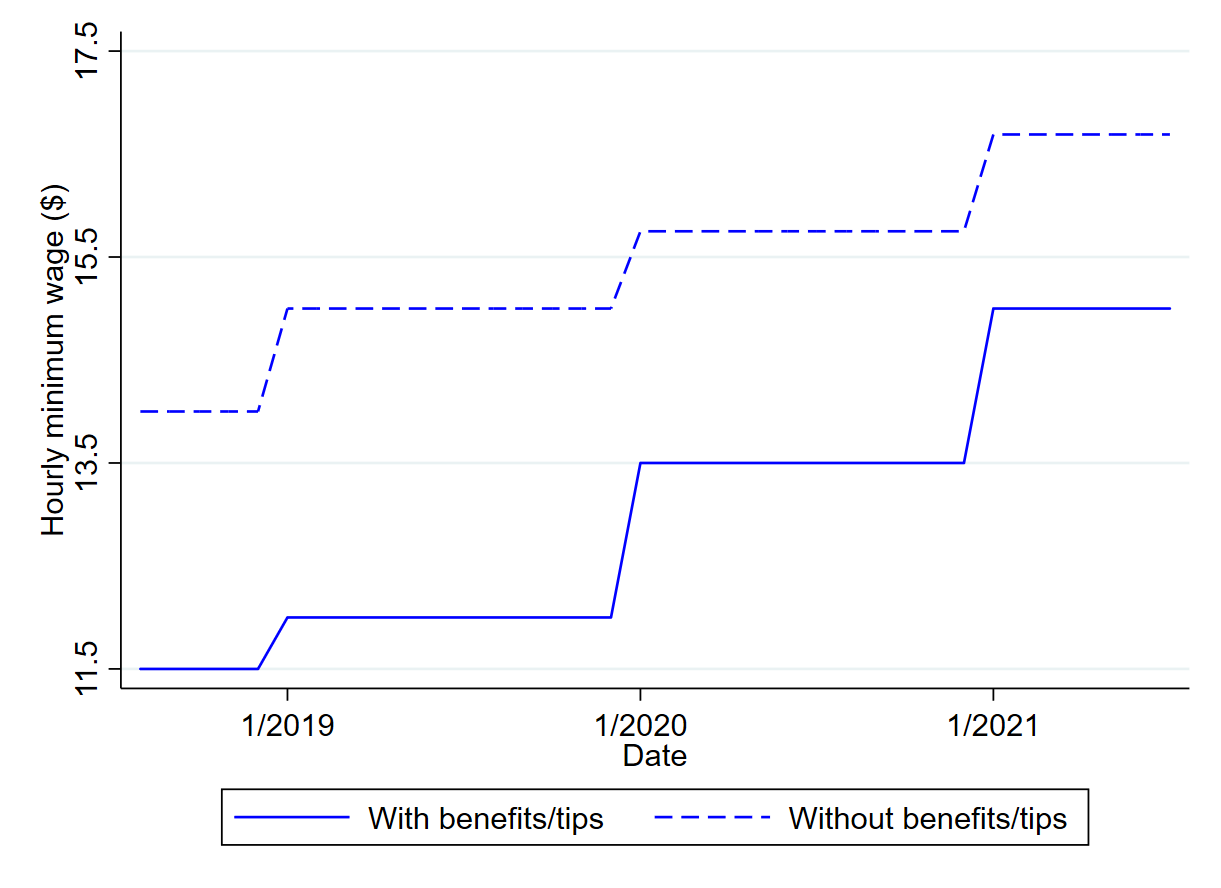}
\label{fig:d.2}
\par \bigskip
\rule{\textwidth}{0.4pt}
\begin{minipage}[h]{\textwidth}
\small \emph{Notes:} The figure shows the schedule for the citywide minimum wage in Seattle. The solid blue line is the minimum wage applicable to employees who receive health benefits or tips, while the dashed line is the minimum wage for employees without benefits or tips. Data source: Washington ESD.
\end{minipage}
\end{figure}

Employment at Seattle establishments is subject to one of two minimum wages, depending on employer contributions to employee medical benefits and whether an employee earns tips.\footnote{Technically, this only applies to small employers (500 or fewer employees), as large firms (over 500 employees) are subject to a separate minimum wage. However, no cannabis firm has more than 500 employees and the average firm size in King county is 10 employees for NAICS 111 and 11.5 employees for NAICS 453 during the sample period. I therefore omit the large firm minimum wage from my analysis.} Employees who receive health benefits or tips are subject to a lower minimum wage than those who do not (Figure \ref{fig:d.2}). For the former, the minimum wage schedule was pre-determined over the sample period, making the hikes contemporaneously exogenous.\footnote{The schedule was determined in 2015.} For the latter group of employees, the hikes for events 1 and 2 (January 1, 2019 and January 1, 2020) were predetermined, while the hike for event 3 was linked to a local CPI. Thus, event 3 may be endogenous for some Seattle establishments, and potentially also for the county Seattle is located in (King county).

Since the treatment variable $\Delta MW_{j,t-l} \times Bite_{k(j),t-l}$ is the product of two parts, it is important to consider how Seattle's minimum wage affects each part in turn. The following assumptions delineate circumstances under which one or both of these parts could be endogenous.

\subsubsection*{Assumption 1 (Exogeneity)}
    \begin{enumerate}
        \item [\textbf{1.A:}] \label{as:1B} All Seattle firms in NAICS 111 (NAICS 453) pay benefits or tips
    \end{enumerate}
Under assumption \textbf{1.A}, $\Delta MW_{j,t-l} \times Bite_{k(j),t-l}$ is contemporaneously exogenous because minimum wage hikes are predetermined for the entire sample period. The results in sections \ref{section:4} and \ref{section:5} are based on assumption 1.

\subsubsection*{Assumption 2 (No spillovers to King county)}
    \begin{enumerate}
        \item [\textbf{2.A:}] No Seattle firms in NAICS 111 (NAICS 453) pay benefits or tips
        \item [\textbf{2.B:}] There are no spillovers from the Seattle minimum wage hike to wages at establishments located outside of Seattle but in King County (applies to event 3 only).
    \end{enumerate}
Under assumption \textbf{2.A}, $\Delta MW_{j,t-l}$ is predetermined (and hence exogenous) for Seattle establishments in events 1 and 2, but it is endogenous for event 3. Thus, Seattle establishments must be dropped from the sample for event 3. Under assumption \textbf{2.B}, Seattle's endogenous hike at event 3 does not affect $Bite_{k(j),t-l(e=3)}$, meaning non-Seattle establishments located in King County can be kept in the sample for that event. $Bite_{k(j),t-l(e=3)}$ will be mismeasured for King County at event 3 which may attenuate estimates.

\subsubsection*{Assumption 3 (Spillovers to King county)}
    \begin{itemize}
        \item [\textbf{3.A:}]\label{as:3A} No Seattle firms in NAICS 111 (NAICS 453) pay benefits or tips
        \item [\textbf{3.B:}]\label{as:3B} There are spillovers from the Seattle minimum wage hike to wages at non-Seattle establishments in King County (applies to event 3 only).
    \end{itemize}
Assumption \textbf{3.A} carries over from \textbf{2.A}, meaning $\Delta MW_{j,t-l}$ is exogenous for Seattle establishments in events 1 and 2 but it is endogenous for event 3. Now however, assumption \textbf{3.B} implies that $Bite_{k(j),t-l(e=3)}$ is also endogenous for event 3, since Seattle's endogenous minimum wage hike spills over to surrounding King county establishments, possibly lowering the King county bite. This means that all King county establishments must be dropped from the sample for event 3.

Table \ref{tab:d.1} reports results from estimating equation \ref{eq:2} under assumptions 2 and 3 for producers. As columns 3 and 4 illustrate, wholesale price effects are very similar to those obtained in the main paper. Table \ref{tab:d.2}  (columns 3 and 4) shows that the same holds for retail price effects. Taken together, these results suggest that reverse causality from Seattle's minimum wage does not drive my main results. 

\begin{table}[!htbp]
\small
\centering
\caption{Robustness checks for indirect retail price effects}
\label{tab:d.3}
\renewcommand{\tabcolsep}{1pt}{
\def\sym#1{\ifmmode^{#1}\else\(^{#1}\)\fi}
\begin{tabular*}{\hsize}{@{\hskip\tabcolsep\extracolsep\fill}l*{7}{c}}
\toprule


\addlinespace

&\multicolumn{3}{c}{Trend-adjusted} &\multicolumn{4}{c}{Unadjusted, establishment FE} \\
            \addlinespace
            \cline{2-4} \cline{5-8}
        \addlinespace
&\multicolumn{1}{c}{(1)}&\multicolumn{1}{c}{(2)}&\multicolumn{1}{c}{(3)}&\multicolumn{1}{c}{(4)}&\multicolumn{1}{c}{(5)} &\multicolumn{1}{c}{(6)} &\multicolumn{1}{c}{(7)} \\
\addlinespace


\midrule
\addlinespace
\multicolumn{1}{l}{\parbox{.5cm}{$E_0$}}       &     0.005* & 0.005* & 0.007*** & 0.005 & 0.005 & 0.006* & 0.005* \\
    &  (0.003) & (0.003) & (0.002) & (0.003) & (0.003) & (0.003) & (0.003) \\

\addlinespace
\multicolumn{1}{l}{\parbox{.5cm}{$E_2$}}       &     0.005 & 0.003 & 0.008* & 0.003 & 0.004 & 0.003 & 0.003   \\ 
            &   (0.005) & (0.005) & (0.004) & (0.006) & (0.006) & (0.006) & (0.005) \\
\addlinespace
\multicolumn{1}{l}{\parbox{.5cm}{$E_4$}}       &    0.005 & 0.002 & 0.007** & -4.72e-04 & 0.004 & 0.003 & -1.95e-04 \\
            &   (0.004) & (0.004) & (0.004) & (0.005) & (0.006) & (0.005) & (0.005)  \\
\midrule
\multicolumn{1}{l}{\parbox{.5cm}{$\sum \text{Pre}$\\ $\text{-event}$}}
            &  1.44e-07 & 1.01e-07 & -3.08e-08 & 0.00427 & 0.000251 & -0.000437 & 0.00470 \\
            &   (0.005) & (0.005) & (0.005) & (0.005) & (0.005) & (0.005) & (0.005) \\
\midrule
\(N\)       &     13,501 & 13,426 & 13,559 & 13,559 & 13,501 & 13,426 & 13,559 \\
Time FE &        YES          &   YES       &   YES                     &   YES                 &         YES &   YES    &   YES \\
Controls &       YES          &   YES       &   YES     &   YES      &      YES   &   YES      &      YES        \\
Trend-adjusted & YES          &   YES       &   YES     &   NO                 &         NO &   NO     &   NO               \\
Establishment trends & NO          &   NO       &   NO     &   YES    &         YES &   YES     &   YES               \\

\bottomrule
\end{tabular*}
\begin{minipage}[h]{\textwidth}
\medskip
\small \emph{Notes:} Dependent variable: monthly establishment-level inflation rate. Listed coefficients are sums $E_L$ of the distributed lag coefficients $\psi_l$, $L$ months after the minimum wage hikes, relative to the baseline period in $t-2$. Standard errors of the sums are clustered at the county level and are shown in parentheses. County level controls are the monthly unemployment rate and average monthly wage. For columns 1-3, establishment-level inflation is trend-adjusted. For columns 4-7, establishment-level inflation is unadjusted but establishment FE are included, which removes the pre-trend to a certain extent. (1) uses leads $t-3$ through $t-1$ to compute average wholesale expenditure shares. (2) uses all pre-treatment leads for average wholesale expenditure shares. (3) uses leads $t-4$ through $t-2$ as in the baseline specification, but jointly weights joint bite: $JB_{r,P,t-l} = \sum_{p = s}^{S} \alpha_{r,p} \Delta MW_{p,t-l} \times Bite_{k(p),t-l}$. (4) uses the baseline specification, with inflation unadjusted and with establishment FE. (5) is the same as (1) but with inflation unadjusted and with establishment FE. (6) is the same as (2) but with the dependent variable unadjusted and with establishment FE. (7) is same as (3) but with the dependent variable unadjusted and with establishment FE. \sym{*} \(p<0.10\), \sym{**} \(p<0.05\), \sym{***} \(p<0.01\). Data from Washington ESD and Top Shelf Data, August 2018-July 2021.
\end{minipage}

}
\end{table}

\subsection{Bite as treatment intensity}

\begin{figure}[!htbp]
\caption{Direct pass-through with bite-only treatment intensity}\label{fig:d.4}
\centering
	\begin{subfigure}{.4\textwidth}
	\centering
		\includegraphics[width=\linewidth]{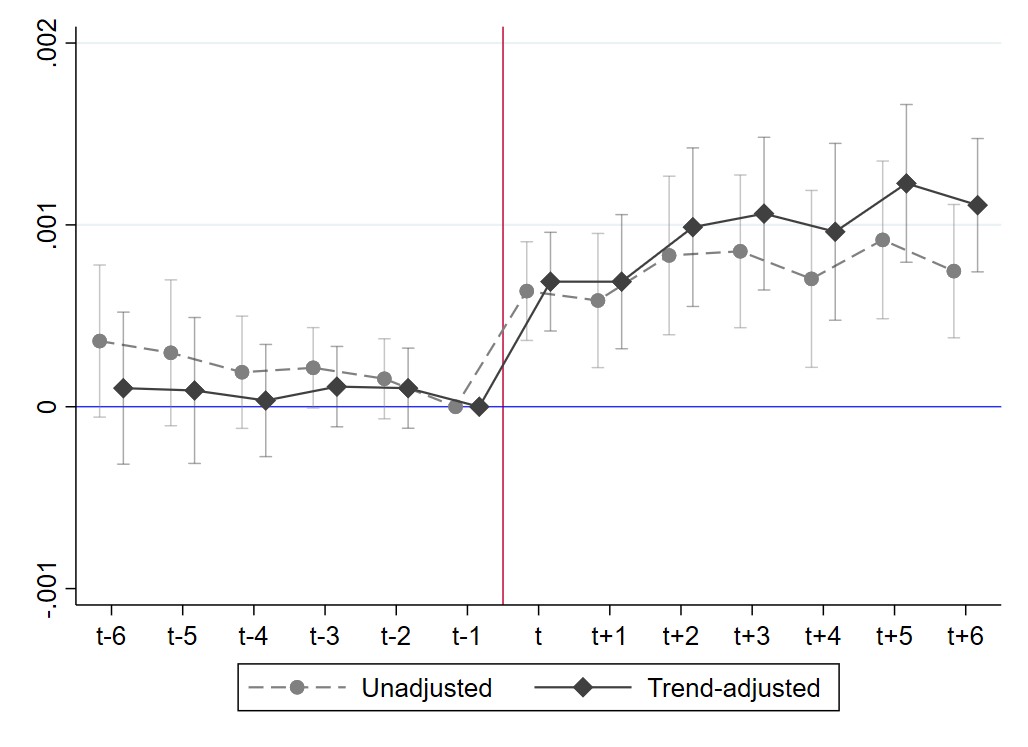}
		\caption{Wholesale prices}
	\end{subfigure}\hfil
	\begin{subfigure}{.4\textwidth}
	\centering
	    \includegraphics[width=\linewidth]{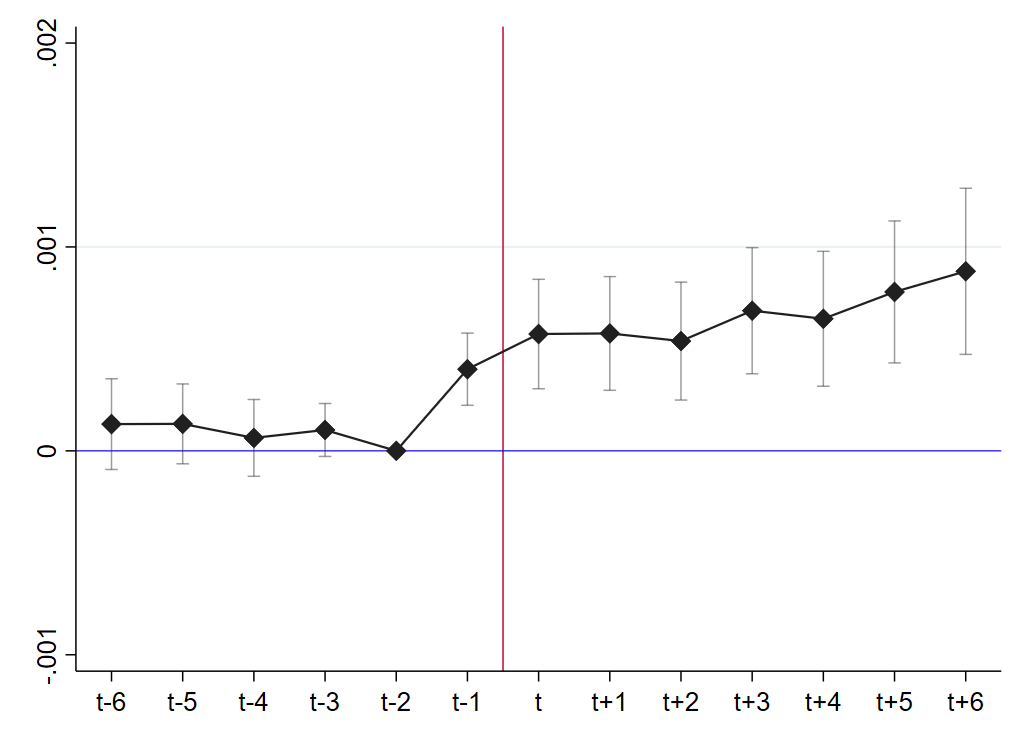}
		\caption{Retail prices}
	\end{subfigure}\hfil
\par \bigskip
\rule{\textwidth}{0.4pt}
\begin{minipage}[h]{\textwidth}
\medskip
\small \emph{Notes:} The figures show cumulative price level effects when the treatment intensity does not include an interaction term for the size of the minimum wage hike. Effects are cumulative relative to the normalized baseline period ($t-1$ for producers, $t-2$ for retailers). Cumulative effects $E_L$ are obtained by summing the distributed lag coefficients to lead or lag $L$ as detailed in section \ref{section:4}. The figures show 90\% confidence intervals of the sums based on SE clustered at the county level. The dependent variable is the establishment-level inflation rate. Estimates are from equation \ref{eq:d.4} with time and county fixed effects, estimated separately for producers and retailers. In panel (a), the dependent variable is adjusted for a bite-specific trend as described in Appendix \ref{appendix:c}. Data source: Top Shelf Data and Washington ESD, August 2018-July 2021.
\end{minipage}
\end{figure}

The main results do not rely on interacting bite with the size of the minimum wage hike. To verify this, I estimate a variation of equation \ref{eq:2}:
\begin{equation}\label{eq:d.4}
   \pi_{j,t} = \sum_{l =-5}^{6} \beta_l Bite_{k(j),t-l} + X_{k(j),t} + \theta_k + \gamma_t + \epsilon_{j,t}.
\end{equation}
Here, the treatment intensity is the minimum wage bite and it is not multiplied with $\Delta MW_{j, t-l}$. Figure \ref{fig:d.4} shows that retail and wholesale price effects follow a very similar time path to those in the main section. Since the treatment intensity variable is defined differently, the coefficients are not directly comparable to those in the main section. However, the relative magnitude of direct pass-through to wholesale and retail prices is the same as in the main part of the paper. Two months after a minimum wage hike, direct pass-through to wholesale prices is approximately twice the size of direct pass-through to retail prices.

\newpage
\subsection{Bite at the detailed industry level (5-digit NAICS)}\label{appendix:d.3}

In this section, I use an alternate bite variable based on more detailed NAICS codes and wage data from the QCEW. In particular, I define bite as the difference between the FTE weekly minimum wage salary and the actual average weekly wage, where the latter is reported by the QCEW on a quarterly basis. This bite variable is similar to that used in other papers on minimum wage effects (see e.g. \cite{renkin2020,leung2021}). I estimate equation \ref{eq:2} with this alternative bite variable in place of the original bite variable in the treatment intensity interaction term $\sum_{l =-5}^{6} \beta_l \Delta MW_{j,t-l} \times Bite_{k(j),t-l}$. However, despite the more granular level of industrial classification, the alternative bite variable carries several disadvantages. First, due to the wage floor imposed by the minimum wage, outliers will pull the mean wage upwards. Thus, a bite variable proportional to the mean wage will likely underestimate true exposure to the minimum wage.\footnote{An alternative would be to use the median wage. Unfortunately, the QCEW does not publish median wages at the detailed industry-by-county level.} Second, while cannabis producer-processors belong to a single three-digit NAICS code (111), they fall under two different four- and five-digit NAICS codes depending on whether they are indoor or outdoor growers (indoor growers belong to NAICS 11141 while outdoor growers belong to NAICS 11199). Producer-processor licenses are based on a three-tier system governing the square footage of plant canopy a producer is permitted to operate. Tiers 1 and 2 permit 2,000 and 10,000 square feet of plant canopy, respectively, and thus largely comprise indoor grow operations \citep{lcbtier1}. Tier 3 producers can operate up to 30,000 square feet of plant canopy, meaning tier 3 comprises more balanced mix of indoor and outdoor grow operations compared to tiers 1 and 2.\footnote{For example, only 10\% of Tier 1 producers grow outdoors \citep{lcbtier1}.} Thus, it is not possible to determine which five-digit NAICS code applies to the majority of tier 3 producers, meaning substantial measurement error will result for tier 3 producers in either case. Therefore, I drop tier 3 producers from the sample and restrict the analysis to tiers 1 and 2 (i.e. indoor growers) and use NAICS 11141 for the bite variable.\footnote{NAICS 11141 corresponds to "Food Crops Grown Under Cover" and includes as a subcategory "Other Food Crops Grown Under Cover, including Marijuana Grown Under Cover" \citep{naics2007}.}

A final disadvantage to the more detailed industry classification is that the QCEW data does not distinguish between full-time and part-time workers, meaning the wage data are not based on FTE. This contrasts to the bite variable in the main specification, which is based on FTE.

\begin{figure}[!htbp]
\caption{Direct pass-through to wholesale prices using 5-digit NAICS bite}
\centering
	\begin{subfigure}{.4\textwidth}
	\centering
		\includegraphics[width=\linewidth]{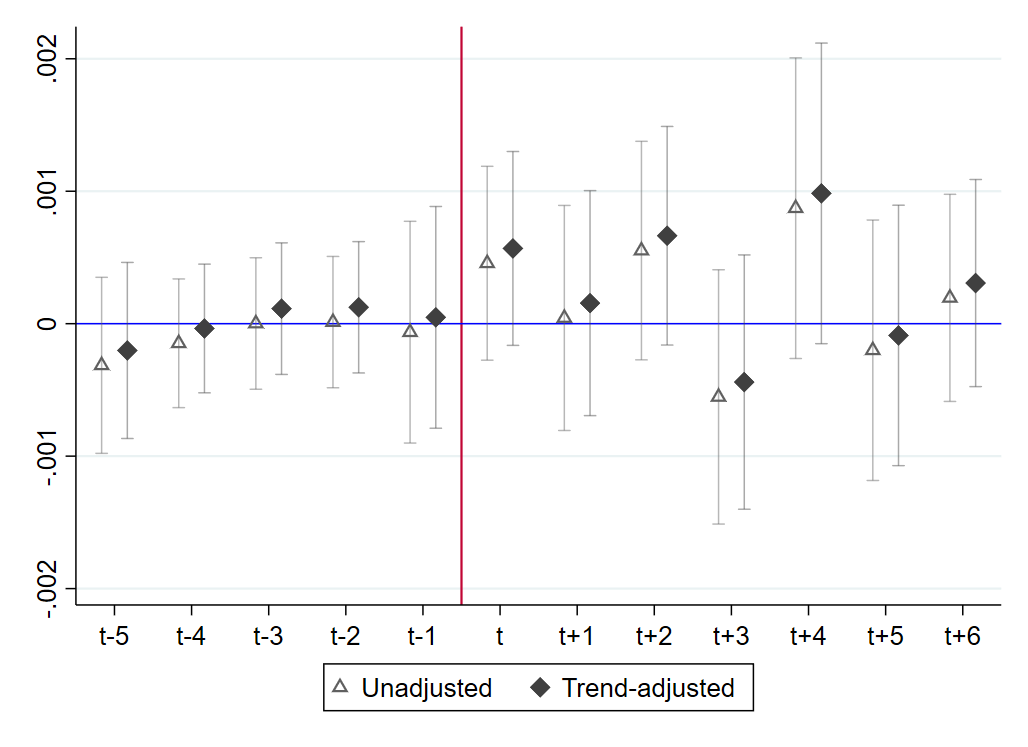}
		\caption{Effect on the inflation rate}
		\label{fig:d.3a1}		
	\end{subfigure}\hfil
	\begin{subfigure}{.4\textwidth}
	\centering
	    \includegraphics[width=\linewidth]{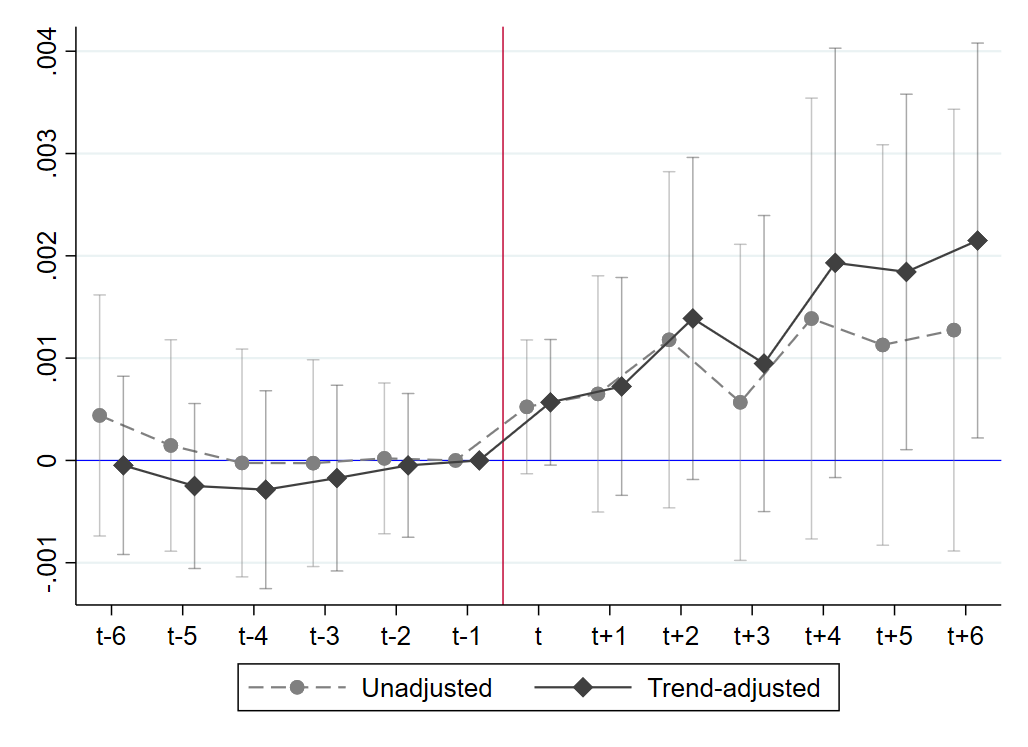}
		\caption{Effect on the price level}
		\label{fig:fig:d.3a2}
	\end{subfigure}\hfil
\label{fig:d.3a}
\par \bigskip
\rule{\textwidth}{0.4pt}
\begin{minipage}[h]{\textwidth}
\medskip
\small \emph{Notes:} The figures show estimates from equation \ref{eq:2} with the bite variable based on NAICS 11141 as described in appendix \ref{appendix:d.3}. Tier 3 producers and producer-processors are omitted from the estimation sample. Equation \ref{eq:2} is estimated with time and county fixed effects. The dependent variable is the establishment-level inflation rate, adjusted for a bite-specific trend as described in section \ref{section:5.1}. The dependent variable is not trimmed. Panel (a) shows the distributed lag coefficients, $\beta_l$, with 90\% confidence intervals based on SE clustered at the county level. Panel (b) depicts cumulative price level effects ($E_L$) relative to the baseline period in $t-1$. Cumulative effects $E_L$ are obtained by summing the distributed lag coefficients to lead or lag $L$ as detailed in section \ref{section:4}. Panel (b) shows 90\% confidence intervals of the sums based on SE clustered at the county level. Data source: Top Shelf Data and Washington ESD, August 2018-July 2021.
\end{minipage}
\end{figure}

\begin{figure}[!htbp]
\caption{Direct pass-through to retail prices using 5-digit NAICS bite}
\centering
	\begin{subfigure}{.4\textwidth}
	\centering
		\includegraphics[width=\linewidth]{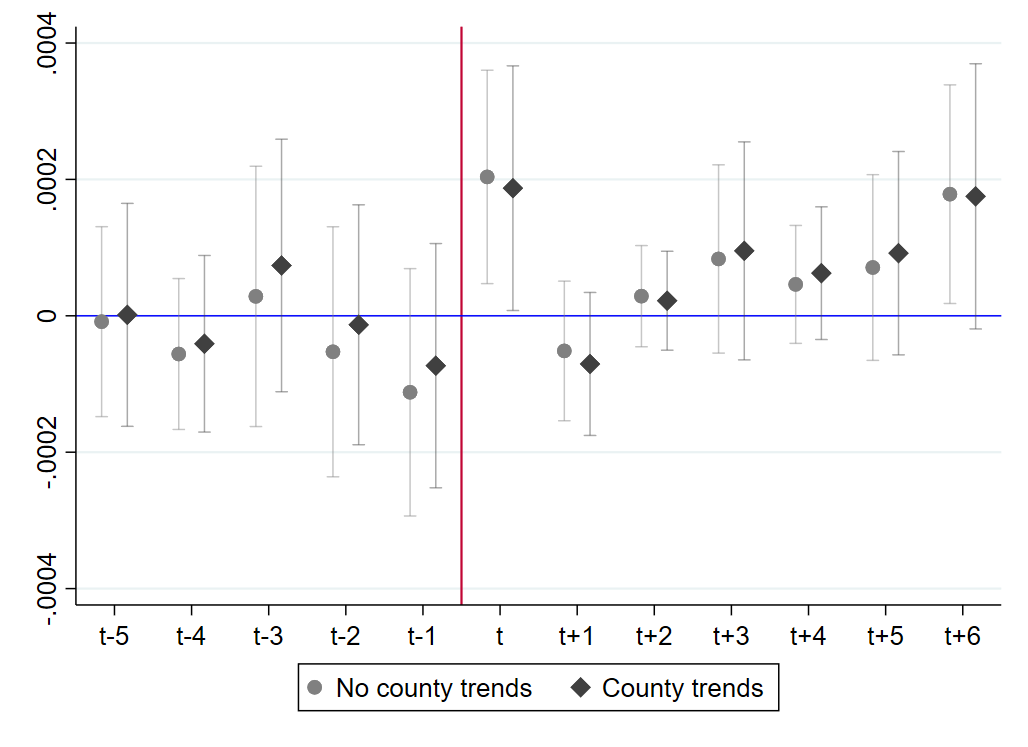}
		\caption{Effect on the inflation rate}
		\label{fig:d.3b1}		
	\end{subfigure}\hfil
	\begin{subfigure}{.4\textwidth}
	\centering
	    \includegraphics[width=\linewidth]{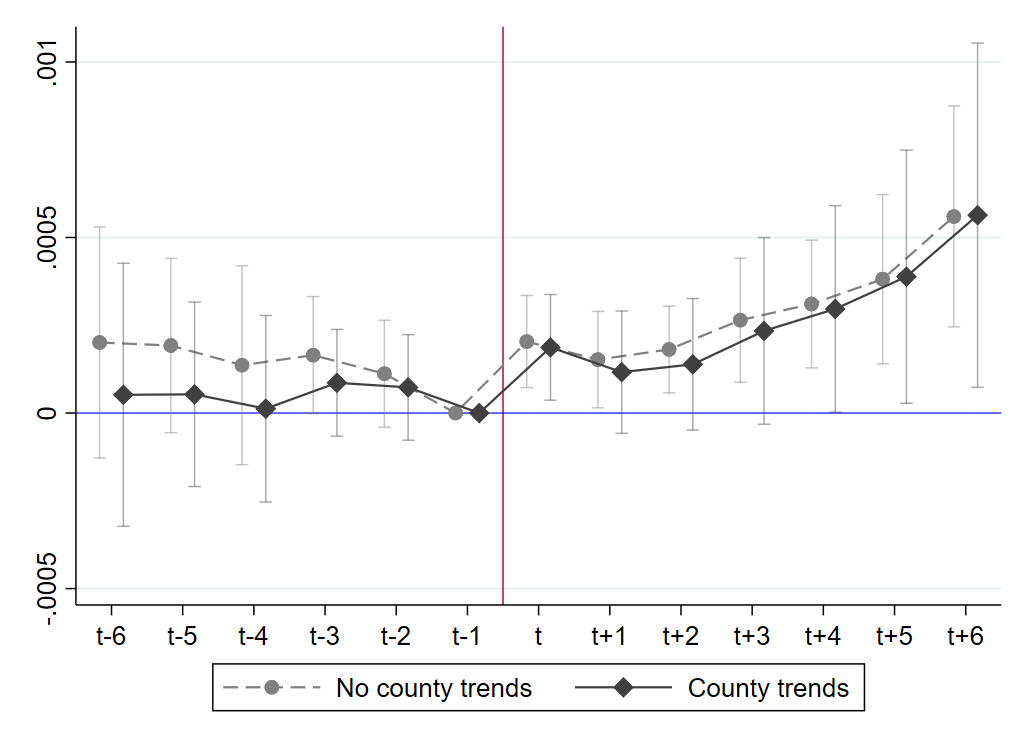}
		\caption{Effect on the price level}
		\label{fig:fig:d.3b2}
	\end{subfigure}\hfil
\label{fig:d.3b}
\par \bigskip
\rule{\textwidth}{0.4pt}
\begin{minipage}[h]{\textwidth}
\medskip
\small \emph{Notes:} The figures show estimates from equation \ref{eq:2} with the bite variable based on NAICS 45399 as described in appendix \ref{appendix:d.3}. Equation \ref{eq:2} is estimated with time and county fixed effects. The dependent variable is the establishment-level inflation rate, which is not trimmed and not adjusted for a bite-specific trend. Panel (a) shows the distributed lag coefficients, $\beta_l$, with 90\% confidence intervals based on SE clustered at the county level. Panel (b) depicts cumulative price level effects ($E_L$) relative to the baseline period in $t-1$. Cumulative effects $E_L$ are obtained by summing the distributed lag coefficients to lead or lag $L$ as detailed in section \ref{section:4}. Panel (b) shows 90\% confidence intervals of the sums based on SE clustered at the county level. Data source: Top Shelf Data and Washington ESD, August 2018-July 2021.
\end{minipage}
\end{figure}

Figures \ref{fig:d.3a} and \ref{fig:d.3b} show sharp inflationary treatment effects at the period of the minimum wage hike for both wholesale and retail cannabis prices, and the effect is statistically significant at the 10\% and 5\% level, respectively.

\newpage
\subsection{Direct retail pass-through with indirect bite}

\begin{figure}[!htbp]
\caption{Direct pass-through with and without indirect bite as a control}
\centering
	\includegraphics[width=0.4\textwidth]{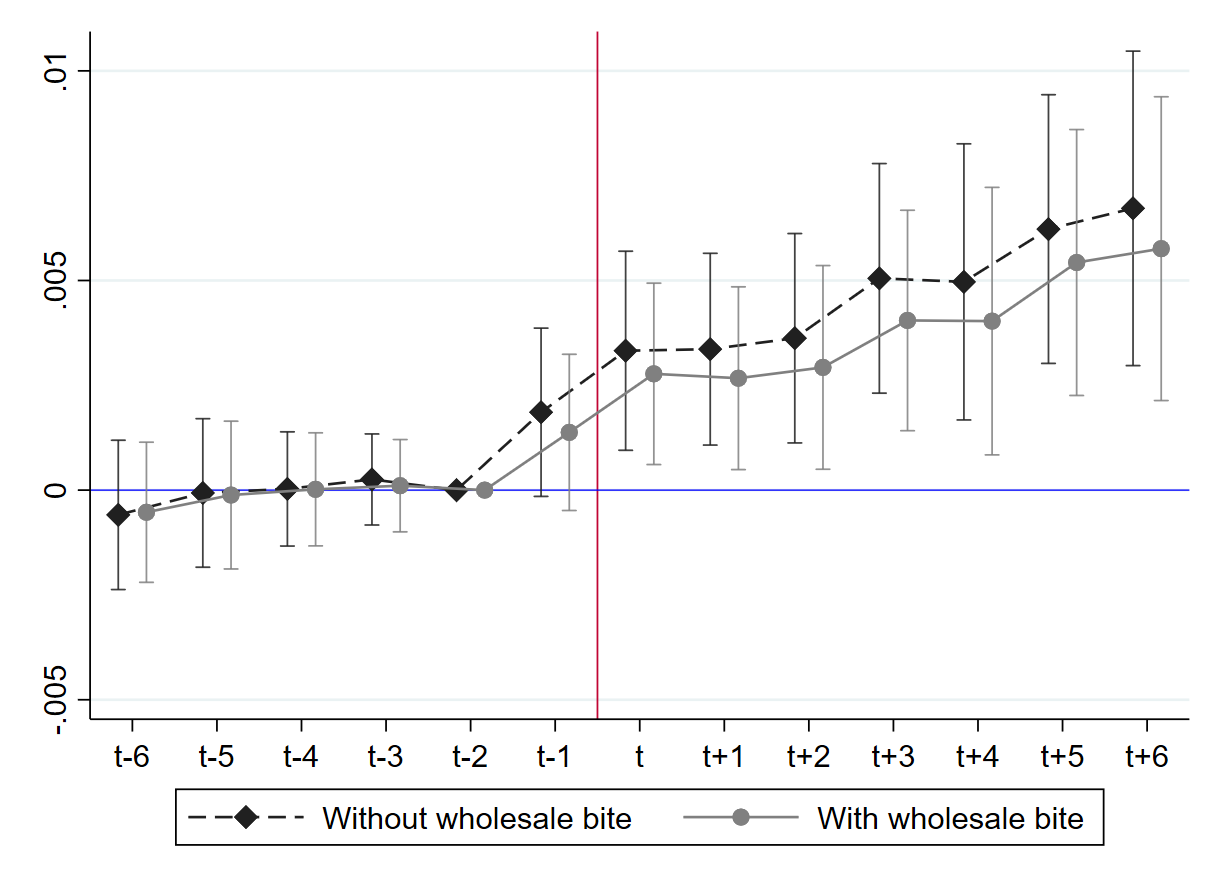}
\label{fig:6.1}
\par \bigskip
\rule{\textwidth}{0.4pt}
\begin{minipage}[h]{\textwidth}
\medskip
\small \emph{Notes:} The figure compares cumulative direct price level effects for retailers when indirect bite is included versus omitted from equation \ref{eq:5}. Both specifications include time fixed effects and county-level controls. The estimated coefficients $\beta_l$ are summed up to cumulative effects $E_L$ relative to the baseline period in $t-2$. The figures show 90\% confidence intervals of the sums based on SE clustered at the county level. Data source: Top Shelf Data and Washington ESD, August 2018-July 2021.
\end{minipage}
\end{figure}

When estimating indirect retail pass-through, one concern is whether the estimates for $\beta_l$---the direct pass-through rate---are affected by the inclusion of indirect bite as an additional variable. If estimates for direct pass-through were to change, this would cast doubt on the main identification strategy and, by extension, the results from section \ref{section:4}. Figure \ref{fig:6.1} compares cumulative direct pass-through from equation \ref{eq:5} with and without indirect bite included. Reassuringly, the figure shows that direct pass-through estimates are unaffected by the inclusion of indirect bite. The pre-treatment period is identical and treatment effects appear in $t-2$ for both specifications. Including indirect bite slightly attenuates the estimates, but the difference is not statistically significant (as evidenced by the overlapping confidence intervals).

\newpage
\subsection{Legislation vs. implementation for event 3}\label{appendix:h.3}
\begin{figure}[!htbp]
\caption{Direct pass-through for event 3}\label{fig:h.4}
\centering
	\begin{subfigure}{.4\textwidth}
	\centering
		\includegraphics[width=\linewidth]{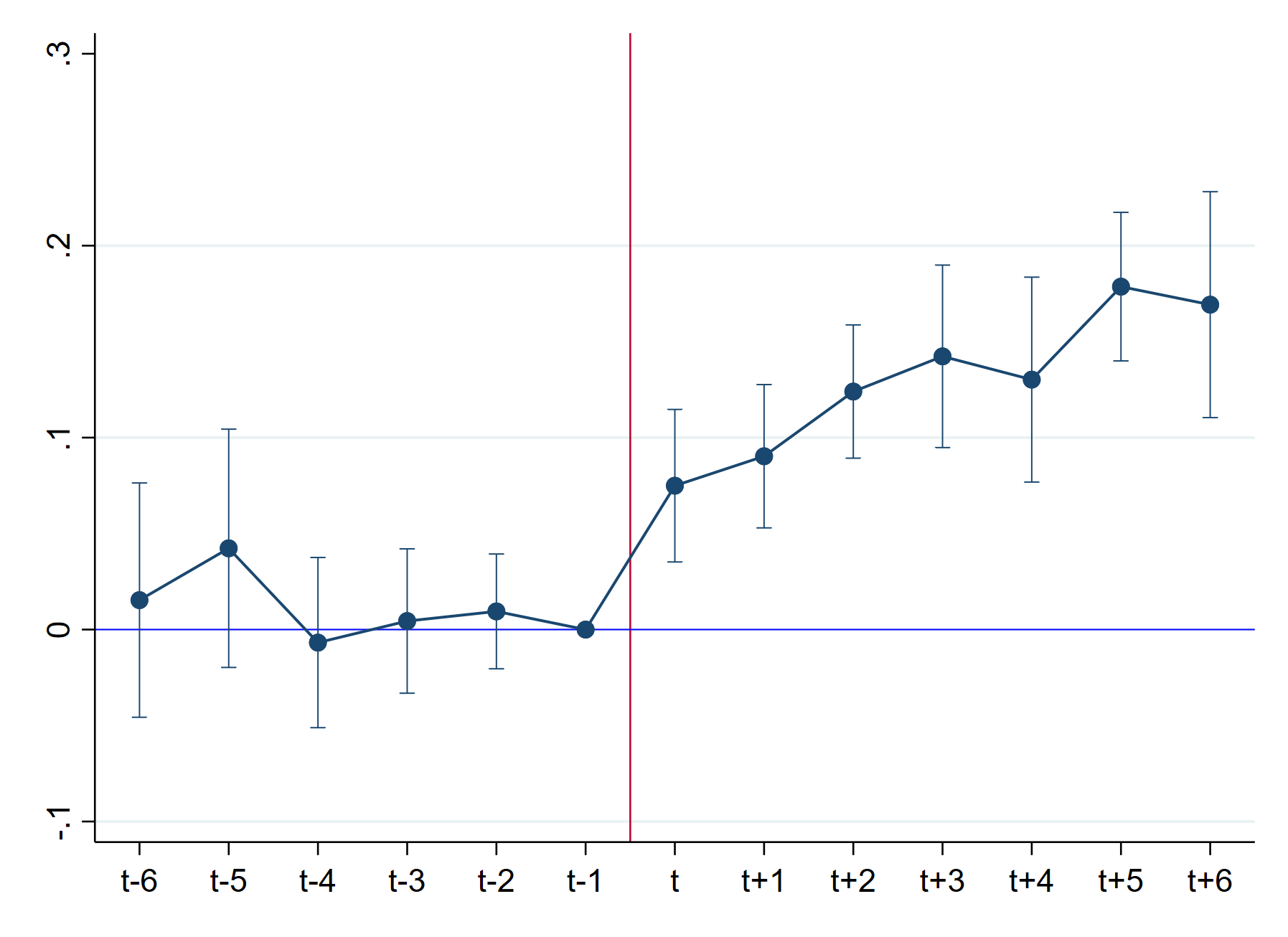}
		\caption{Wholesale price level effects}
	\end{subfigure}\hfil
	\begin{subfigure}{.4\textwidth}
	\centering
	    \includegraphics[width=\linewidth]{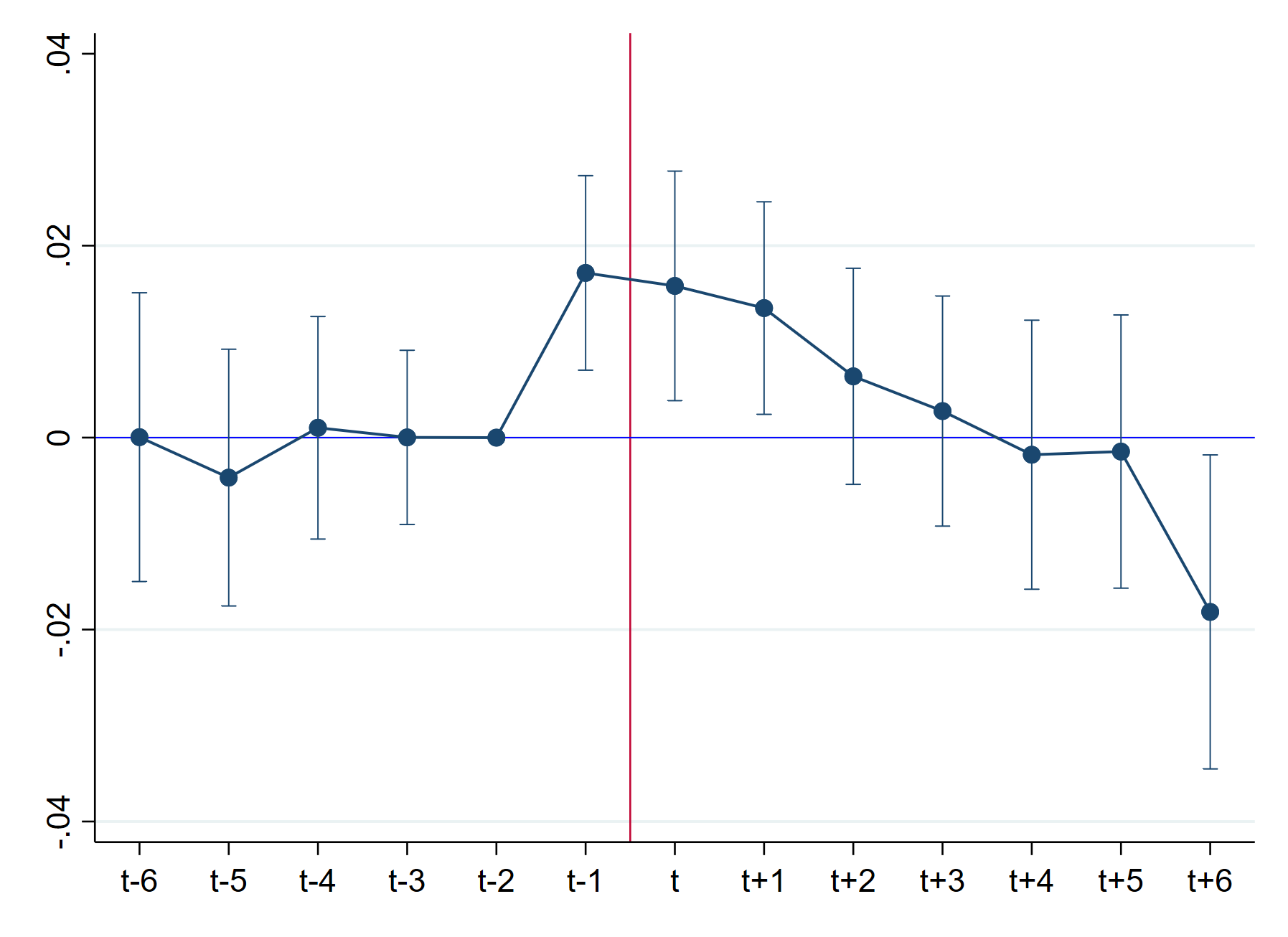}
		\caption{Retail price level effects}
	\end{subfigure}\hfil
\par \bigskip
\rule{\textwidth}{0.4pt}
\begin{minipage}[h]{\textwidth}
\medskip
\small \emph{Notes:} The figures show cumulative price level effects for event 3 only. Effects are cumulative relative to the normalized baseline period ($t-1$ for producers, $t-2$ for retailers). Cumulative effects $E_L$ are obtained by summing the distributed lag coefficients to lead or lag $L$ as detailed in section \ref{section:4}. The figures show 90\% confidence intervals of the sums based on SE clustered at the county level. The dependent variable is the establishment-level inflation rate. For the retail price level regression (panel b), the dependent variable is adjusted for an event-specific bite-specific trend as described in section \ref{section:5.1}. Data source: Top Shelf Data and Washington ESD, August 2020-July 2021.
\end{minipage}
\end{figure}

For event 3, the magnitude of the new minimum wage hike was announced in September 2020, three months before implementation on January 1st, 2021. In this section, I test whether price effects emerge at the time that the hike size was made public ($t-4$) versus when it was implemented ($t$). To do this, I estimate equation \ref{eq:2} for event 3 only. Figure \ref{fig:h.4} shows no evidence of price level effects in $t-4$ for wholesale and retail prices. Instead, treatment effects appear in period $t-1$ for wholesale prices and $t-2$ for retail prices, which is identical to the results in the main part of the paper. Note that, in contrast to the main results, retail price effects for event 3 are undone in later periods and return to zero by $t+4$.

\newpage
\subsection{Region-time FE}\label{appendix:i}

\begin{figure}[!htbp]
\caption{Socioeconomic regions of Washington state}
\centering
	\includegraphics[width=0.6\textwidth]{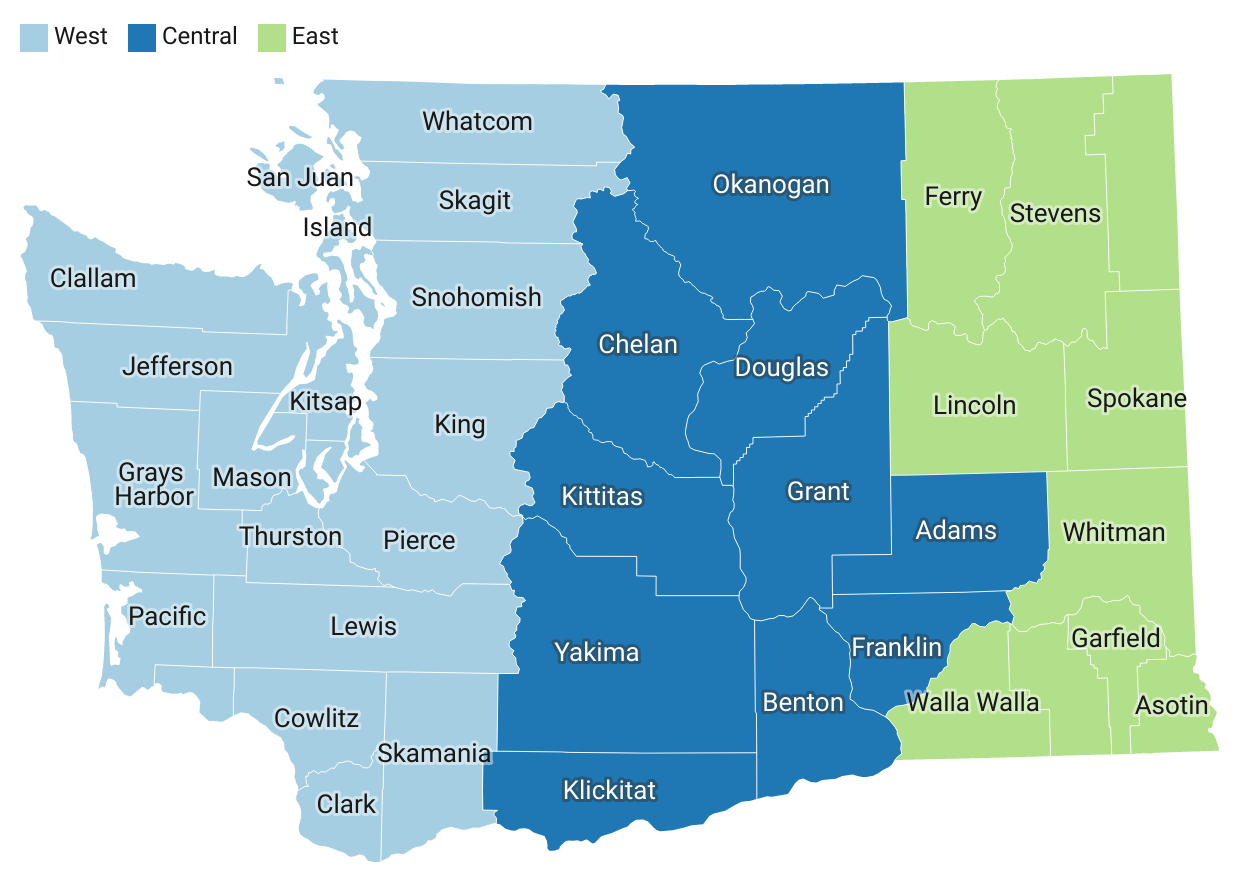}
\label{fig:i.1}
\par \bigskip
\rule{\textwidth}{0.8pt}
\begin{minipage}[h]{\textwidth}
\medskip
\small \emph{Notes:} This figure shows the three major socioeconomic regions in Washington state and the counties within each region.
\end{minipage}
\end{figure}

The Cascade mountain range and a semiarid shrub-steppe create three distinct socioeconomic, political, and topographic regions in Washington state (West, Central, and East), depicted in Figure \ref{fig:i.1}. To account for time-variant unobserved heterogeneity across these regions I estimate equation \ref{eq:2} with region-time FE (i.e. region $\times$ time interaction terms). To ensure the robustness of the region $\times$ time specification, I define regions in two different ways. The first is the specification presented in section \ref{section:5}, which defines three regions (West, Central, East) based on well-established topographic and economic boundaries in Washington state. For the second version, I collapse the latter two regions (Central and East) into a single region (East), creating two distinct regions (West and East). This corresponds to the boundary specified by the Washington state legislature in repeated attempts to create two separate states \cite{hallenberg2017}. Note that this boundary is clearly visible in the average bite depicted in Figure \ref{fig:2a} in the main part of the paper. Results for the two-region version are not shown here; they are virtually identical to the three-region version and available upon request. 

\newpage
\section{Other margins of firm adjustment}\label{appendix:j}
\setcounter{table}{0}
\setcounter{figure}{0}

This section reports results on employment effects and demand feedback discussed in section \ref{section:7}. Employment estimates are from equation \ref{eq:8} while effects on retail cannabis consumption are from equation \ref{eq:9}. Note that column 1 shows a positive effect that is significant at the 10\% level. However, this effect disappears when including region-time fixed effects, which is an important robustness check in the minimum wage-employment context \citep{reich2018}.

\begin{table}[!htbp] 
\centering
\caption{Minimum wage effects on employment, by industry}
\label{tab:j.1}
\renewcommand{\tabcolsep}{1pt}{
\def\sym#1{\ifmmode^{#1}\else\(^{#1}\)\fi}
\begin{tabular*}{\hsize}{@{\hskip\tabcolsep\extracolsep\fill}l*{6}{c}}
 \toprule
 \addlinespace
            &\multicolumn{3}{c}{Indoor crops} &\multicolumn{3}{c}{Retailers} \\
            \addlinespace
            \cline{2-4} \cline{5-7}
            \addlinespace
            &\multicolumn{1}{c}{(1)} &\multicolumn{1}{c}{(2)} & \multicolumn{1}{c}{(3)} &\multicolumn{1}{c}{(4)} &\multicolumn{1}{c}{(5)} &\multicolumn{1}{c}{(6)} \\
\addlinespace
\midrule
\midrule
\addlinespace
$E_0$       &     0.0659* & -0.0108 & -0.00927 & 0.00659 & 0.0264* & 0.0192 \\
            &   (0.0361) & (0.0256) & (0.0290) & (0.0160) & (0.0139) & (0.0160)  \\
\addlinespace
$E_2$       &     0.0853* & 0.00679 & 0.00635 & 0.000232 & 0.0145 & 0.000603 \\
            &   (0.0483) & (0.0294) & (0.0402) & (0.0187) & (0.0136) & (0.0182) \\
\addlinespace
$E_4$       &     0.0739 & -0.0341 & -0.0364 & 0.0254 & 0.0326 & 0.0116 \\
            &   (0.0708) & (0.0320) & (0.0435) & (0.0283) & (0.0240) & (0.0274) \\
\midrule
$\sum \text{Pre-event}$   
            & -0.0113 & -0.0239 & -0.0269 & 0.0255 & 0.0196 & 0.0342 \\ 
            &    (0.0344) & (0.0276) & (0.0411) & (0.0196) & (0.0239) & (0.0228) \\ 
\midrule
\(N\)       &       603 & 603 & 603 & 851 & 851 & 851  \\
Time FE     &        YES          &   YES               &   YES                     &   YES                 &         YES               &   YES  \\
County FE   &  YES    &   NO  & YES &  YES  &  NO  & YES \\

Region-time FE   &  NO    &   YES  & YES &  NO  &  YES  & YES \\

\bottomrule
\end{tabular*}
\begin{minipage}[h]{\textwidth}
\medskip
\small The table reports cumulative employment effects $E_L$ relative to the normalized baseline period in $t-2$, as described in equation \ref{eq:8}. The dependent variable is classified at the 5-digit NAICS level. Indoor crops corresponds to NAICS 11141, and retailers corresponds to NAICS 45399. The treatment intensity is defined as in the previous sections. Standard errors are clustered by county.
\sym{*} \(p<0.10\), \sym{**} \(p<0.05\), \sym{***} \(p<0.01\). Data from the QCEW and Washington ESD, 2018-2021.
\end{minipage}
}

\end{table}

\begin{table}[!htbp] 
\centering
\caption{Effect of minimum wage hikes on retail cannabis consumption}
\label{tab:j.2}
\renewcommand{\tabcolsep}{1pt}{
\def\sym#1{\ifmmode^{#1}\else\(^{#1}\)\fi}
\begin{tabular*}{\hsize}{@{\hskip\tabcolsep\extracolsep\fill}l*{6}{c}}
\toprule
            &\multicolumn{1}{c}{(1)}&\multicolumn{1}{c}{(2)}&\multicolumn{1}{c}{(3)}&\multicolumn{1}{c}{(4)}&\multicolumn{1}{c}{(5)}&\multicolumn{1}{c}{(6)}\\
            
            &\multicolumn{1}{c}{\parbox{1cm}{Baseline}}&\multicolumn{1}{c}{\parbox{1cm}{No controls}}&\multicolumn{1}{c}{\parbox{1cm}{County FE}}&\multicolumn{1}{c}{\parbox{1cm}{Reg.-time FE}}&\multicolumn{1}{c}{\parbox{1cm}{Winsor-ized}}&\multicolumn{1}{c}{\parbox{1cm}{Outliers}} \\
\midrule
\midrule
\addlinespace
$E_0$       &     -0.00905 & -0.0109 & -0.00603 & -0.00180 & -0.00481 & -0.00829 \\
            &   (0.00774) & (0.00811) & (0.00896) & (0.00798) & (0.00828) & (0.00834) \\
\addlinespace
$E_2$       &    -0.00365 & -0.00430 & 0.0000823 & -0.000420 & 0.00242 & -0.000948 \\ 
            &   (0.00955) & (0.00977) & (0.0101) & (0.0118) & (0.0114) & (0.0106)   \\
\addlinespace
$E_4$       &     -0.0137 & -0.00717 & -0.00919 & -0.00704 & -0.00578 & -0.00197 \\
            &   (0.0131) & (0.0130) & (0.0142) & (0.0165) & (0.0106) & (0.00965) \\
\midrule
$\sum \text{Pre-event}$   
            &  -0.0394*** & -0.0397*** & -0.0464*** & -0.0377*** & -0.0448*** & -0.0434*** \\
            &   (0.0114) & (0.0106) & (0.0139) & (0.0123) & (0.0126) & (0.0131)  \\ 
\midrule
\(N\)       &       13922 & 13922 & 13922 & 13922 & 14189 & 14067 \\
Time FE     &        YES          &   YES               &   YES                     &   YES                 &         YES               &   YES  \\
Controls    &        YES          &   NO               &   YES                      &   YES                  &         YES                &   YES \\
County FE   &        NO           &   NO                &   YES                      &   NO                 &         NO                &   NO  \\
Region-time FE    &        NO           &   NO                &   NO                     &   YES                  &         NO                &   NO  \\
Trimmed     &        YES          &   YES               &   YES                     &   YES                 &         NO                &   NO  \\
Winsorized &        NO           &   NO                &   NO                      &   NO                  &   YES                     &   NO  \\

\bottomrule
\end{tabular*}
\begin{minipage}[h]{\textwidth}
\medskip
\small \emph{Notes:} The dependent variable is the natural logarithm of the retail establishment-level quantity index for the "Usable Marijuana" product category, with all product quantities converted into grams. The listed coefficients are the sum of the distributed lag coefficients $E_L$, $L$ months after the minimum wage hikes, relative to the normalized baseline period in $t-2$. The distributed lag coefficients are estimated from equation \ref{eq:8}. The baseline specification in (1) includes as controls the monthly unemployment rate and monthly average wage, both at the county level. (2) excludes county controls. (3) controls for county-level price trends. (4) includes region-time FE but not county FE. (5) uses a winsorized outcome (99\% windsorization). (6) does not trim or winsorize the outcome. Standard errors are clustered at the county level and are shown in parentheses. \sym{*} \(p<0.10\), \sym{**} \(p<0.05\), \sym{***} \(p<0.01\). Data from Washington ESD and Top Shelf Data, July 2018-August 2021.
\end{minipage}

}

\end{table}

\end{document}